\documentclass[12pt]{article}
\usepackage{verbatim} 
\usepackage{jheppub} 

\usepackage[T1]{fontenc} 
\usepackage[utf8]{inputenc}
\usepackage{amsfonts,amssymb,amsmath,amsthm}
\usepackage{graphicx}
\usepackage{slashed}
\usepackage{tikz}

\newcommand{\dd}{\mathrm{d}}

\newcommand{\eq}[2]{\begin{equation} #1 \label{#2} \end{equation}}

\title{Equivalences between 2D dilaton gravities, their asymptotic symmetries, and their holographic duals}

\author[a]{Florian Ecker,}
\emailAdd{fecker@hep.itp.tuwien.ac.at}
\author[a]{Daniel Grumiller,}
\emailAdd{grumil@hep.itp.tuwien.ac.at}
\author[b]{Carlos Valc\'arcel}
\emailAdd{valcarcel.flores@gmail.com}
\author[c]{and Dmitri Vassilevich}
\emailAdd{dvassil@gmail.com}

\affiliation[a]{Institute for Theoretical Physics, TU Wien, Wiedner Hauptstr.~8-10/136, A-1040 Vienna, Austria}
\affiliation[b]{Instituto de F\'isica - Universidade Federal da Bahia, C\^ampus Universit\'ario de Ondina, 40210-340, Salvador, B.A. Brazil}
\affiliation[c]{CMCC-Universidade Federal do ABC, \\ Avenida dos Estados 5001, CEP 09210-580, Santo Andr\'e, S.P. Brazil}

\abstract{
Dilaton gravities in two dimensions can be formulated as particular Poisson sigma models. Target space diffeomorphisms map different models to each other and establish a one-to-one correspondence between their classical solutions. We obtain a general form of such diffeomorphisms in Lorentzian and Euclidean signatures and use them to extend known holographic results, including the Schwarzian action on the asymptotic boundary, from JT to a large class of dilaton gravity models.}

\begin{document}
\maketitle

\section{Introduction}\label{sec:intro}

In recent years, we saw two significant proposals for holographic duals of two-dimensional (2D) gravity models. In one of them \cite{Maldacena:2016hyu, Maldacena:2016upp, Jensen:2016pah}, the holographic dual is the Sachdev--Ye--Kitaev (SYK) model \cite{Sachdev:1992fk, Kitaev:15ur} with a Schwarzian boundary-action as an intermediate step in this construction. The other proposal \cite{Saad:2019lba} posits that the dual theory is a random matrix integral. Both proposals focus on the Jackiw--Teitelboim (JT) model \cite{Jackiw:1984, Teitelboim:1983ux} as bulk theory.

Apart from JT, there is a plethora of dilaton gravity models in 2D.
They describe a metric $g_{\mu\nu}$ on a manifold $\mathcal{M}$ and a scalar field $X$, called the dilaton. These models have no local physical degrees of freedom but can allow boundary excitations. Thus, they are tailor-made for holography.

The second-order bulk action of a large class of 2D dilaton gravity models, 
\begin{equation}\label{D08}
 I_{\mathrm{2nd}}=\pm \frac{k}{4\pi}\int\mathrm{d}^{2}x\;\sqrt{ |g|}\big(XR-U(X)g^{\mu\nu}(\partial_\mu X)(\partial_\nu X)-2V(X)\big) 
\end{equation}
contains the curvature scalar $R$ of the metric $g_{\mu\nu}$, the kinetic potential $U$, and the dilaton potential $V$. The overall factor is the gravitational coupling constant $k=1/(4G)$, where $G$ is the 2D Newton constant. The upper (positive) sign in front of the action \eqref{D08} corresponds to Lorentzian signature, while the lower (negative) sign is used for Euclidean signature.  For the JT model (with unit AdS radius), these potentials are given by
\begin{equation}
    U_{\textrm{\tiny JT}}(X)=0\qquad \qquad  V_{\textrm{\tiny JT}}(X)=-X\,.\label{JTpot}
\end{equation}
For a review of dilaton gravity theories in 2D we refer the reader to \cite{Grumiller:2002nm}.

Dilaton gravities in 2D can be formulated as (in general non-linear) gauge theories \cite{Ikeda:1993fh}, known as Poisson sigma models (PSMs) \cite{Schaller:1994es}. In these PSMs, the (three-dimensional) target space is equipped with a degenerate Poisson structure. PSMs are rigid, in the sense that the most general consistent deformation of a PSM is another PSM with the same dimension of the target space \cite{Izawa:1999ib}. (To be interpretable as a Lorentzian gravity theory, the Poisson tensor has to be compatible with Lorentz boosts,  see e.g.~\cite{Grumiller:2021cwg}.) 

The PSM perspective shows that all 2D dilaton gravity models are related to each other by consistent deformations. The broad goal of the present work is to exploit this general perspective for holographic applications.

The technical key idea that we employ is straightforward: Diffeomorphisms of the target space map one PSM model to another. In particular, we may relate classes of models to JT gravity \eqref{JTpot} in this way. Since diffeomorphisms are changes of variables, at first glance any two models related by a target space diffeomorphism should be classically equivalent to each other. However, as we shall elaborate, we keep fixed a  map between PSM and Cartan variables, so that the action of target space diffeomorphisms is non-trivial from the 2D dilaton gravity perspective.

The sharp goal of our paper is to relate Lorentzian and Euclidean JT gravity to other 2D dilaton gravity models by virtue of target space diffeomorphisms, such that we can make statements about the asymptotic symmetries and the boundary actions for these models. In this way, known holographic aspects of JT gravity can be applied to new models. So in simple terms, our procedure provides a shortcut to holography. We follow the approach suggested in \cite{Valcarcel:2022zqm} and extend the results of that paper in an essential way.

This paper is organized as follows. In section \ref{sec:dil}, we review the PSM formulation of 2D dilaton gravity and apply it specifically to JT gravity, where we also recall the analysis of asymptotic symmetries and boundary charges. In section \ref{sec:td}, we discuss target space diffeomorphisms, first in full generality and then restricted to a useful class that preserves the Lorentz structure. In section \ref{sec:U0}, we apply the target space diffeomorphisms to map JT gravity to other models without kinetic potential, with the conformally transformed CGHS model \cite{Callan:1992rs} as an example. In sections \ref{sec:Edil}, we transit to Euclidean signature, again reviewing some basic aspects of the JT model, including the Schwarzian action.  In section \ref{sec:Wick}, we map between Lorentzian and Euclidean models using target space diffeomorphisms and compare them with Wick rotations. In section \ref{sec:Eucl_td}, we apply target space diffeomorphisms to the Euclidean case. In section \ref{sec:final} we conclude.

\section{2D dilaton gravities and Poisson sigma models}\label{sec:dil}

We start our analysis focussing on Lorentzian signature. The dilaton gravity action \eqref{D08} has an equivalent first-order form obtained by introducing zweibein one-forms $e^{\pm}$, with the metric given by
\eq{
g_{\mu\nu} = e^a_\mu e^b_\nu\,\eta_{ab}
}{eq:metric}
where $\eta_{\pm\pm}=0$ and $\eta_{\pm\mp}=1$, and a Lorentz connection one-form $\omega$, together with two auxiliary fields $X^{\pm}$ generating the torsion constraints,
\begin{equation}
I_{\mathrm{1st}}= \frac{k}{2\pi}\int_{\mathcal{M}}\big( X\dd \omega +X^+(\dd -\omega )\wedge e^- +X^-(\dd +\omega )\wedge e^+ +\mathcal{V}\, e^-\wedge e^+  \big)\,. \label{1st}
\end{equation}
The quantity $\mathcal{V}=-U(X)X^{+}X^{-}+V(X)$ contains the kinetic and the dilaton potential. The EOM descending from the action \eqref{1st},
\begin{align}
\mathrm{d}X-X^{+}e^{-}+X^{-}e^{+}	&=	0\label{D03a}\\
\mathrm{d}X^{\pm}\pm X^{\pm}\omega\pm\mathcal{V}e^{\pm}	&= 0\label{D03b}\\
\mathrm{d}\omega+\frac{\partial \mathcal{V}}{\partial X}\, e^{-}\wedge e^{+} &= 0\label{D03c}\\
\left(\mathrm{d}\pm\omega\right)\wedge e^{\pm}+\frac{\partial \mathcal{V}}{\partial {X^\mp}}\,e^{-}\wedge e^{+} &=0    \label{D03d}\end{align}
are first order in derivatives.

Dilaton gravities in 2D are invariant under local Lorentz transformations parametrized by $\sigma$ and under world-sheet diffeomorphisms parametrized by $\xi^\mu$,
\begin{align}
    \delta_{\xi,\sigma}\omega_{\mu}&=\xi^{\nu}\partial_{\nu}\omega_{\mu}+\omega_{\nu}\partial_{\mu}\xi^{\nu}-\partial_{\mu}\sigma & \delta_{\xi,\sigma}X&=\xi^{\mu}\partial_{\mu}X \label{grag1} \\
    \delta_{\xi,\sigma}e_{\mu}^{\pm}&=\xi^{\nu}\partial_{\nu}e_{\mu}^{\pm}+e_{\nu}^{\pm}\partial_{\mu}\xi^{\nu}\pm e_{\mu}^{\pm}\sigma & \delta_{\xi,\sigma}X^{\pm}&=\xi^{\mu}\partial_{\mu}X^{\pm}\pm X^{\pm}\sigma ~. \label{grag2}
\end{align}
All of these models are integrable, i.e., they admit solutions in closed form for any choice of the potentials $U$ and $V$. First, we identify a Casimir function
\begin{equation} \label{D04}
\mathcal{C}\equiv e^{Q\left(X\right)}X^{+}X^{-}+w\left(X\right)=e^{Q\left(X\right)}Y+w\left(X\right) 
\end{equation}
where $Y\equiv X^{+}X^{-}$ and 
\begin{equation}\label{D05}
Q\left(X\right)=\int^{X}\mathrm{d}y\;U\left(y\right)\qquad \qquad w\left(X\right)=-\int^{X}\mathrm{d}y\;e^{Q\left(y\right)}V\left(y\right) ~.
\end{equation}
The Casimir function is absolutely conserved on-shell,
\begin{equation}
\dd\mathcal{C}=0\label{dC} ~.
\end{equation}
By using $\mathcal{C}$ as an integration constant and taking $X$ as one of the coordinates one may obtain the metric
\begin{equation}\label{D06}
\mathrm{d}s^{2}=-2e^{Q}\mathrm{d}u\,\mathrm{d}X+2e^{Q}\left[\mathcal{C}-w\left(X\right)\right]\mathrm{d}u^{2}.    
\end{equation}
At the Killing horizons $X=X_h$ the $\dd u^2$ part of the metric vanishes, $w(X_h)=\mathcal{C}$. 

Using the map
\begin{equation}\label{D01}
X^{I}=\left(X,X^{+},X^{-}\right)\qquad \qquad A_{I}=\left(\omega,e^{-},e^{+}\right) 
\end{equation}
between PSM and Cartan variables, and the Poisson tensor
\begin{equation}\label{D02}
P^{X\pm}=\mp X^{\pm}\qquad \qquad P^{+-}=\mathcal{V}  
\end{equation}
one can relate the action \eqref{1st} to the PSM action,
\begin{equation}\label{PSM02}
I\mathrm{_{PSM}}=I_{\mathrm{1st}}-\frac{k}{2\pi}\int_{\mathcal{M}}\mathrm{d}\big(X^{I}A_{I}\big)  \end{equation}
which reads
\begin{equation}\label{PSM01}
I\mathrm{_{PSM}}=\frac{k}{2\pi}\int_{\mathcal{M}}\big(A_{I}\wedge\mathrm{d}X^{I}+\frac{1}{2}P^{IJ}A_{I}\wedge A_{J}\big)    ~.
\end{equation}
The PSM can be formulated with an arbitrary Poisson manifold $\mathcal{P}$ being the 
target space. The fields $X^I$ are coordinates on $\mathcal{P}$ and scalar fields on $\mathcal{M}$,  $A_I$ are components of one-forms\footnote{%
To make this manifest in \eqref{PSM01} we added the boundary term in \eqref{PSM02}.
} on $\mathcal{P}$ and one-forms on $\mathcal{M}$, and $P^{IJ}$ is a Poisson tensor satisfying the nonlinear Jacobi identities
\begin{equation}
P^{IL}\partial_L P^{JK}+P^{JL}\partial_L P^{KI}+P^{KL}\partial_L P^{IJ} =0 ~. \label{Jac}
\end{equation}
One can easily check that for $P^{IJ}$ defined in \eqref{D02} this identity is indeed satisfied. In this formalism one can write the EOM \eqref{D03a}-\eqref{D03d} compactly as 
 \begin{equation}\label{eq:PSM_EOMs}
     \dd X^I+P^{IJ}A_J=0 \qquad \qquad \dd A_I+\partial _IP^{JK}A_J\wedge A_K=0 ~.
\end{equation} 
The Poisson tensor $P^{IJ}$ defines a Poisson bracket of any two smooth functions $F(X)$ and $G(X)$ on $\mathcal{P}$ as
\begin{equation}
\{ F, G\}=P^{IJ}\partial_J F \partial_J G ~.\label{Pbr}
\end{equation}
The Casimir function $\mathcal{C}$ has a vanishing Poisson bracket with any function,
\begin{equation}
\{ \mathcal{C},F\} =0 ~.\label{Cbra}
\end{equation}
Another useful equation involving $\mathcal{C}$ is
\begin{equation}\label{T04}
\left(\partial_X+\mathcal{V}\,\partial_Y\right)\mathcal{C}=0   
\end{equation}
where we introduced $Y:=X^+X^-$. In the PSM formulation of dilaton gravity the gauge transformations \eqref{grag1}-\eqref{grag2} are encoded in a certain way which will be made precise below.

Our intention is to perform target space diffeomorphisms on the PSM side while keeping the map \eqref{D01} fixed. To still have a dilaton gravity interpretation available we will need to take care that in the transformed PSM the Poisson tensor still has the form defined in \eqref{D02}, i.e., the transformed PSM needs to be in the image of the map \eqref{D01}. It follows that from the PSM perspective, a target space diffeomorphism is merely a change of variables whereas the physical content is changed on the dilaton gravity side. For example, the world-sheet metric defined by \eqref{eq:metric} is not invariant under target space diffeomorphisms. We will also explore the possibility to map a part of the target space of one model to (a part of) the target space in another one. Physically, it is typically justified to restrict the dilaton to non-negative values, $X\geq 0$.

\subsection{Structure of PSM gauge transformations}\label{sec:PSM_gauge_trafos}

Before describing the PSM symmetries in more detail, let us take a step back and summarize some important general concepts for gauge theories \cite{Henneaux:1992}. Consider a gauge theory with some field content $\phi _i$. Gauge transformations are given by expressions of the form
\begin{align}\label{eq:gen_gaugetrafo}
    \delta _\epsilon \phi _i=R^\alpha _i\epsilon _\alpha  
\end{align}
with arbitrary spacetime dependent parameters $\epsilon_\alpha $ and in general field-dependent coefficients $R^\alpha _i$. We use $\alpha $ not necessarily as an element of a finite index set but make it include a continuous part as well if the gauge transformations depend on derivatives of the gauge parameters. The theory is described by an action $I[\phi ^i]$ which is gauge invariant up to boundary terms, i.e.
\begin{align}
    \frac{\delta I}{\delta \phi _i}\,\delta _\epsilon \phi _i =0 ~.
\end{align}
If we use \eqref{eq:gen_gaugetrafo} and vary $I$ with respect to $\epsilon _\alpha $ we find the Noether identities 
\begin{align}
    N^\alpha =\frac{\delta I}{\delta \phi _i}R^\alpha _i =0 
\end{align}
which are a manifestation of the gauge redundancy present in the theory: They make the equations of motion (EOM) dependent on each other and thus lead to arbitrary parameters appearing in the solutions. While the set of all (infinitesimal) gauge transformations always forms a Lie algebra, one usually works with a minimal subset, just large enough to exhaust all the possible Noether identities. To illustrate this, consider another gauge transformation for the same theory,
\eq{
    \delta_\eta \phi_i=R^\alpha _iA_\alpha^\beta \eta_\beta 
}{eq:alt_gen_gaugetrafo}
with some field-dependent matrix $A_\alpha ^\beta $. The action is still invariant under these new transformations but it can be checked easily that there are no new Noether identities generated. So, to exhaust the Noether identities one could either take \eqref{eq:gen_gaugetrafo} or \eqref{eq:alt_gen_gaugetrafo} but does not need both. 

This defines a \textit{generating set} of gauge transformations as a minimal set containing all the information about the Noether identities. Let us assume that we have such a generating set $R^\alpha _i$. Then any gauge transformation $\delta _\epsilon \phi _i$ can be written as
\begin{equation}
    \delta_\epsilon \phi_i=R^\alpha _i \Bar{\epsilon} _\alpha (\epsilon )+\mu _{ij}\frac{\delta I}{\delta \phi _j} \qquad \qquad \mu _{ij}=(-1)^{\mathrm{deg(\phi _i)deg(\phi _j)}+1} \mu _{ji} 
\end{equation}
for some possibly field dependent $\Bar{\epsilon }_\alpha $ and $\mu _{ij}$. The symmetry of $\mu _{ij}$ depends on the form degree of the respective fields $\phi _i$ denoted by $\mathrm{deg}(\phi_i)$. It encodes trivial gauge transformations, i.e., gauge transformations that vanish on-shell and that are not generated by any constraints. It can be shown that they form an ideal $\mathcal{N}$ in the algebra of all gauge transformations and one often only works with the reduced algebra, having identified by elements of $\mathcal{N}$. If one is just interested in on-shell descriptions of gauge symmetries this is a convenient step but for our purposes, it is crucial to work with the full algebra. As the commutator of two elements of the generating set has to be again a gauge transformation we must have 
\begin{align}
    [\delta _\epsilon ,\delta _\eta ]\phi _i&=\Big(R^\alpha _j \frac{\delta R^\beta _i}{\delta \phi _j}-R^\beta _j\frac{\delta R^\alpha _i}{\delta \phi _j}\Big)\epsilon _\alpha \eta _\beta +R^\gamma _i\delta _\epsilon \eta _\gamma -R^\gamma _i\delta _\eta \epsilon _\gamma \label{eq:var1}\\
    &\overset{!}{=}\Big(C^{\alpha \beta }{}_\gamma \,\epsilon _\alpha \eta _\beta +\delta _\epsilon \eta _\gamma -\delta _\eta \epsilon _\gamma \Big)R^\gamma _i+M_{ij}^{\alpha \beta }\frac{\delta I}{\delta \phi _j}\epsilon _\alpha \eta _\beta \\
    &=:\delta _{[\epsilon ,\eta ]^\ast }\phi _i+M_{ij}^{\alpha \beta }\frac{\delta I}{\delta \phi _j}\epsilon _\alpha \eta _\beta \label{eq:var3}
\end{align}
where
\begin{align}
    M^{\alpha \beta }_{ij}=(-1)^{\mathrm{deg(\phi _i)deg(\phi _j)}+1} M^{\alpha \beta }_{ji}~
\end{align}
with the same sign choice as above and $[\cdot ,\cdot ]^\ast $ denoting a Lie-bracket on the space of gauge parameters. Like previously considered in \cite{Barnich:2010xq}, we additionally took into account a possible field-dependence of the gauge parameters, which is frequently encountered for theories on manifolds with (asymptotic) boundary. Equations \eqref{eq:var1}-\eqref{eq:var3} then allow an on-shell interpretation of the space of gauge parameters $A$ as a Lie algebroid $A\to \mathcal{F}$ over field space $\mathcal{F}$ with the bracket between its sections given by 
\begin{align}
    [\epsilon ,\eta ]^\ast _\alpha =C^{\beta \gamma }{}_\alpha \,\epsilon _\beta \eta _\gamma +\delta _\epsilon \eta _\alpha -\delta _\eta \epsilon _\alpha 
\end{align}
and an anchor $\epsilon \mapsto \delta _\epsilon $. Following \cite{Barnich:2010xq} we refer to this as the gauge algebroid associated with the generating set and the given boundary conditions. The functions $C^{\alpha \beta }{}_\gamma$ have been defined as the part of the prefactor of $R^\gamma_i$ on the right-hand side of \eqref{eq:var1} that does not depend on variations of the gauge parameters. However, even for field-independent gauge parameters, $C^{\alpha \beta }{}_\gamma $ can depend on the fields, which spoils the Lie algebra property \cite{Bojowald:2003pz}.

The possibility of a trivial gauge transformation on the right-hand side of \eqref{eq:var3} shows that in general, i.e., for non-zero functions $M_{ij}^{\alpha \beta }$, the generating set does not close off-shell. Whether this is the case depends on the theory at hand as well as on the chosen generating set $R_i^\alpha $. Generating sets are far from unique; one can relate $R^\alpha _i$ to another set $\tilde{R}^\alpha _i$ by a field dependent transformation $t^\alpha _\beta $,
\eq{
    R^\alpha _i=t^\alpha _\beta\,  \tilde{R}^\beta _i+M ^\alpha _{ij}\,\frac{\delta I}{\delta \phi _j} \qquad\qquad M^\alpha _{ij}=(-1)^{\mathrm{deg(\phi _i)deg(\phi _j)}+1} M^\alpha _{ji} ~.
}{eq:changeofset}
This can have notable effects on the algebraic properties of the generating set. For the same gauge theory different generating sets can have different structure functions, and it might even happen that one set does not form a closed algebroid off-shell while another one does. 

Coming back to the PSM, the action \eqref{PSM01} is invariant under the gauge transformations 
\begin{equation}
\delta_{\lambda}X^{I} = P^{IJ}\lambda_{J} \qquad \qquad \delta_{\lambda}A_{I} = -\mathrm{d}\lambda_{I}-\partial_{I}P^{JK}A_{J}\lambda_{K}   \label{PSM04a}
\end{equation}
with parameters $\lambda_I$. On-shell, the gauge transformations are equivalent to \eqref{grag1}-\eqref{grag2} with 
\begin{align}\label{eq:FO_gaugetrafos}
\lambda_{I}&=\lambda_{I}\left(\xi\right)+\lambda_{I}\left(\sigma\right)
\end{align} 
and
\begin{equation}
\lambda_{I}\left(\xi\right)=-A_{\mu I}\xi^{\mu} \qquad \quad  \lambda_{X}\left(\sigma\right)=\sigma  \qquad \quad \lambda_{\pm}\left(\sigma\right)=0 ~.
\end{equation}
The transformations \eqref{PSM04a} form a generating set, which for generic $P^{IJ}$ only closes on solutions of the EOM like in the generic example above. Indeed, one can show that off-shell
\begin{align}
    [\delta _{\lambda _1},\delta _{\lambda _2}]X^I&= P^{IJ}[\lambda _1,\lambda _2]^\ast_J \label{eq:gaugecomm1}\\
    [\delta _{\lambda _1},\delta _{\lambda _2}]A_I&= -\dd ([\lambda _1,\lambda _2]^\ast_I)-\partial _IP^{KJ}A_K[\lambda _1,\lambda _2]^\ast_J \label{eq:gaugecomm2}\\[.5em]
    &\qquad \qquad +(\lambda _1)_K(\lambda _2)_L\partial _I\partial _JP^{KL}(\dd X^J+P^{JM}A_{M}) \nonumber
\end{align}
where the bracket $[\cdot,\cdot]^\ast $ is defined by
\begin{align}\label{eq:mod_brack}
    [\lambda _1,\lambda _2]^\ast _I:=-\partial _IP^{KJ}(\lambda _1)_K(\lambda _2)_J+\delta _{\lambda _1}(\lambda _2)_I-\delta _{\lambda _2}(\lambda _1)_I ~.
\end{align}
By comparing \eqref{eq:gaugecomm2} with \eqref{eq:PSM_EOMs}, one can see that the closure is spoilt by the term in the second line proportional to an EOM. One can directly read off the functions 
\begin{equation}
    C^{IJ}{}_K=-\partial _KP^{IJ} \qquad \qquad M_{IJ}^{KL}=\partial _I\partial _JP^{KL} ~.
\end{equation}
These functions are field-dependent for a generic Poisson tensor. We stress again that this is a property of the chosen generating set. Later on, we shall find that target space diffeomorphisms of PSMs do not preserve generating sets. This is the decisive feature for finding an off-shell closed generating set for general models.  

In some special cases, i.e., if the Poisson tensor is linear in the target space coordinates, the generating set given here closes off-shell because $M_{IJ}^{KL}=0$. Therefore, for field-independent gauge parameters, the structure functions just reduce to structure constants and one is back in the realm of Lie algebras. Among the few examples is the JT model\footnote{This goes hand in hand with being able to write the JT model as a nonabelian $BF$-theory \cite{Fukuyama:1985gg,Isler:1989hq,Chamseddine:1989yz}.}.

\subsection{Lorentzian JT gravity}\label{sec:JT}

For the JT model $U=0$ and $V=-X$, so that 
$Q\left(X\right)=0$, $w\left(X\right)=\tfrac{1}{2}X^{2}$
and the Casimir function is given by 
\eq{
\mathcal{C}=X^{+}X^{-}+\frac{1}{2}X^{2}=Y+\frac{1}{2}X^{2}\,.  
}{JT02}

Here, we rewrite the most general asymptotic conditions for JT gravity in generalized Bondi gauge obtained 
previously in \cite{Ruzziconi:2020wrb} in their first-order form. We start with the line element
\eq{
\mathrm{d}s^{2}=-2e^{\Theta(u)}\mathrm{d}u\,\mathrm{d}r-2B\left(u,r\right)\mathrm{d}u^{2} 
}{eq:whatever}
where the asymptotic boundary is situated at $r\to\infty$, and $u$ is a coordinate along this boundary. In comparison to the usual Bondi gauge, the off-diagonal component of the metric is not constant but depends on an arbitrary function $\Theta(u)$. In terms of first-order variables, the metric \eqref{eq:whatever} can be described by partially fixing the gauge to
\begin{equation}\label{eq:lorentzianJT_gauge}
    e_{r}^{+}=0 \qquad \qquad e_{r}^{-}=1 \qquad \qquad \omega _r=0
\end{equation}
and solving the EOM containing derivatives with respect to $r$ up to several arbitrary functions of $u$. We obtain 
\begin{align}
    e_{u}^{+}=-e^{\Theta (u)} \qquad \qquad e_{u}^{-}=e^{-\Theta (u)}B(u,r) \qquad \qquad \omega_{u}=-e^{-\Theta(u)}\partial_{r}B(u,r)
\end{align}
where
\begin{equation}\label{eq:B_sol}
B=\frac{1}{2}e^{2\Theta}r^{2}-e^{\Theta}\mathcal{P}r+\mathcal{T}+\frac{1}{2}\mathcal{P}^2 ~.
\end{equation}
Here, $\mathcal{P}$ and $\mathcal{\mathcal{T}}$ are arbitrary functions of $u$. Thus, we arrive at
\begin{equation}
 e_{u}^{+}=-e^{\Theta}\qquad\quad e_{u}^{-}=\frac{1}{2}e^{\Theta}r^{2}-\mathcal{P}r+e^{-\Theta}\Big(\mathcal{T}+\frac{1}{2}\mathcal{P}^2\Big) \quad \qquad \omega_{u}=-e^{\Theta}r+\mathcal{P} ~.    
\end{equation}
Solutions of the radial EOM for $X$ and $X^\pm$ are
\begin{equation}
X=e^{\Theta}\varphi_{1}r+\varphi_{0}\quad \qquad X^{+}=e^{\Theta}\varphi_{1}\quad \qquad X^{-}=-\frac{1}{2}e^{\Theta}\varphi_{1}r^{2}-\varphi_{0}r-e^{-\Theta}\varphi_{-1}    
\end{equation}
and depend on three other arbitrary functions, $\varphi_1(u)$, $\varphi_0(u)$, and $\varphi_{-1}(u)$.

On-shell the arbitrary functions in time are further constrained by the temporal EOM following from \eqref{D03a} and \eqref{D03b},
\begin{align}
\mathcal{E}_1&:=\partial_{u}\varphi_{1}+\varphi_{0}+\left(\mathcal{P}+\Theta '\right)\varphi_{1} =0	\label{ueom1}\\
\mathcal{E}_0&:=\partial_{u}\varphi_{0}+\varphi_{-1}-\varphi_{1}\Big(\mathcal{T}+\frac{1}{2}\mathcal{P}^2\Big)=0 \label{ueom2}\\
\mathcal{E}_{-1}&:=\partial_{u}\varphi_{-1}-\varphi_{0}\Big(\mathcal{T}+\frac{1}{2}\mathcal{P}^2\Big) -\left(\mathcal{P}+\Theta '\right)\varphi_{-1} =0 ~.\label{ueom3}    
\end{align}
A complete solution can be specified by providing three functions $\mathcal{T}(u)$, $\mathcal{P}(u)$, $\Theta (u)$ together with three initial conditions for the boundary EOM. The Casimir function is given by 
\begin{equation}
\mathcal{C}=\frac{1}{2}\varphi_{0}^{2}-\varphi_{1}\varphi_{-1} 
\end{equation}
and does not depend on $u$ when evaluated on solutions of \eqref{ueom1}-\eqref{ueom3}, i.e., $\partial_u\mathcal{C}=0$. To set up the variational principle, we choose boundary conditions
\begin{align}
     X&=e^\Theta \varphi _1r+\varphi _0+\mathcal{O}(r^{-1}) & \omega_u&=-\partial_r e_u^- + {\cal O}(r^{-1})\\
    X^+&=\partial _r X+\mathcal{O}(r^{-2}) & e^+_u&=-e^\Theta +\mathcal{O}(r^{-1})\\
    \partial _r X^-&=- X+\mathcal{O}(r^{-1}) & e^-_u&=e^{-\Theta }B(u,r)+\mathcal{O}(r^{-1})
\end{align}
with $B$ given by \eqref{eq:B_sol}. The theory is described by the action
\begin{align}\label{eq:full_action_lorentz}
    \Gamma =\frac{k}{2\pi}\int_{\mathcal{M}}\Big ( X^{I} &\mathrm{d}A_{I}+\frac{1}{2}P^{IJ}A_{I}\wedge A_{J}\Big ) \\
    &-\frac{k}{2\pi }\int_{\partial \mathcal{M}}\Big (X^IA_I+\mathcal{C}\,\dd f+\frac{X}{2}\,\dd \ln \Big |\frac{X^+}{X^-}\Big \vert \Big ) \nonumber
\end{align}
where $\mathcal{C}$ is given by \eqref{JT02}. The last term renders the action Lorentz-invariant \cite{Bergamin:2007sm,Bergamin:2005pg}. The boundary one-form $\dd f$ is given in terms of the field variables as
\begin{align}\label{eq:bound_oneform}
    \dd f=f'\dd u =\frac{1}{\varphi _1}\dd u ~.
\end{align}
 In the first variation of the action,
\begin{align}\label{eq:full_variation_lorentz}
    \delta \Gamma = (\textrm{EOM})\,&- \frac{k}{2\pi }\int _{\partial \mathcal{M}}\big(\pi _X \delta X+\pi _+\delta X^++\pi _-\delta X^-+\mathcal{C}\delta \dd f\big)\\
    &-\frac{k}{2\pi }\int _{\partial \mathcal{M}}\dd \left(\frac{X}{2}\delta \ln \Big \vert \frac{X^+}{X^-}\Big \vert \right)
\end{align}
all variations with respect to the gauge field cancel. The variations of the scalar fields come with coefficients $\pi_i$, given by 
\begin{align}
    \pi_X&=\omega +X\dd f+\frac{1}{2}\dd \ln \Big \vert \frac{X^+}{X^-}\Big \vert \\
    \pi _+&=e^-+X^-\dd f-\frac{1}{2X^+}\dd X \\
    \pi _-&=e^++X^+\dd f+\frac{1}{2X^-}\dd X \,.
\end{align}
To make the variational principle well-defined, the on-shell variation of the action needs to vanish up to corner terms, see e.g. \cite{Harlow:2019yfa}. Therefore, the second part of \eqref{eq:full_variation_lorentz} on-shell needs to be a total boundary derivative contributing to the corner term in the second line. Inserting the boundary conditions, the first variation of the action evaluated on the EOM reads
\begin{align}\label{eq:cornerterms}
    \delta \Gamma &\approx -\frac{k}{2\pi }\int_{\partial \mathcal{M}}\dd \left(\frac{X}{2}\delta \ln \Big \vert \frac{X^+}{X^-}\Big \vert+\mathcal{C}\delta f\right)+\mathcal{O}(r^{-1})\\
    &=\frac{k}{2\pi }\bigg(\frac{\varphi _0}{e^\Theta \varphi _1}\delta \big(e^\Theta \varphi _1\big)-\delta \varphi _0+\mathcal{C}\delta f\bigg)\bigg|^{u_1}_{u_0}+\mathcal{O}(r^{-1})
\end{align}
which indeed vanishes up to corner terms. Turning to the on-shell action, the bulk contribution vanishes for the JT model  while the boundary part gives 
\begin{align}
    \Gamma &=\frac{k}{2\pi }\int _{\partial \mathcal{M}}\dd u \, \left(\mathcal{T}\varphi _1-(\mathcal{P}+\Theta ')\varphi _0-\frac{\varphi _0^2}{2\varphi _1}+\varphi _0'-\varphi _0\frac{\varphi _1'}{\varphi _1}\right)+\mathcal{O}(r^{-1}) \label{eq:bound_action}\\
    &\approx \frac{k\,\mathcal{C}}{2\pi }\int _{\partial \mathcal{M}}\dd u\, \frac{1}{\varphi _1} +\mathcal{O}(r^{-1})~.
\end{align}
For arriving at the last line all the evolution equations \eqref{ueom1}-\eqref{ueom3} were used.

\subsection{Residual gauge transformations of Lorentzian JT}\label{sec:res_gauge_lorntz}
Preserving the gauge conditions \eqref{eq:lorentzianJT_gauge} leads to three differential equations restricting the gauge parameters
\begin{equation}\label{eq:gauge_equs}
    \partial _r\lambda _-=0 \qquad \qquad  \partial _r\lambda _++\lambda _X=0 \qquad \qquad  \partial _r\lambda _X-\lambda _-=0 ~.
\end{equation}
The general solution to \eqref{eq:gauge_equs} contains three arbitrary functions which are denoted by $(\varepsilon (u),\gamma (u),\eta (u))$. In a convenient parametrization it reads
\begin{subequations}\label{eq:asymp_PSM}
\begin{align} 
    \lambda _X&=e^\Theta \varepsilon r+\gamma -\varepsilon \mathcal{P} \\
    \lambda _-&=e^\Theta \varepsilon \\
    \lambda _+&=-\frac{1}{2}e^\Theta \varepsilon r^2-\left(\gamma -\varepsilon \mathcal{P}\right)r-e^{-\Theta } \Big(\eta +\mathcal{T}\varepsilon +\frac{1}{2}\mathcal{P}^2\varepsilon \Big) 
\end{align}
\end{subequations}
and the infinitesimal action on the free functions is 
\begin{subequations}\label{eq:trafo_metr}
\begin{align}
 \delta _\lambda \mathcal{T}&=\varepsilon \mathcal{T}'+2\varepsilon '\mathcal{T}+\mathcal{P}\gamma '-\Theta '\eta +\eta ' & \delta _\lambda \varphi _1&=\varepsilon \varphi _1'-\varepsilon '\varphi _1-\varepsilon \,\mathcal{E}_1 \\
    \delta _\lambda \mathcal{P}&=\varepsilon \mathcal{P}'+\varepsilon '\mathcal{P}-\eta -\gamma ' & \delta _\lambda \varphi _0&=\varepsilon \varphi _0'+\eta \varphi _1-\varepsilon \,\mathcal{E}_0\\
    \delta _\lambda \Theta &=\varepsilon \Theta '+\varepsilon '+\gamma  & \delta _\lambda \varphi _{-1}&=\varepsilon \varphi _{-1}'+\varepsilon '\varphi _{-1}+\eta \varphi _0-\varepsilon \, \mathcal{E}_{-1}~.
\end{align}
\end{subequations}
Interpreting the transformations generated by $\varepsilon$ as holomorphic conformal transformations we can read off the conformal weights of all quantities in the left equations \eqref{eq:trafo_metr}: $\cal T$ has weight 2, like a chiral stress tensor (but without anomalous term); $\cal P$ has weight 1, like a 1-form; $\Theta$ has weight 0, like a scalar, and transforms anomalously. On-shell, the quantities $\varphi_n$ acquire conformal weights $-n$.

The definition \eqref{eq:bound_oneform} does not impose further restrictions since the variation
\begin{align}
    \delta_\lambda \frac{1}{\varphi _1}= \left(\frac{\varepsilon }{\varphi_1}\right)'
\end{align}
is again a total boundary derivative on-shell. As we are working off-shell, the transformations \eqref{eq:trafo_metr}, in general, map between off-shell configurations and do not necessarily have to form a closed algebra. Given a certain reference configuration, we think of them as infinitesimal spectrum-generating transformations. Once restricted to the solution space, these transformations are true symmetries, i.e., they map solutions to solutions. 

As explained in the previous section, the JT model allows for a closed generating set of gauge transformations off-shell. Computing the brackets \eqref{eq:mod_brack} with \eqref{eq:asymp_PSM} obtains 
\begin{align}\label{eq:sym_bracket}
    \Big [\lambda [\varepsilon ,\gamma ,\eta ],\lambda [\bar{\varepsilon},\bar{\gamma },\bar{\eta }]\Big ]_I^\ast =\lambda _I[\varepsilon \bar{\varepsilon }'-\bar{\varepsilon }\varepsilon ',\varepsilon \bar{\gamma }'-\bar{\varepsilon }\gamma ',(\varepsilon \bar{\eta })'-(\bar{\varepsilon }\eta )'] ~,
\end{align}
where the functions $\lambda _I[\varepsilon ,\gamma ,\eta ]$ are defined such that \eqref{eq:asymp_PSM} is recovered for $\lambda [-\varepsilon ,-\gamma ,-\eta ]$. We highlight two properties. First, as the PSM parameters $\lambda _I$ depend linearly on the gauge parameters $(\varepsilon, \gamma, \eta)$, the functions $\lambda_I$ provide a Lie algebra homomorphism $\lambda_I: \mathfrak{g}\to\Gamma (A)$ from the space of parameters to sections of the gauge algebroid where the bracket on the space of parameters is given by
\begin{align}
    \big[(\varepsilon ,\gamma ,\eta ),(\Bar{\varepsilon },\Bar{\gamma },\Bar{\eta })\big ]_{\mathfrak{g}}=\big(\varepsilon \Bar{\varepsilon }'-\Bar{\varepsilon }\varepsilon ',\varepsilon \bar{\gamma }'-\bar{\varepsilon }\gamma ',(\varepsilon \bar{\eta })'-(\bar{\varepsilon }\eta )'\big) ~.
\end{align}
Second, the bracket $[\cdot ,\cdot ]_{\mathfrak{g}}$ defines the Lie algebra $\mathfrak{g}$ of residual gauge transformations. Its associated group is
\begin{align}
   \mathcal{G}=\text{Diff}(\mathbb{R})\ltimes \left(C^\infty (\mathbb{R})\times \Omega ^1(\mathbb{R})\right) ~.
\end{align}
This follows from the weights of the parameters under transformations generated by $\varepsilon $: $\varepsilon $ is a vector field, $\gamma $ is a scalar function, and $\eta $ is a one-form. The relation \eqref{eq:sym_bracket} thus implies that the functions $\lambda _I$ form a representation of this algebra. In Laurent modes 
\begin{align}
    T_n:=\lambda _I(\varepsilon =u^{n+1},0,0) && P_n:=\lambda _I (0,\gamma =u^n,0) && Q_n:=\lambda _I (0,0,\eta =u^n)
\end{align}
this algebra is given by
\begin{align}
    [T_n,T_m]^\ast &=(n-m)T_{n+m} & [Q_n,Q_m]^\ast&=0\\
    [T_n,P_m]^\ast&=-mP_{m+n} & [P_n,P_m]^\ast&=0\\
    [T_n,Q_m]^\ast&=-(m+n)Q_{m+n} & [Q_n,P_m]^\ast&=0 ~.
\end{align}
The transformations of $(\mathcal{T},\mathcal{P},\Theta )$ in \eqref{eq:trafo_metr} can be identified as the infinitesimal coadjoint action of a certain central extension $\Hat{\mathcal{G}}$. As this property will reappear in the Euclidean case we refer to section \ref{subsec:resEucl} for a more detailed discussion. There are several possible restrictions of these boundary conditions leading to other familiar algebras. 
\begin{itemize}
    \item Fixing $\Theta =0$ and defining $\mathcal{L}=\mathcal{T}+\frac{1}{2}\mathcal{P}^2$ leads to a BMS$_2$ algebra like it was recently studied in the context of flat JT-gravity \cite{Afshar:2021qvi}. The coadjoint action of its central extension is realized on the fields $(\mathcal{L},\mathcal{Q})$ with $\mathcal{P}=\mathcal{Q}'+c$. The tower $Q_n$ is not present in this case.
    \item Taking the same definitions and fixing additionally $\eta =\sigma '$  leads to a warped conformal algebra realized on $(\mathcal{L},\mathcal{P})$ like it was obtained in \cite{Afshar:2015wjm,Godet:2020xpk,Hofman:2011zj,Detournay:2012pc,Afshar:2019tvp}. 
    \item The choice $\mathcal{P}=0=\Theta $ implies $\eta =-\gamma '$ and $\gamma =-\varepsilon '$ such that one is left with a single Virasoro algebra realized on $\mathcal{T}$ \cite{Grumiller:2017qao,Godet:2020xpk}. 
\end{itemize}
Up to this point, it is still unclear whether these residual transformations are proper gauge transformations. To answer this question we look at the associated canonical boundary charges derived in appendix \ref{app:gauge_structure}. Evaluating them yields
\begin{align}
    \slashed \delta Q _\lambda =\frac{k}{2\pi }\lambda _I\delta X^I&= \frac{k}{2\pi }\Big[-\varepsilon \Big(\mathcal{P}\delta \varphi _0+e^\Theta \delta \big(\varphi _{-1}e^{-\Theta }\big)+e^{-\Theta }\Big(\mathcal{T}+\frac{1}{2}\mathcal{P}^2\Big)\delta \big(\varphi _1e^\Theta  \big )\Big) \nonumber \\[.5em]
    &\qquad \quad +\gamma \delta \varphi _0-\eta e^{-\Theta }\delta \big(\varphi _1e^\Theta \big)\Big]\Big \vert_{\textrm{EOM}}\Big \vert _{u=u_0} \\
    &=\frac{k}{2\pi }\Big[-\delta _\lambda \varphi _0\,\delta \ln \big(e^\Theta \varphi _1\big)+\delta _\lambda \ln \big(e^\Theta \varphi _1\big)\delta \varphi _0+\frac{\varepsilon }{\varphi _1}\delta \mathcal{C} \Big]\Big \vert _{u=u_0}
\end{align}
which is a radially independent  expression. Generically, the boundary conditions lead to field dependence in the $\lambda _I$ which makes the charges non-integrable in field space, as denoted by $\slashed \delta $. Despite this non-integrability, one can see that the three residual gauge parameters are associated to improper gauge symmetries and thus transform between physically distinct configurations. The non-integrability is usually associated with non-vanishing flux at the boundary which cannot be present in this theory because of the lack of propagating modes in the bulk. As explained in \cite{Grumiller:2019fmp,Adami:2020ugu,Adami:2021nnf,Adami:2022ktn,Geiller:2021vpg} one rather has to think of the non-integrability as a consequence of `fake'-flux, associated to a choice of phase space slicing. 
As previously shown in \cite{Ruzziconi:2020wrb}, an integrable slicing always exists in 2D dilaton gravity. It can be reached directly by defining new fields
\begin{align}
    q=\varphi _0 && p=\ln \big(e^\Theta \varphi _1\big)
\end{align}
and new field-independent gauge parameters $(\lambda _q,\lambda _p,\lambda _{\mathcal{C}})$ by
\begin{align}
    \lambda _q=\delta _\lambda \ln \big(e^\Theta \varphi _1\big) && \lambda _p=-\delta _\lambda \varphi _0 && \lambda _{\mathcal{C}}=\frac{\varepsilon }{\varphi _1} ~.
\end{align}
The charges can then be integrated to
\begin{align}
    Q_\lambda =\frac{k}{2\pi }\Big(\lambda _q \, q+\lambda _p\, p+\lambda _{\mathcal{C}}\mathcal{C}\Big)
\end{align}
and represent a Heisenberg loop algebra through their canonical Poisson brackets (see also appendix \ref{app:gauge_structure}),
\begin{align}
    -\delta _\eta Q_\lambda =\{Q_\lambda ,Q_\eta \}=\frac{k}{2\pi }\big(\lambda _q\eta _p-\lambda _p\eta _q\big) ~.
\end{align}

\section{Target space diffeomorphisms in the Lorentzian case}\label{sec:td}

Let us take a closer look at Poisson diffeomorphisms on target space and under which conditions they can be interpreted as mapping two dilaton gravity models to each other. We need some elementary facts from Poisson geometry \cite{Crainic:2021}.
Poisson manifolds admit foliations by symplectic leaves. The latter ones are defined as submanifolds to which Hamiltonian vector fields $(H_\phi)^J:=P^{JK}\partial_K \phi(X)$ span the tangent space at each point. Due to the property \eqref{Cbra}, the Casimir function $\mathcal{C}$ is constant on any symplectic leaf. The symplectic leaves are symplectic manifolds. Thus, the dimension of a symplectic leaf is even. For any classical solution of a PSM, the coordinates $X^I(x)$ have values within a single symplectic leaf.

 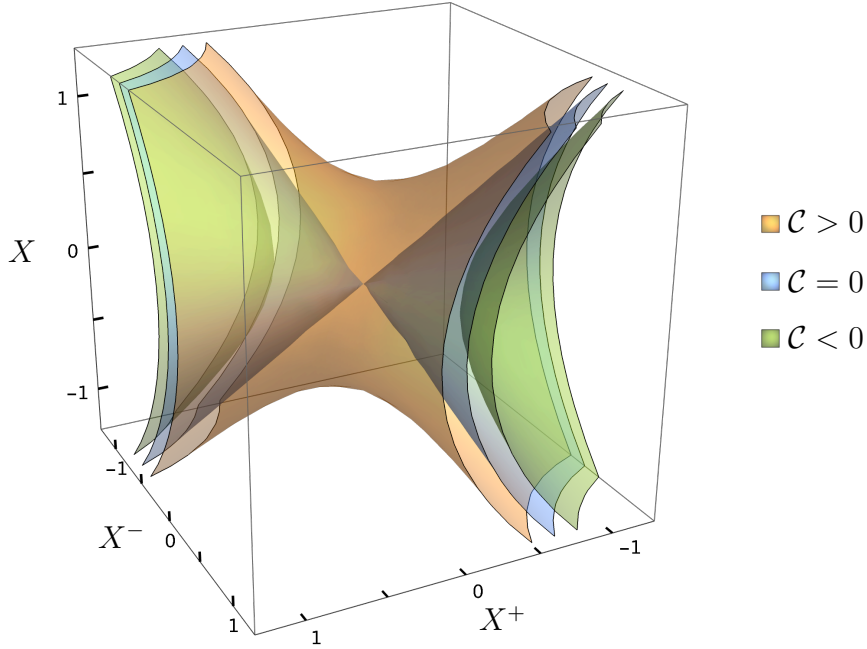
\begin{figure}
 \begin{center}
\begin{tikzpicture}
    \node[anchor=south west,inner sep=0] (image) at (0,0) {\includegraphics[width=11cm]{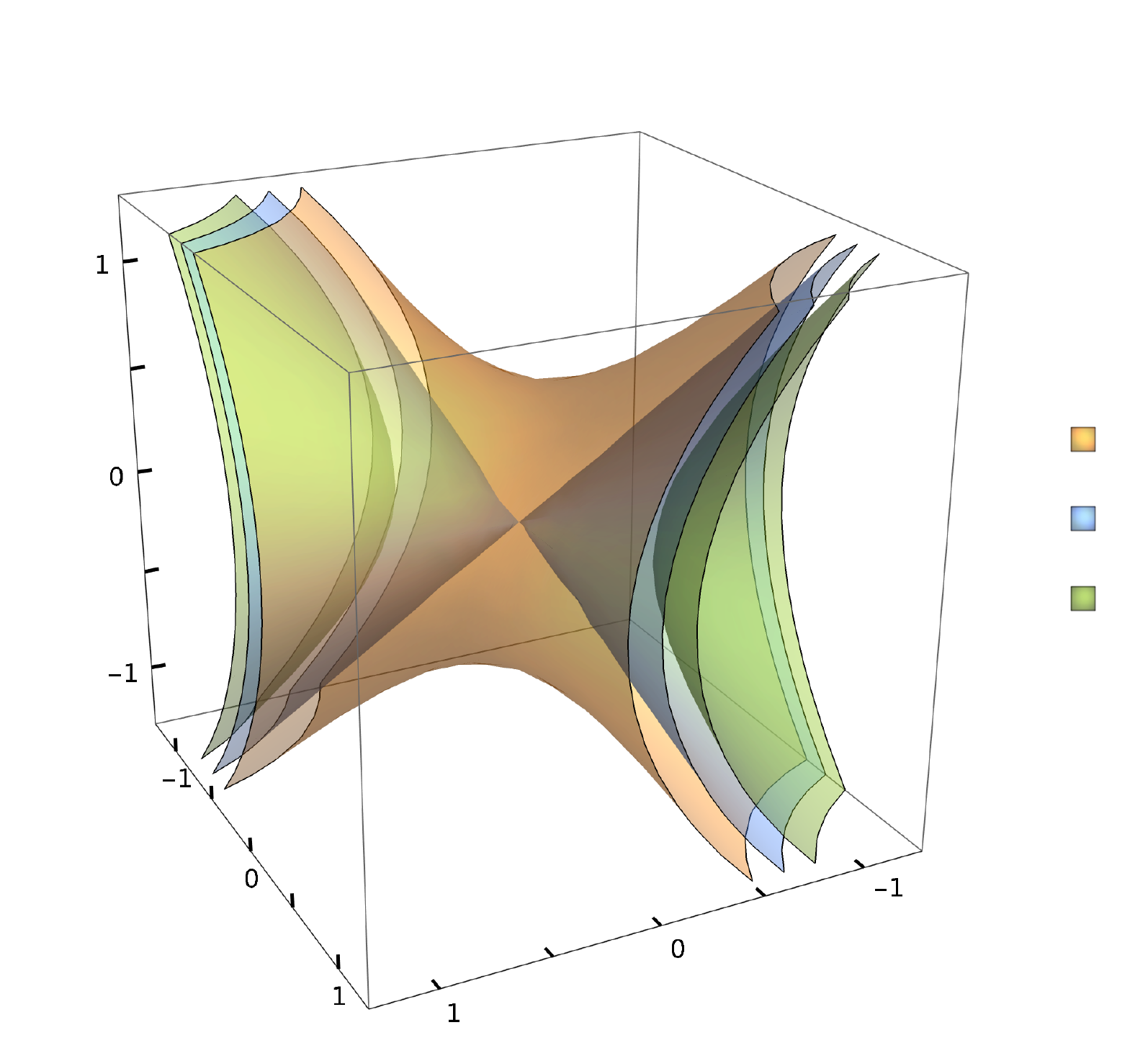}};
    \begin{scope}[x={(image.south east)},y={(image.north west)}]
        \node[label=above:$X^-$] (A) at (0.16,0.12) {};
         \node[label=above:$X^+$] (A) at (0.62,0.01) {};
          \node[label=above:$X$] (A) at (0.04,0.5) {};
          \node[label=above:$\mathcal{C}>0$] (A) at (1.01,0.538) {};
           \node[label=above:${\mathcal{C}=0}$] (A) at (1.01,0.46) {};
          \node[label=above:$\mathcal{C}<0$] (A) at (1.01,0.384) {};
    \end{scope}
\end{tikzpicture}
\caption{Symplectic leaves of the JT model. The leaves with $\mathcal{C}>0$ contain world-sheet horizons at the target space loci $X^+X^-=0$, referred to as ``horizon points''.}
\label{fig:JT_targetspace}
\end{center}
 \end{figure}
The global structure of the symplectic foliation varies considerably from model to model. For JT gravity, this structure is relatively simple as can be seen in figure \ref{fig:JT_targetspace}. For $\mathcal{C}>0$ there is a single symplectic leaf, a one-sheet hyperboloid. This leaf includes the points of the horizon where $X^+X^-=0$. For $\mathcal{C}<0$ there are two symplectic leaves that, taken together, form a two-sheet hyperboloid. These leaves do not contain points corresponding to horizons. For $\mathcal{C}=0$ there are three symplectic leaves: two cones without the tip, and a zero-dimensional leaf $X=X^+=X^-=0$.

Under the target space diffeomorphisms $X^{I}\rightarrow\bar{X}^{I}
=\bar{X}^{I}\left(X^{J}\right)$ with
\begin{equation}  \label{PSM05}
\bar{A}_{I}=A_{J}\frac{\partial X^{J}}{\partial\bar{X}^{I}}\qquad
\qquad\bar{P}^{IJ}(\bar X)=P^{KL}(X(\bar X))\frac{\partial\bar{X}^{I}}{\partial X^{K}}\frac{\partial\bar{X}^{J}}{\partial X^{L}}    
\end{equation}
the PSM action remains invariant, $\bar{I}\mathrm{_{PSM}}=I_{\mathrm{PSM}}$. This implies that the 
first-order dilaton gravity action receives a boundary term
\begin{equation} \label{PSM07}
\bar{I}_{\mathrm{1st}}=I_{\mathrm{1st}}-\frac{k}{2\pi}\int_{\mathcal{M}}\mathrm{d}\left(X^{I}A_{I}\right)+\frac{k}{2\pi}\int_{\mathcal{M}}\mathrm{d}\left(\bar{X}^{I}\bar{A}_{I}\right) ~.    
\end{equation}
These transformations map one 2D dilaton gravity model to another one.

The target space diffeomorphisms locally respect symplectic foliations. However, the possibly very different topologies of the latter make it hard to expect general statements on the diffeomorphisms that map the whole Poisson manifold $\mathcal{P}$ to $\bar{\mathcal{P}}$. More general and interesting results may be achieved by mapping a part of $\mathcal{P}$ to a part $\bar{\mathcal{P}}$. In \cite{Valcarcel:2022zqm}, maps between asymptotic regions were considered. In the present paper, we extend this analysis further into the bulk.

We are interested only in those Poisson diffeomorphisms that map one dilaton gravity model to another dilaton gravity model. Such diffeomorphisms should preserve the structure of the Poisson tensor given in \eqref{D02}. 

Since the Poisson tensor in the coordinates $\bar X$, $\bar X^\pm$ is requested to have the form \eqref{D02}, the corresponding PSM possesses local Lorentz symmetry with $\bar X$ being a Lorentz scalar and $\bar X^\pm$ transforming exactly as $X^\pm$ in the initial model. To respect the Lorentz symmetry, the target space diffeomorphism should be of the form
\begin{eqnarray}
\bar{X} &=& \bar{X}\left(X,Y\right) \label{T01a}\\
\bar{X}^{+}	&=& X^{+}f^{\left(+\right)}\left(X,Y\right) \label{T01b}\\
\bar{X}^{-} &=&X^{-}f^{\left(-\right)}\left(X,Y\right) \label{T01c} 
\end{eqnarray}
with some undetermined functions $f^{(\pm)}$ and $\bar X$. By using \eqref{T01a}-\eqref{T01c} in \eqref{PSM05} together with \eqref{D02}, we obtain
\begin{eqnarray}
\bar{P}^{\bar X\pm} &=&	\mp\bar{X}^{\pm}\left(\partial_{X}\bar{X}+\mathcal{V}\partial_{Y}\bar{X}\right) \label{T02a}\\
\bar{P}^{+-} &=&	\left(Y\partial_{X}+\mathcal{V}Y\partial_{Y}+\mathcal{V}\right)f=\left(\partial_{X}+\mathcal{V}\partial_{Y}\right)\tilde{f}    \label{T02b}
\end{eqnarray}
where we defined $f\equiv f^{\left(+\right)}f^{\left(-\right)}$ and $\tilde{f}\equiv Yf$. Thus, to maintain the form \eqref{D02} we must request
\begin{eqnarray}
\left(\partial_{X}+\mathcal{V}\partial_{Y}\right)\bar{X}	&=&	1 \label{T03a}\\
\left(\partial_{X}+\mathcal{V}\partial_{Y}\right)\tilde{f}	&=&	\bar{\mathcal{V}}\left(\bar{X},\bar{Y}\right) ~.   \label{T03b} 
\end{eqnarray}
Equation \eqref{T03a} is linear and quite easy to solve. Taking into account \eqref{T04} we obtain
\begin{equation}\label{T06}
\bar{X}=X+g_{1}\left(\mathcal{C}\right)  
\end{equation}
where $g_1$ is an arbitrary function of the Casimir. Since $\tilde{f}=\bar Y$, \eqref{T03b} is non-linear and more complicated. To solve this equation, we substitute
\begin{equation}\label{T07}
\tilde{f}\left(X,Y\right)=e^{-\bar{Q}\left(\bar{X}\right)}\left[-\bar{w}\left(\bar{X}\right)+\tilde{G}\left(X,Y\right)\right]    
\end{equation}
where $\tilde{G}$ is a new unknown function. Equation \eqref{T03b} reduces to
\begin{equation}\label{T08}
\left(\partial_{X}+\mathcal{V}\partial_{Y}\right)\tilde{G}\left(X,Y\right)=0
\end{equation}
so that the general solution is $\tilde{G}\left(X,Y\right)=g_{2}\left(\mathcal{C}\right)$, where $g_{2}\left(\mathcal{C}\right)$ is another function of the Casimir. Thus, we have
\begin{equation}
   \tilde{f}\left(X,Y\right) = e^{-\bar{Q}\left(\bar{X}\right)}\left[-\bar{w}\left(\bar{X}\right)+g_{2}\left(\mathcal{C}\right)\right] ~.\label{T09a} 
\end{equation}
Equations \eqref{T06} and \eqref{T09a} describe general Poisson diffeomorphisms that map one 2D dilaton gravity model to another.

It is convenient to fix the Lorentz gauge by the condition $f^{(+)}=1$ to obtain the target space diffeomorphism
\begin{eqnarray}
\bar{X}	&=&	X+g_{1}\left(\mathcal{C}\right)\label{T10a}\\
\bar{X}^{+}	&=&	X^{+} \label{T10b}\\
\bar{X}^{-}	&=& \frac{1}{X^{+}}e^{-\bar{Q}\left(\bar{X}\right)}\left[g_{2}\left(\mathcal{C}\right)-\bar{w}\left(\bar{X}\right)\right] \label{T10c}    
\end{eqnarray}
between a model with the coordinates $X^I$ and the Casimir function $\mathcal{C}$, and another model with coordinates $\bar X^I$ with dilaton potentials $\bar V$ and $\bar U$ defining $\bar w$ and $\bar Q$ through the equations \eqref{D05}. So far, the functions $g_1$ and $g_2$ are arbitrary. Equation \eqref{T10c} implies 
\begin{equation}
    \bar{\mathcal{C}}=g_2(\mathcal{C}) ~. \label{barCC}
\end{equation}
To avoid a singularity in \eqref{T10c} at $X^+=0$, one needs to impose a relation
between $g_1$ and $g_2$,
\begin{equation}
    g_2(w(X_h))=\bar w(X_h+g_1(w(X_h)))\label{g2g1}
\end{equation}
where $X_h$ is a solution of the equation
\begin{equation}
    w(X_h)=\mathcal{C} ~.\label{wXh}
\end{equation}

\subsection{Poisson automorphisms}\label{sec:aut}

If there is a Poisson diffeomorphism mapping (a sector of) one PSM to 
(a sector of) another PSM then such a diffeomorphism is not unique. We consider now the consequences of this non-uniqueness. Suppose there are two Poisson 
diffeomorphisms, $\Phi_1$ and $\Phi_2$, mapping PSM$_1$ to PSM$_2$. Then, $\Phi_1\circ \Phi_2^{-1}$ is an 
automorphism of PSM$_2$. It is also clear, that any Poisson diffeomorphism from PSM$_1$ to PSM$_2$ is 
a composition of a fixed diffeomorphism from PSM$_1$ to PSM$_2$ and an arbitrary automorphism of any 
of these models.

The Poisson automorphisms are generated by the vector fields $\xi$ on $\mathcal{P}$ that leave the Poisson tensor invariant,  
\eq{
 \mathcal{L}_\xi \, P^{IJ}=P^{IK}\partial_K\xi^J +P^{KJ}\partial_K\xi^{I} - 
 \bigl( \partial_K P^{IJ}\bigr)\xi^K =0\,.
}{delxiP}
Such vector fields are called Poisson vector fields. Their structure can be easily understood in Casimir--Darboux coordinates where the non-zero components of the Poisson tensor are 
$P^{12}=-P^{21}=1$. Equation \eqref{delxiP} has two types of solutions. Solutions of the 
first type are Hamiltonian vector fields 
\begin{equation}
\xi^I=P^{IJ}\partial_J \lambda ~. \label{hamvec}
\end{equation}
These vector fields do not change the Casimir function\footnote{Target space Hamiltonian vector fields can be related to on-shell gauge transformations, see \cite{Cattaneo:2000iw,Baulieu:2001fi,Bojowald:2003pz,Salzer:2018zlv}.} and map each symplectic leaf to itself. The other type of solution is given by vectors 
$\xi=(0,0,\xi^3(X^3))$ that map the Casimir $\mathcal{C}=X^3$ to an arbitrary function
of $\mathcal{C}$. These Poisson vector fields form the first Poisson cohomology group 
$H^1_P(\mathcal{P})$. We call these automorphisms essential automorphisms of a given PSM. According to \eqref{barCC}, the transformations \eqref{T10a}-\eqref{T10c} with $\bar{U}=U$ and $\bar{V}=V$ under condition \eqref{g2g1} do exactly what essential automorphisms have to do. Thus, we may use these transformations as representatives of equivalence classes in $H^1_P(\mathcal{P})$. Of course, there may be some global issues that should be resolved separately in each particular model.

Since according to \eqref{barCC} the function $g_2$ changes the Casimir, one can expect that $g_1$ corresponds to Hamiltonian vector fields \eqref{hamvec}. To see this, let us take $g_2(\mathcal{C})=\mathcal{C}$ and $g_1(\mathcal{C})=\delta g_1(\mathcal{C})$ being an infinitesimal parameter. Then the corresponding infinitesimal automorphism reads
\begin{equation}
\delta X=\delta g_1(\mathcal{C}) \qquad \qquad \delta X^+=0 \qquad \qquad \delta X^-=\frac{\delta g_1(\mathcal{C})\, \mathcal{V}}{X^+} ~.\label{infaut}    
\end{equation}
One can easily check that these transformations are reproduced by $\delta X^I=\xi^I$ with $\xi^I$ given by \eqref{hamvec} and
\begin{equation}
\lambda= -\ln (X^+)\, \delta g_1(\mathcal{C}) ~.\label{hamg1}
\end{equation}
These transformations are generated by a Hamiltonian vector field having a singularity at $X^+=0$. Their role is to compensate the singularity of $g_2$ transformations at the same points.

Let us see how the automorphisms work for the JT model. The map $\mathcal{C}\to g_2(\mathcal{C})$ should be invertible and has to preserve the topological structure of the symplectic leaves. Therefore, $g_2$ has to be a monotonously increasing function with $g_2(0)=0$. Considering $\mathcal{C}>0$ and using $w_{\mathrm{JT}}=\tfrac 12 X^2$, we find horizon points at $X_h=\pm \sqrt{2\mathcal{C}}$. In this case \eqref{wXh} yields
\begin{equation}
    \bigl(\sqrt{2\mathcal{C}} + g_1(\mathcal{C})\bigr)^2=\bigl(-\sqrt{2\mathcal{C}} + g_1(\mathcal{C})\bigr)^2  \label{sqrt2}
\end{equation}
which only has the solution $g_1=0$. As a consequence, \eqref{g2g1} implies that $g_2$ is an identity function for this range of $\mathcal{C}$. Alternatively, we may discard the parts of symplectic leaves corresponding to negative (or positive) values of $X$. In any case, the regularity on symplectic leaves with $\mathcal{C}\leq 0$ does not impose any restrictions on $g_2$. 

\subsection{Mapping the variational principle}\label{sec:var_prin}
Suppose we have a dilaton gravity model in its PSM formulation with some boundary conditions and the associated variational principle. The space of kinematically allowed field configurations is then described by a manifold $\mathcal{F}$ while the submanifold $\mathcal{S}\subset \mathcal{F}$ describes the classical solutions of the model.
Our prototypical example is the JT model with boundary conditions and variational principle described in section \ref{sec:JT}. Under a target space diffeomorphism\footnote{Note the slight abuse of notation: Before $\Phi $ was a target space diffeomorphism while here it denotes the thereby induced map between the spaces of field configurations.} $\Phi :\mathcal{F}\to \Bar{\mathcal{F}}$ the asymptotic conditions of JT are mapped to asymptotic conditions in another model described by a manifold $\Bar{\mathcal{F}}$. Similarly, the solution space $\mathcal{S}$ is mapped to a submanifold $\Bar{\mathcal{S}}$. By applying the target space diffeomorphism to the PSM variables, the action $\Bar{\Gamma }$ of the new model is given by
\begin{align}\label{eq:actionmap}
    \Bar{\Gamma }[\Bar{X}^I,\Bar{A}_I]:=\Gamma \Big[X^I,\frac{\partial \Bar{X}^J}{\partial X^I}\bar{A}_J\Big]\Big \vert _{X^I=X^I(\Bar{X}^K)} 
\end{align}
which implies that the on-shell value of the barred action is exactly the same as the one of the unbarred action.
Moreover, the covariance of the PSM EOM implies that solutions of the classical EOM are again mapped to solutions. Using these two facts we infer
\begin{align}
    \frac{\delta \Bar{\Gamma }}{\delta \Bar{X}^I}\Big \vert _{\Bar{\mathcal{S}}}\delta \Bar{X}^I+\frac{\delta \Bar{\Gamma }}{\delta \Bar{A}_I}\Big \vert _{\Bar{\mathcal{S}}}\delta \Bar{A}_I=\frac{\delta \Gamma }{\delta X^I}\Big \vert _{\mathcal{S}}\delta X^I+\frac{\delta \Gamma }{\delta A_I}\Big \vert _{\mathcal{S}}\delta A_I =0
\end{align}
up to corner terms. In the last equality the well-definedness of the variational principle for JT was used and the field variations are restricted to obey the boundary conditions. This shows that stationary points are also mapped to stationary points and the submanifold $\Bar{\mathcal{S}}$ indeed describes the classical solution space of the new model.

To give an explicit form, transforming the JT action \eqref{eq:full_action_lorentz} with some target space diffeomorphism leads to the new action
\begin{align}
     \Bar{\Gamma }[\Bar{X}^I,\Bar{A}_I]&=\frac{k}{2\pi}\int_{\mathcal{M}}\left(\Bar{X}^{I}\mathrm{d}\Bar{A}_{I}+\frac{1}{2}\Bar{P}^{IJ}\Bar{A}_{I}\wedge \Bar{A}_{J}\right)\\
     &\hspace{2cm}-\frac{k}{2\pi }\int_{\partial \mathcal{M}}\left(\Bar{X}^I\Bar{A}_I+g_2^{-1}(\Bar{\mathcal{C}})\,\dd f+\Big(\frac{X}{2}\dd \ln \Big |\frac{X^+}{X^-}\Big \vert \Big)\Big \vert _{X^I=X^I(\Bar{X}^K)}\right) \nonumber
\end{align}
where most of the terms take again the same form as in JT because of their covariance. The last term, however, differs in general.

\subsection{The fate of asymptotic symmetries}\label{sec:fate}
Let us now look at how asymptotic symmetries transform. In general, boundary condition preserving gauge transformations are vector fields on $\mathcal{F}$,
\begin{align}\label{eq:gauge_vfs}
    V_\lambda \in \Gamma (T\mathcal{F}) \qquad \qquad V_\lambda =\int _{\mathcal{M}}\dd ^2x \,\Big(\delta _\lambda X^I \frac{\delta }{\delta X^I}+\delta _\lambda A_I \frac{\delta }{\delta A_I}\Big) 
\end{align}
with their components given by the transformations \eqref{PSM04a} and appropriate gauge parameters $\lambda _I$. A consequence of the discussion in section \ref{sec:PSM_gauge_trafos} is that they in general are not in involution, i.e.
\begin{align}
    [V_\lambda ,V_\eta ]=V_{[\lambda ,\eta ]^\ast }+\int _{\mathcal{M}}\dd ^2x \,\Big(\lambda _K\eta _L\partial _I\partial _JP^{KL}\big(\dd X^J+P^{JM}A_M\big)\Big)\frac{\delta }{\delta A_I}
\end{align}
only defines a closed generating set if the EOM for the gauge fields are imposed. However, we have seen that for the JT model this is not necessary as the term spoiling closedness vanishes by linearity of the Poisson tensor. Let us therefore again pick the JT model as presented in sections \ref{sec:JT}-\ref{sec:res_gauge_lorntz} as a starting point and map the asymptotic conditions to some new model. 

The new field configurations $\Bar{\mathcal{F}}$ are still given in terms of the old field variables, which still transform like in \eqref{eq:trafo_metr}. Therefore, on this level, we trivially already know the new boundary condition preserving transformations. In the following, we construct the action of these transformations on the level of the PSM variables of the new model showing that they are still given by gauge transformations. 

For this, we note that the vector fields \eqref{eq:gauge_vfs} are pushed forward to the new field space $\Bar{\mathcal{F}}$, implying that they automatically preserve the boundary conditions obtained for that model. More explicitly, the basis transforms as
\begin{align}
    \frac{\delta }{\delta X^I}=\frac{\partial \Bar{X}^J}{\partial X^I}\frac{\delta }{\delta \Bar{X}^J}+\partial _I\Big(\frac{\partial X^K}{\partial \Bar{X}^J}\Big)\frac{\partial \Bar{X}^M}{\partial X^K}\Bar{A}_M\frac{\delta }{\delta \Bar{A}_J} && \frac{\delta }{\delta A_I}=\frac{\partial \Bar{X}^I}{\partial X^J}\frac{\delta }{\delta \Bar{A}_J} 
\end{align}
while the components are related by
\begin{align}
    \delta _{\Bar{\lambda }}\Bar{X}^I=\frac{\partial \Bar{X}^I}{\partial X^J}\delta _\lambda X^J && \delta _{\Bar{\lambda }}\Bar{A}_I=\frac{\partial X^J}{\partial \Bar{X}^I}\delta _\lambda A_J +\delta _\lambda X^J\partial _J\Big(\frac{\partial X^K}{\partial \Bar{X}^I}\Big)A_K ~.
\end{align}
Writing out the latter expressions yields
\begin{align}
    \delta _{\Bar{\lambda }} \Bar{X}^I&=\Bar{P}^{IJ}\frac{\partial X^K}{\partial \Bar{X}^J}\lambda _K\label{eq:new_gaugetrafo1}\\
    \delta _{\Bar{\lambda }} \Bar{A}_I&=-\dd \Big(\frac{\partial X^J}{\partial \Bar{X}^I}\lambda _J\Big)-\partial _{\Bar{I}}\Bar{P}^{JK}\Bar{A_J}\frac{\partial X^M}{\partial \Bar{X}^K}\lambda _M+\lambda _M\frac{\partial ^2X^M}{\partial \Bar{X}^I\partial \Bar{X}^J}\Big(\dd \Bar{X}^J+\Bar{P}^{JK}\Bar{A}_K\Big) ~,\label{eq:new_gaugetrafo2}
\end{align}
where the $\Bar{\lambda }_I$ on the left-hand side are understood to depend implicitly on the $\lambda _I$ on the right-hand side in some yet-to-be-determined way. These transformations leave the action of the new model invariant up to boundary terms so they are indeed still gauge transformations. Moreover, because the push-forward is compatible with the Lie-brackets on $\mathcal{F}$ and $\Bar{\mathcal{F}}$,
\begin{align}\label{eq:new_closedness}
    \Phi _\ast [V_\lambda ,V_\eta ]=[\Phi _\ast V_\lambda ,\Phi _\ast V_\eta ]=\Phi _\ast V_{[\lambda ,\eta ]^\ast }
\end{align}
the vector fields are still in involution off-shell and form a representation of the same algebra as for the JT model. If we define the gauge parameters of the new model as
\begin{equation}
     \bar{\lambda}_{I}=\frac{\partial X^J}{\partial\bar{X}^{I}}\, \lambda_J \label{barlambda}
\end{equation}
the transformations \eqref{eq:new_gaugetrafo1}, \eqref{eq:new_gaugetrafo2} read
\begin{align}
    \delta _{\Bar{\lambda }}\Bar{X}^I&=\Bar{P}^{IJ}\Bar{\lambda }_J\label{eq:new_gaugetrafo3}\\
    \delta _{\Bar{\lambda }}\Bar{A}_I&=-\dd \Bar{\lambda }_I-\partial _{\Bar{I}}\Bar{P}^{JK}\Bar{A}_J\Bar{\lambda }_K +\Bar{\lambda }_K M^K_{IJ}\Big(\dd \Bar{X}^J+\Bar{P}^{JL}\Bar{A}_L\Big) \label{eq:new_gaugetrafo4}
\end{align}
where
\begin{align}
    M^K_{IJ}=\frac{\partial ^2 X^M}{\partial \Bar{X}^I\partial \Bar{X}^J}\frac{\partial \Bar{X}^K}{\partial X^M}=M^K_{JI} ~.
\end{align}
The transformations \eqref{eq:new_gaugetrafo3}-\eqref{eq:new_gaugetrafo4} can be identified as just a different generating set from \eqref{PSM04a}. This is the reason why \eqref{eq:new_closedness} does not contradict the closedness properties of the PSM gauge transformations, as different generating sets can have very different algebraic properties \cite{Henneaux:1992}.

Moreover, the transformation behavior \eqref{barlambda} implies that the charge variations are invariant under the target space diffeomorphism, 
\begin{align}
    \slashed \delta Q_\lambda =\frac{k}{2\pi }\lambda_I\,\delta X^I=\frac{k}{2\pi }\Bar{\lambda}_I\,\delta \Bar{X}^I 
\end{align}
excluding the possibility that improper gauge transformations are mapped to proper ones. This means that target space diffeomorphisms can be interpreted as changes of state space slicings.

\section{Maps from Lorentzian JT to models with \texorpdfstring{$\boldsymbol{\bar{U}=0}$}{U=0}}\label{sec:U0}
In this section, we study global aspects of target space diffeomorphisms. In particular, we shall see that the topology of the symplectic leaves does not always allow a globally well-defined map. For simplicity, we restrict to the case $\bar{U}(\bar X)=0$.

\subsection{Conformally transformed CGHS}\label{sec:CGHS}
There is a model that is even simpler than JT. It corresponds to $\bar U=0$ and $V=-\alpha$, where $\alpha$ is a constant. This model can be obtained from CGHS by a dilaton-dependent rescaling of the metric. Then,
\begin{equation}
    \bar w(\bar X)=\alpha \bar X \qquad \qquad \bar{\mathcal{C}}=\bar X^+\bar X^- +\alpha \bar{X}\label{CGHSw}
\end{equation}
so that classical solutions are Rindler spacetimes with the Unruh temperature $T\propto\alpha$ and the ADM energy $\propto \bar{\mathcal{C}}$ (precise coefficients may be found e.g.~in \cite{Liebl:1996ti}).

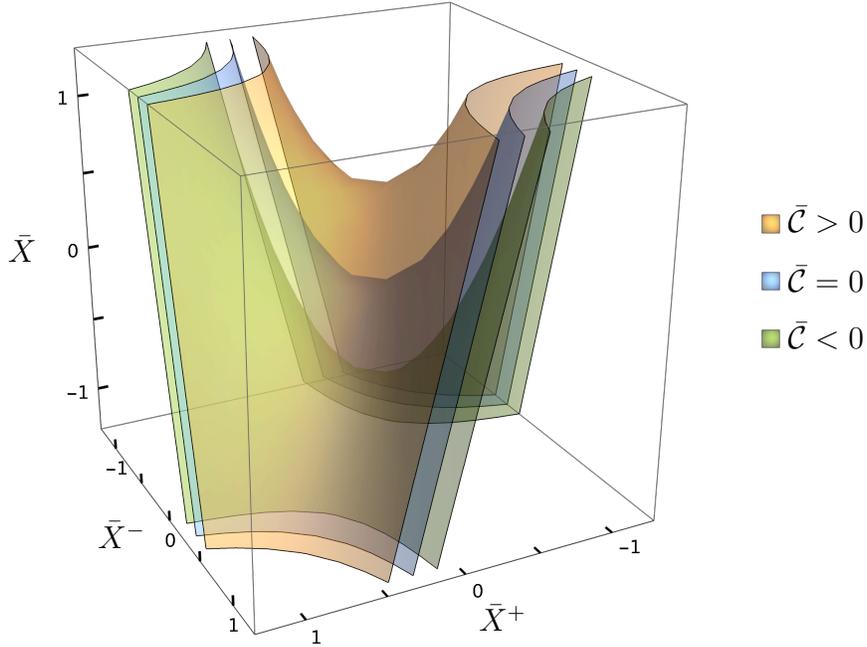
\begin{figure}
 \begin{center}
\begin{tikzpicture}
    \node[anchor=south west,inner sep=0] (image) at (0,0) {\includegraphics[width=11cm]{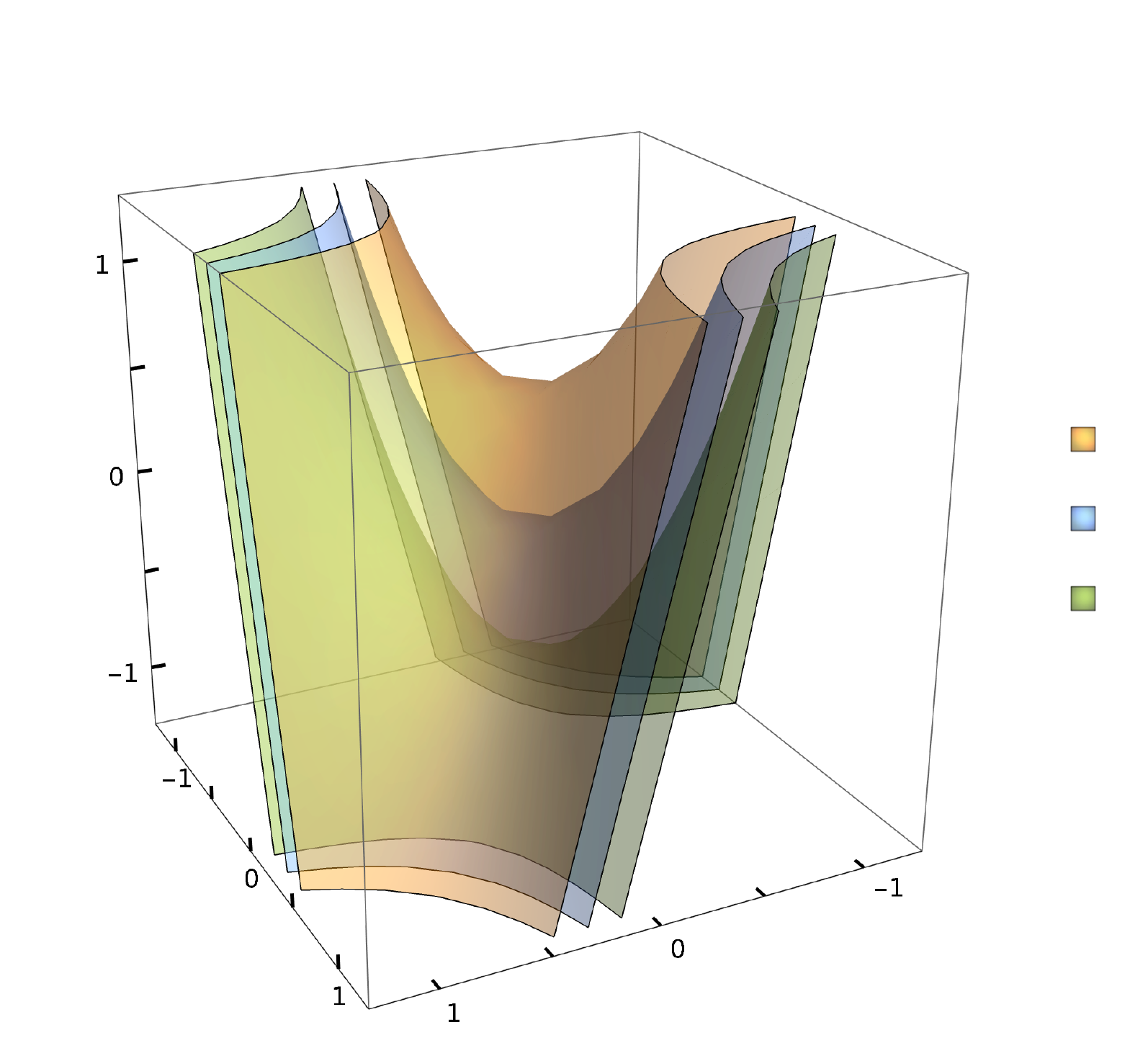}};
    \begin{scope}[x={(image.south east)},y={(image.north west)}]
        \node[label=above:$\bar X^-$] (A) at (0.16,0.12) {};
         \node[label=above:$\bar X^+$] (A) at (0.62,0.01) {};
          \node[label=above:$\Bar{X}$] (A) at (0.04,0.5) {};
          \node[label=above:$\Bar{\mathcal{C}}>0$] (A) at (1.01,0.538) {};
           \node[label=above:${\Bar{\mathcal{C}}=0}$] (A) at (1.01,0.46) {};
          \node[label=above:$\Bar{\mathcal{C}}<0$] (A) at (1.01,0.384) {};
    \end{scope}
\end{tikzpicture}
\caption{Symplectic leaves of the conformally rescaled CGHS model with $\alpha =\frac{1}{2}$. The symplectic leaves for $\Bar{\mathcal{C}}>0$ are simply connected as opposed to the leaves with $\mathcal{C}>0$ of JT.}
\label{fig:CGHShat_targetspace}
\end{center}
 \end{figure}

In this model, the Poisson tensor has always rank 2. The symplectic leaves are paraboloids of a constant $\bar{\mathcal{C}}$ as can be seen in figure \ref{fig:CGHShat_targetspace}. 

Let us try to map symplectic leaves of JT with $\mathcal{C}>0$ to these symplectic leaves. The transformation \eqref{T10a}-\eqref{T10c} is nonsingular if 
\begin{equation}
    g_2(\mathcal{C})\mp \alpha \sqrt{2\mathcal{C}}-\alpha g_1(\mathcal{C})=0 ~.\label{CGHSg2}
\end{equation}
(The sign factor $\mp$ appears in this equation since for a positive $\mathcal{C}$ the symplectic leaves of JT contain two horizon points.)
Obviously, this equation cannot be satisfied simultaneously for both signs in front of the square root. This is an expected result since symplectic leaves of JT and conformally transformed CGHS have different topologies. On the other hand, if we restrict ourselves to positive values of $X$ in JT (corresponding to the upper sign in \eqref{CGHSg2}), the equation always has a solution for any choice of $g_2$. To perform the most general (up to a gauge transformation) Poisson diffeomorphism between the selected part of the JT target space and the target space of the conformally transformed CGHS model one has two options. One can make an essential Poisson automorphism on the JT side and compose it with the transformation to CGHS with arbitrary but fixed $g_2$. Alternatively, one can fix the essential automorphism of JT to the identity but consider any choice of $g_2$ in the map to CGHS.

\subsection{Global equivalence to JT}\label{sec:glob_JT}
In this subsection, we describe the dilaton gravity models with $\bar U=0$ whose target spaces $\bar{\mathcal{P}}$ are globally Poisson diffeomorphic to the target space $\mathcal{P}_{\mathrm{JT}}$ of JT gravity. Let us start with symplectic leaves in $\mathcal{P}_{\mathrm{JT}}$ corresponding to $\mathcal{C}>0$. We have to find two functions, $g_1$ and $g_2$, such that \eqref{g2g1} holds for a given $\bar w$ and two values of $X_h=\pm\sqrt{2\mathcal{C}}$ for each positive $\mathcal{C}$. Thus, similarly to \eqref{sqrt2}, we need 
\begin{equation}
    \bar w\bigl(g_1(\mathcal{C}) +\sqrt{2\mathcal{C}}\bigr)=\bar w\bigl( g_1(\mathcal{C}) -\sqrt{2\mathcal{C}}\bigr) ~.
    \label{barwbarw}
\end{equation}
Since all functions involved in this equation are at least differentiable, we can extend \eqref{barwbarw} to $\mathcal{C}=0$.
The shifted function $\bar w_{g_1}\bigl(z\bigr):=\bar w\bigl(z+g_1(z^2/2)\bigr)$ is an even function on $\mathbb{R}$. If there is a function $g_1$ that solves \eqref{barwbarw}, then for non-negative values of $\mathcal{C}$
\begin{equation}
    g_2(\mathcal{C})=\bar w_{g_1}\bigl(\sqrt{2\mathcal{C}}\bigr) ~.\label{g2C}
\end{equation}
Since $g_2$ defines a diffeomorphism, it has to be smooth and invertible. For all this to work, $\bar w$ needs to  be a smooth function on $\mathbb{R}$ which takes exactly twice all values in its image (except for one value which is taken once and corresponds to $\mathcal{C}=0$). A suitable function should start with some value $\bar w(-\infty)$, monotonously decrease (respectively, increase) till some point $\bar X=\bar X_0$, and afterward monotonously increase (respectively, decrease) reaching the value $\bar w(+\infty)=\bar w(-\infty)$. The limiting values may be infinite. We call such functions \textit{admissible}\footnote{If one restricted the target space manifold to $X>0$ from the beginning, this condition was relaxed such that more functions would become admissible. Then, it is sufficient to have a function $\bar w$ that monotonously increases (decreases) from $\bar X_0$ on.}. (We shall add another restriction on $\bar w$ and thus complete this definition below.)

The admissibility of $\bar w$ is clearly necessary for the existence of a global Poisson diffeomorphism between  $\mathcal{P}_{\mathrm{JT}}$ and $\bar{\mathcal{P}}$. Let us check if it is also sufficient. Any admissible function $\bar w$ defines a set of pairs $\{ (\bar X_{(+)},\bar X_{(-)})\}$ such that $\bar w(\bar{X}_{(+)})=\bar w(\bar{X}_{(-)})$ and $\bar{X}_{(+)}>\bar{X}_{(-)}$ (except for a single degenerate pair with $\bar{X}_{(+)}=\bar{X}_{(-)}=\bar X_0$). If we define
\begin{equation}
    \sqrt{2\mathcal{C}}=\frac 12 \left(  \bar{X}_{(+)}-\bar{X}_{(-)} \right) 
    \label{CXpm}
\end{equation}
it is easy to see that one has a one-to-one correspondence between non-negative $\mathcal{C}$ and the set of pairs. Then, 
\begin{equation}
    g_1(\mathcal{C}) =\frac 12 \left(  \bar{X}_{(+)}+\bar{X}_{(-)} \right) \label{g1C}
\end{equation}
defines a unique solution of \eqref{barwbarw}. The function $g_2$ is uniquely defined by \eqref{g2C}.

The function $\bar w$ should have an extremum at the point $\bar X_0$. Let us assume for simplicity that this point is a minimum $\bar{\mathcal{C}}_0$. The other case is treated similarly. Let $\bar{\mathcal{C}}_{\mathrm{max}}\equiv \bar w(\pm\infty)$. Then, $g_2$ maps the interval $[0,+\infty)$ to $[\bar{\mathcal{C}}_0,\bar{\mathcal{C}}_{\mathrm{max}})$. The space $\bar{\mathcal{P}}$ contains symplectic leaves with any values of $\bar{\mathcal{C}}\in (-\infty,+\infty)$. Therefore, to really have a global identification, the function $g_2$ has to map the interval $(-\infty,0)$ to the rest of the allowed values of $\bar{\mathcal{C}}$, i.e., to $(-\infty,\bar{\mathcal{C}}_0)\,\cup\, (\bar{\mathcal{C}}_{\mathrm{max}},+\infty)$. Since $g_2$ is continuous and even smooth, this is not possible unless $\bar{\mathcal{C}}_{\mathrm{max}}=+\infty$. We shall call admissible only those functions $\bar w$ for which $\bar w(\pm\infty)=+ \infty$. Now, one can check that there are no more obstructions to the existence of a Poisson isomorphism between $\mathcal{P}_{\mathrm{JT}}$ and $\bar{\mathcal{P}}$. Thus, we have the following result:

\medskip

\noindent \emph{
The target space of the JT model is Poisson isomorphic to the target space of another dilaton gravity with $\bar U=0$ if and only if the function $\bar w$ is admissible. 
}

\paragraph{Example 1: Perturbations of the JT model.}
Recently, various perturbations of the JT model were considered in the literature \cite{Witten:2020wvy,Momeni:2020tyt,Johnson:2020lns,Turiaci:2020fjj,Alishahiha:2020jko}.
In particular, the paper \cite{Witten:2020wvy}
studied the models with
\begin{equation}\bar V(\bar X)=-\bar X-\beta \,e^{-\alpha \bar X} \label{VWitten}
\end{equation}
where $\beta$ and $\alpha$ are some constants, $\alpha>0$, and $\beta$ is a perturbation parameter. The potential function reads
\begin{equation}
    \bar w(\bar X)=\frac 12 \bar{X}^2 -\frac {\beta}{\alpha} e^{-\alpha \bar X}
\end{equation}
and is admissible iff $\beta\leq 0$. To find the precise form of $g_1$ and $g_2$ for this example one has to solve transcendental equations.

\paragraph{Example 2: A map from AdS$_{\mathbf{2}}$ to dS$_{\mathbf{2}}$.} Let us consider a dilaton potential 
\begin{equation}
    \bar V=+\bar X \label{minusJT}
\end{equation}
which is obtained by an inversion of the sign in front of $V_{\mathrm{JT}}$. All classical solutions for this model have a constant positive scalar curvature $R=2$ and thus locally correspond to dS$_{2}$. The potential $\bar w= -\tfrac 12 \bar{X}^2 $ is admissible. A Poisson isomorphism between $\mathcal{P}_{\mathrm{JT}}$ and the target space of this model corresponds to
\begin{equation}
    g_1=0 \qquad \qquad g_2(\mathcal{C})=-\mathcal{C}
\end{equation}
or
\begin{equation}
    \bar X=X \qquad \qquad \bar{X}^+=X^+ \qquad \qquad \bar{X}^-=-X^- \label{AdSdS}
\end{equation}
and $\bar e^-=-e^-$, so that the sign of the metric is inverted and AdS$_2$ is mapped to dS$_2$. The dS$_2$ horizons appear at negative values of $\bar{\mathcal{C}}$. The transformation \eqref{AdSdS} maps horizons to horizons.

\section{Euclidean dilaton gravities in 2D}\label{sec:Edil}

Let us describe the structure of 2D dilaton gravity with Euclidean signature. Here, we mostly follow the conventions of \cite{Bergamin:2004pn} but slightly change the notations. 
As in Lorentzian signature, the action \eqref{D08} can be converted to a first-order form
\begin{equation}
\mathrm{I_{1st}}=\frac{k}{2\pi}\int_{\mathcal{M}}\left(X\mathrm{d}\omega+X^{a}\left(\mathrm{d}e_{a}+\epsilon_{a}^{\;b}\omega\wedge e_{b}\right)-\frac{1}{2}\mathcal{V}\epsilon^{ab}e_{a}\wedge e_{b}\right)\label{E1st}
\end{equation}
where the Latin indices $a,b$ take values $\underline{1},\underline{2}$. They are raised and lowered with the Euclidean metric $\delta_{ab}=\mathrm{diag}\left(1,1\right)$ and $\epsilon_{ab}$ is the Levi--Civit\'a tensor, $\epsilon_{\underline{1}\underline{2}}=1$. The potentials are now given by $\mathcal{V}=\frac{1}{2}U\left(X\right)\delta_{ab}X^{a}X^{b}+V\left(X\right)$.
After an integration by parts, the action \eqref{E1st} becomes that of a PSM \eqref{PSM02} with $X^I=(X,X^{\underline{1}},X^{\underline{2}})$ and $A_I=(\omega,e_{\underline{1}},e_{\underline{2}})$. The Poisson tensor is defined by
\begin{equation}
    P^{ab}=-\epsilon^{ab}\mathcal{V} \qquad \qquad P^{Xb}=X^a\epsilon_a^{\ b} ~.\label{EPSM}
\end{equation}
It is convenient to introduce complex fields
\begin{equation}
X_{\mathrm{E}}^{\pm}\equiv\frac{1}{\sqrt{2}}\left(X^{\underline{1}}\pm iX^{\underline{2}}\right) \qquad \qquad e^{\pm}_{\mathrm{E}}\equiv\frac{1}{\sqrt{2}}\left(e^{\underline{1}}\pm ie^{\underline{2}}\right)  
\end{equation}
that are analogs of the light-cone variables in Lorentzian signature. In terms of these variables,
\begin{align}
\mathrm{I_{1st}}=\frac{k}{2\pi}\int_{\mathcal{M}}\big (X\mathrm{d}\omega &+X_{\mathrm{E}}^{+}\big (\mathrm{d}e^{-}_{\mathrm{E}}+i\omega\wedge e^{-}_{\mathrm{E}}\big )\\
&+X_{\mathrm{E}}^{-}\big (\mathrm{d}e^{+}_{\mathrm{E}}-i\omega\wedge e^{+}_{\mathrm{E}}\big )+i\mathcal{V}e^{-}_{\mathrm{E}}\wedge e^{+}_{\mathrm{E}}\big ) \nonumber ~.  
\end{align}
The complexified target space coordinates are $X^{I}=\left(X,X_{\mathrm{E}}^{+},X_{\mathrm{E}}^{-}\right)$ and the one-forms are grouped as $A_{I}=\left(\omega ,e^-_{\mathrm{E}},e^+_{\mathrm{E}}\right)$. The corresponding Poisson tensor has components
\begin{equation}
P^{X\pm}=\pm iX_{\mathrm{E}}^{\pm} \qquad \qquad P^{+-}=i\mathcal{V} ~.    \label{complexP}
\end{equation}
We stress that the target space remains real and the complex variables are introduced just for technical convenience. The EOM are solved more easily if one considers the $+$ and $-$ components as independent fields, though at the end one has to impose the reality conditions
\begin{equation}
X_{\mathrm{E}}^{-}	=	\left(X_{\mathrm{E}}^{+}\right)^{*} \qquad \qquad
e^{-}_{\mathrm{E}}	=	\left(e^{+}_{\mathrm{E}}\right)^{*}.  \label{CD05}  
\end{equation}
The functions $w(X)$ and $Q(X)$ are defined exactly as in the Lorentzian case, see \eqref{D05}. The Casimir function
\begin{equation}
\mathcal{C}\equiv w-e^{Q}X_{\mathrm{E}}^{+}X_{\mathrm{E}}^{-}    
\end{equation}
is absolutely conserved on-shell, $\dd\mathcal{C}=0$, and the identity
\begin{equation}
\left(\partial_{X}-\mathcal{V}\partial_{Y}\right)\mathcal{C}=0    
\end{equation}
holds true. By taking $X$ as one of the coordinates, one can write the line element of a generic solution as \cite{Bergamin:2004pn}
\begin{equation}
\mathrm{d}s^{2}=2e^{Q}\left(w-\mathcal{C}\right)\left(\mathrm{d}\theta\right)^{2}+\frac{e^{Q}}{2\left(w-\mathcal{C}\right)}\left(\mathrm{d}X\right)^{2} ~. \label{Emet}  
\end{equation}
This line element corresponds to a positive definite metric as long as
\begin{equation}
    w(X)-\mathcal{C}\geq 0 ~.
\end{equation}
The points where 
\begin{equation}
    w(X_h)=\mathcal{C} \qquad \qquad X_{\mathrm{E}}^{+}=X_{\mathrm{E}}^{-}=0 \label{Esing}
\end{equation}
are coordinate singularities of \eqref{Emet}. They are Euclidean horizons (tips) of the geometries. To avoid conical singularities at these points, the coordinate $\theta$ has to be periodic with 
\begin{equation}
    \theta \sim \theta +\beta _\theta \qquad \qquad \beta _\theta ^{-1}=\frac{|w'|}{2\pi }\Big \vert _{X=X_h} ~.
\end{equation}

\subsection{Euclidean JT gravity}\label{sec:EJT}
Euclidean JT gravity describes spaces with constant negative curvature $R=-2$ and is given by the potentials
\begin{equation}
U\left(X\right)=0 \qquad \qquad V\left(X\right)=-X ~.    
\end{equation}
Therefore, one has
\begin{equation}
Q=0 \qquad \qquad w=\frac{X^{2}}{2} \qquad \qquad\mathcal{C}=\frac{1}{2}\left(X^{2}-\left(X^{\underline{1}}\right)^{2}-\left(X^{\underline{2}}\right)^{2}\right) ~.    
\end{equation}
Euclidean horizons are located where $X^{\underline{1}}=X^{\underline{2}}=0$ and thus $X=X_{h}=\pm\sqrt{2\mathcal{C}}$. 

\subsection{Boundary conditions and solution space for Euclidean JT}\label{sec:bcsp}
We are ultimately interested in identifying the black hole sectors of two dilaton gravity theories, so let us look at solutions with temperature $1/\beta $ and compactify Euclidean time $\tau $ to a circle with $\tau \sim \tau +\beta $. Labeling the radial coordinate on the world-sheet by $\rho $ we choose a gauge
\begin{equation}\label{eq:gauge_conditions}
e_{\rho}^{\underline{1}}=0\qquad \qquad e_{\rho}^{\underline{2}}=1\qquad \qquad \omega_{\rho}=0 
\end{equation}
and define $e_{\tau}^{\underline{1}}=h$, $e_{\tau}^{\underline{2}}=j$, where $h$ and $j$ are arbitrary functions of $\left(\rho,\tau\right)$.
In a second-order formulation, this corresponds to a line element in generalized Fefferman--Graham gauge
\begin{equation}\label{eq:line_element_Eucl}
\mathrm{d}s^{2}=\mathrm{d}\rho^{2}+2j\mathrm{d}\tau\mathrm{d}\rho+\left(h^{2}+j^{2}\right)\mathrm{d}\tau^{2}    
\end{equation}
like it was previously considered in \cite{Grumiller:2017qao}. Let us briefly recapitulate the solution space as it is obtained in that paper. In the chosen gauge, the EOM involving the Cartan variables reduce to
\begin{equation}
\partial_{\rho}h-\omega_{\tau}=0\qquad\qquad \partial_{\rho}j=0\qquad \qquad \mathrm{\partial_{\rho}}\omega_{\tau}-h=0 
\end{equation}
and have the general solution 
\begin{equation}\label{eq:EJTsol1}
h=\mathcal{L}^{+}e^{\rho}-\mathcal{L}^{-}e^{-\rho}\qquad \qquad j=\mathcal{L}^{0} \qquad \qquad \omega_{\tau}=\mathcal{L}^{+}e^{\rho}+\mathcal{L}^{-}e^{-\rho} 
\end{equation}
where $\mathcal{L}^{\pm,0}=\mathcal{L}^{\pm,0}\left(\tau\right)$. The radial EOM for the scalar fields are
\begin{equation}
\partial_{\rho}X+X^{\underline{1}}=0 \qquad \qquad\partial_{\rho}X^{\underline{1}}+X=0 \qquad \qquad\partial_{\rho}X^{\underline{2}}=0
\end{equation}
and are solved analogously by
\begin{equation}\label{eq:EJTsol2}
X=x^{+}e^{\rho}+x^{-}e^{-\rho}\qquad X^{\underline{1}}=-x^{+}e^{\rho}+x^{-}e^{-\rho}\qquad X^{\underline{2}}=x^{0}    
\end{equation}
where $x^{\pm,0}=x^{\pm,0}\left(\tau\right)$. Inspired by these solutions, we choose boundary conditions for the fields at $\rho\rightarrow\infty$,
\begin{subequations}\label{eq:asymp_cond_Eucl_JT}
\begin{align}
    X&=x^+\,e^\rho +x^-\,e^{-\rho }+\mathcal{O}(e^{-2\rho }) & \omega _\tau &=\partial_\rho e_\tau^{\underline{1}}+{\cal O}(e^{-2\rho})\\
    X^{\underline{1}}&=-\partial _\rho X & e^{\underline{1}}_\tau &=\mathcal{L}^+e^\rho -\mathcal{L}^-e^{-\rho }+\mathcal{O}(e^{-2\rho })\\
X^{\underline{2}}&=x^0+\mathcal{O}(e^{-\rho}) & e^{\underline{2}}_\tau &=\mathcal{L}^0 +\mathcal{O}(e^{-2\rho })~.
\end{align}
\end{subequations}
On-shell, the fields in \eqref{eq:EJTsol1} and \eqref{eq:EJTsol2} are further related by the temporal EOM
\begin{eqnarray}\label{eq:boundary_eom}
\partial_{\tau}x^{\pm}\mp x^{\pm}\mathcal{L}^{0}\mp x^{0}\mathcal{L}^{\pm}	=	0\qquad \qquad
\partial_{\tau}x^{0}-2x^{-}\mathcal{L}^{+}+2x^{+}\mathcal{L}^{-}	=	0
\end{eqnarray}
such that a solution can be entirely specified by providing three functions $(\mathcal{L}^\pm (\tau),\mathcal{L}^0(\tau))$ together with three initial conditions for the boundary EOM. In this parametrization, the Casimir reads 
 \eq{
     \mathcal{C}=2x^+x^--\frac{(x^0)^2}{2} 
 }{eq:neweq}
 and satisfies $\partial_\tau \,\mathcal{C}=0$ if all three equations \eqref{eq:boundary_eom} are imposed. 

To have a well-defined variational principle, we need to augment the bulk action $\mathrm{I_{1st}}$ by a boundary term at a $\rho =\text{const.}$ surface. Slightly rewriting the appropriate boundary terms constructed in \cite{Grumiller:2017qao} yields
\begin{align}\label{eq:nolabel}
    \Gamma ^E=\frac{k}{2\pi}\int_{\mathcal{M}}\Big (X^{I}&\mathrm{d}A_{I}+\frac{1}{2}P^{IJ}A_{I}\wedge A_{J}\Big )\\
    &-\frac{k}{2\pi }\int_{\partial \mathcal{M}}\Big (X^IA_I-\mathcal{C}\,\dd f-X\dd \tan ^{-1}\frac{X^{\underline{2}}}{X^{\underline{1}}} \Big ) \nonumber
\end{align}
where the last term is just the analytic continuation of the corresponding counterterm in the Lorentzian action \eqref{eq:full_action_lorentz}. The field $f(\tau)$ is related to the other field variables by
\begin{align}
    \dd f=f'\,\dd \tau = \frac{\mathcal{L}^+}{x^+}\,\dd \tau 
\end{align}
and we impose as an additional boundary condition that the zero mode 
\begin{align}
   \frac{1}{\hat{y}}:= \frac{1}{\beta }\int\limits_0^\beta \frac{\mathcal{L}^+}{x^+}\dd \tau \label{eq:zeromode:bc}
\end{align}
remains fixed, i.e., $\delta \hat{y}=0$ \cite{Grumiller:2017qao,Godet:2020xpk}. This makes the first variation of the action vanish on-shell. The action then evaluates to
\begin{align}
    \Gamma ^E&\approx -\frac{k}{2\pi }\int\limits_0^\beta \dd \tau \, \left(2\mathcal{L}^-x^++\mathcal{L}^0x^0+\frac{(x^0)^2\mathcal{L}^+}{2x^+}+{x^0}'-\frac{x^0{x^+}'}{x^+} \right)+\mathcal{O}(e^{-\rho })\label{action_on_constr}\\
    &\approx -\frac{k}{2\pi }\int\limits_0^\beta \dd \tau \, \frac{\mathcal{L}^+}{x^+}\,\mathcal{C}=-\frac{k\beta}{2\pi \hat{y}}\,\mathcal{C} \label{eq:Eucl_osaction}
\end{align}
where the first line is obtained without imposing the temporal EOM and the last line is the full on-shell result. 

 \subsection{Residual gauge transformations for Euclidean JT}\label{subsec:resEucl}
The PSM gauge parameters preserving the conditions \eqref{eq:gauge_conditions} are
\begin{equation}\label{eq:Eucl_PSM_parameters}
\lambda_{X}=\varepsilon^{+}e^{\rho}+\varepsilon^{-}e^{-\rho} \qquad \qquad\lambda_{\underline{1}}=\varepsilon^{+}e^{\rho}-\varepsilon^{-}e^{-\rho} \qquad\qquad\lambda_{\underline{2}}=\varepsilon^{0}    
\end{equation}
where $\varepsilon^{\pm,0}=\varepsilon^{\pm,0}\left(\tau\right)$ are independent functions. They act on the boundary fields via
\begin{align}
\delta_{\lambda }x^{\pm}&=\mp x^{\pm}\varepsilon^{0}\mp x^{0}\varepsilon^{\pm} & \delta_{\lambda}\mathcal{L}^{\pm}&=-\partial_{\tau}\varepsilon^{\pm}\mp\mathcal{L}^{\pm}\varepsilon^{0}\pm\mathcal{L}^{0}\varepsilon^{\pm} \\	
\delta_{\lambda}x^{0}&=-2x^{-}\varepsilon^{+}+2x^{+}\varepsilon^{-} & \delta_{\lambda}\mathcal{L}^{0}&=-\partial_{\tau}\varepsilon^{0}+2\mathcal{L}^{-}\varepsilon^{+}-2\mathcal{L}^{+}\varepsilon^{-} ~.
\end{align}
In this slicing, the canonical boundary charges are integrable and form a representation of the centerless $\mathfrak{sl}(2)$ current algebra \cite{Grumiller:2017qao}. Here we choose a different slicing such that we get the same symmetry group as in the Lorentzian section. This also assures the presence of a subsector with Schwarzian dynamics.

First, we express the gauge parameters in terms of new field-independent parameters $(\varepsilon, \eta, \gamma)$
\begin{equation}\label{eq:COS_params}
\varepsilon^{+}=-\mathcal{L}^+\varepsilon \qquad \qquad\varepsilon^{-}=-\frac{\eta }{2\mathcal{L}^+}-\mathcal{L}^-\varepsilon \qquad \qquad\varepsilon^{0}=-\gamma-\mathcal{L}^0\varepsilon ~.
\end{equation}
We redefine the fields as
\begin{subequations}\label{eq:COS_fields}
\begin{align}
    \mathcal{L}^{+}&=e^{\Theta} & x^{+}&=e^{\Theta}y^{+}\\
    \mathcal{L}^{-}&=e^{-\Theta}\Big (\frac{1}{2}\mathcal{T}+\frac{1}{4}\mathcal{P}^2\Big ) & x^{-}&=e^{-\Theta}y^{-} \\
    \mathcal{L}^{0}&=-\mathcal{P} & x^{0}&=-y^{0} 
\end{align}
\end{subequations}
which makes the boundary EOM take the form
\begin{align}
E_+&:=\partial_{\tau}y^{+}+\big (\mathcal{P}+\partial _\tau \Theta \big ) y^{+}+y^{0} =0\label{eq:boundaryeom1}\\
E_0&:=\partial_{\tau}y^{0}+2y^{-}-y^{+}\Big (\mathcal{T}+\frac{1}{2}\mathcal{P}^2\Big )=0 \label{eq:boundaryeom2} \\
E_{-}&:=\partial_{\tau}y^{-}-y^{0}\Big (\frac{1}{2}\mathcal{T}+\frac{1}{4}\mathcal{P}^2\Big )-\big (\mathcal{P}+\partial _\tau \Theta \big )y^{-}=0\label{eq:boundaryeom3}~.
\end{align}
The new fields then transform as
\begin{align}
 \delta _\lambda \mathcal{T}&=\varepsilon \mathcal{T}'+2\varepsilon '\mathcal{T}+\mathcal{P}\gamma '-\Theta '\eta +\eta ' & \delta _\lambda y^+&=\varepsilon y^+{}'-\varepsilon 'y^+-\varepsilon \,E_+ \label{eq:trafo1}\\
    \delta _\lambda \mathcal{P}&=\varepsilon \mathcal{P}'+\varepsilon '\mathcal{P}-\eta -\gamma ' & \delta _\lambda y^0&=\varepsilon y^0{}'+\eta y^+-\varepsilon \,E_0 \label{eq:trafo2}\\
    \delta _\lambda \Theta &=\varepsilon \Theta '+\varepsilon '+\gamma  & \delta _\lambda y^-&=\varepsilon y^-{}'+\varepsilon 'y^-+\frac{\eta }{2}y^0-\varepsilon \, E_{-}\label{eq:trafo3}~.
\end{align}
On-shell $y^+$ transforms as a boundary vector field, $y^0$ as a scalar, and $y^-$ as a one-form. We again use the brackets \eqref{eq:mod_brack}, explicitly taking into account variations of the gauge parameters. This leads to the relation
\begin{align}
    \Big [\lambda [\varepsilon _1,\gamma _1,\eta _1],\lambda [\varepsilon _2,\gamma _2,\eta _2]\Big ]_I^\ast =\lambda _I\big [[\varepsilon _1,\varepsilon _2],\varepsilon _1\gamma _2'-\varepsilon _2\gamma _1',(\varepsilon _1\eta _2)'-(\varepsilon _2\eta _1)'\big ] 
\end{align}
with the functions $\lambda _I$ defined such that \eqref{eq:Eucl_PSM_parameters} is obtained for $\lambda _I[-\varepsilon ,-\eta ,-\gamma ]$ after using \eqref{eq:COS_params}-\eqref{eq:COS_fields}. One can see that the free functions have definite weights under the transformations generated by $\varepsilon $. While $\varepsilon $ itself transforms like a vector field, $\gamma $ transforms like a scalar function and $\eta $ transforms like a one-form. Very similarly to section \ref{sec:res_gauge_lorntz} the $\lambda _I$ form a representation of a Lie algebra whose associated group is
\begin{align}\label{eq:non_ext_group}
    \mathcal{G}=\text{Diff}(S^1)\ltimes \left(C^\infty (S^1)\times \Omega ^1(S^1)\right) ~.
\end{align}
The multiplication between two elements $(f_1,g_1,h_1), (f_2,g_2,h_2) \in \mathcal{G}$ is given by
\begin{align}
    (f_1,g_1,h_1)\cdot (f_2,g_2,h_2)=(f_1\circ f_2,g_1+g_2\circ f_1^{-1},h_1+(f_1^{-1})^\ast h_2) ~.
\end{align}
One-forms $h(\tau )\dd \tau \in \Omega ^1(S^1)$ are acted upon by pullback, $(f^\ast h)(\tau )\dd \tau =h(f(\tau ))f'(\tau )\dd \tau$, while diffeomorphisms act on functions $g(\tau )\in C^\infty (S^1)$ via composition. Moreover, the elements need to satisfy the (quasi-)periodicity conditions
\eq{
    f(\tau +\beta )=f(\tau )+\beta \qquad\qquad g(\tau +\beta )=g(\tau ) \qquad\qquad h(\tau +\beta )=h(\tau ) ~.
}{eq:ok}
The left side of \eqref{eq:trafo1}-\eqref{eq:trafo3} can be identified as the infinitesimal coadjoint action of the asymptotic symmetry group, which is a certain central extension $\hat{\mathcal{G}}$ of \eqref{eq:non_ext_group}. Referring to appendix \ref{app:coadjoint} for the details, one finds the finite transformations 
\begin{align}
    \Tilde{\mathcal{T}}(f(\tau ))&=\frac{1}{(f')^2}\Big (\mathcal{T}-\mathcal{P}(g\circ f)'+\Theta 'f'(h\circ f)\label{eq:coadjT}\\
    &\qquad \qquad-\Big ((h\circ f)f'\Big )'-f'(g\circ f)'(h\circ f)-\frac{1}{2} (g\circ f)'^2  \Big )\nonumber \\
    \Tilde{\mathcal{P}}(f(\tau ))&=\frac{1}{f'}\Big (\mathcal{P}+ (h\circ f)f'+ (g\circ f)'\Big )\label{eq:coadjP}\\
    \Tilde{\Theta }(f(\tau ))&=\Theta - \ln f'- g\circ f ~. \label{eq:coadjTheta}
\end{align}
Choosing Fourier modes 
\begin{align}\label{eq:fourier_modes}
    T_n&:=\lambda _I(\varepsilon =\frac{i\beta }{2\pi }e^{i\frac{2\pi }{\beta }n\tau },0,0) \\ P_n&:=\lambda _I (0,\gamma =e^{i\frac{2\pi }{\beta }n\tau },0) \\
    Q_n&:=\lambda _I (0,0,\eta =e^{i\frac{2\pi }{\beta }n\tau })
\end{align}
one can express the brackets of the corresponding centrally extended algebra $\hat{\mathfrak{g}}$ as
\begin{subequations}\label{eq:ext_algebra}
\begin{align}
    [T_n,T_m]^\ast&=(n-m)L_{n+m} & [P_n,P_m]^\ast&=\nu  n\,\delta _{n+m,0} \\
    [T_n,P_m]^\ast&=-mP_{m+n} & [Q_n,Q_m]^\ast&=0\\
    [T_n,Q_m]^\ast&=-(m+n)Q_{m+n}-\lambda n\,\delta _{n+m,0} &  [Q_n,P_m]^\ast&=\mu \,\delta _{n+m,0} 
\end{align}
\end{subequations}
where three central terms $(\lambda, \mu, \nu)$ are switched on. We can use the finite coadjoint transformations \eqref{eq:coadjT}-\eqref{eq:coadjTheta} to find a rough classification of the gravitational phase space by observing that each coadjoint orbit of $\hat{\mathcal{G}}$ is identified by a single real number. Indeed, one can show that for any given coadjoint vector $(\mathcal{T}, \mathcal{P}, \Theta; \mathbf{c})$ one can always find a group element mapping this to a vector $(\tilde{\mathcal{T}},0,0;\mathbf{c})$. One is left with residual transformations acting on $\mathcal{T}$ as
\begin{align}
    \tilde{\mathcal{T}}(f(\tau ))&=\frac{1}{(f')^2}\Big(\mathcal{T}-\{f,\tau \}\Big) ~.
\end{align}
But this is just the coadjoint action of the Virasoro group, the orbits of which are labeled by constant representatives\footnote{There are also orbits without such constant representatives which we discard here.} $\mathcal{T}_0$ \cite{Oblak:2016eij,Salzer:2018zlv,Grumiller:2017qao}. The choice of constant representative determines the stabilizer of the given orbit. In order to have a single cover of an AdS$_2$ black hole as a bulk geometry, the zero mode has to be chosen as
\begin{align}\label{eq:const_rep}
    \mathcal{T}_0=\frac{2\pi ^2}{\beta ^2}
\end{align}
in which case the stabilizer is $SL(2,\mathbb{R})$, corresponding to the local isometries of solutions of the JT model. As these boundary conditions are very similar to the ones presented in the Lorentzian section, we again expect all these transformations to be associated with non-trivial charges. Indeed, using appendix \ref{app:gauge_structure} we find the variations
\begin{align}\label{eq:JT_charges}
    \slashed \delta Q_\lambda =\frac{k}{2\pi }\Big(-\delta _\lambda y^0 \,\delta \ln (e^\Theta y^+)+\delta _\lambda \ln (e^\Theta y^+) \,\delta  y^0 -\frac{\varepsilon }{y^+}\delta \mathcal{C}\Big) \Big \vert _{\tau =\tau _0}
\end{align}
which again is a non-integrable but finite expression. One can find an integrable slicing along the same lines as in section \ref{sec:res_gauge_lorntz}. 

\subsection{Schwarzian action}\label{sec:schwarzian}
Similarly to \cite{Grumiller:2017qao}, the gravitational phase space can be described by an effective holographic action. It is obtained by imposing two boundary EOM \eqref{eq:boundaryeom1},\eqref{eq:boundaryeom2} on the action evaluated on allowed configurations \eqref{action_on_constr}. One arrives at
\begin{align}
    \Gamma ^E\big \vert _{\mathrm{pEOM}}&= -\frac{k}{2\pi }\int\limits_0^\beta \dd \tau \, \left(y^+M -\frac{({y^+}')^2}{2y^+}+{y^+}'' \right) 
\end{align}
where we introduced a mass function 
\begin{align}
    M=\mathcal{T}-\mathcal{P}\Theta '-\frac{1}{2}{\Theta '}^2+(\mathcal{P}+\Theta ')' 
\end{align}
and pEOM denotes a partially on-shell evaluation.
Using the boundary condition \eqref{eq:zeromode:bc}, this can be recast in the form
\begin{align}
    \Gamma ^E\big \vert _{\mathrm{pEOM}}&= -\frac{k\hat{y}}{2\pi }\int\limits_0^\beta  \, \frac{\dd \tau}{f'}\big(M -\{f,\tau \} \big) && \{f,\tau \}=\frac{f'''}{f'}-\frac{3}{2}\Big(\frac{f''}{f'}\Big)^2~.\label{eq:schwarzian_action}
\end{align}
If we choose a constant representative \eqref{eq:const_rep} such that $M=\frac{2\pi ^2}{\beta ^2}$ and use the cocycle condition $\{f\circ g,\tau \}=(g')^2\{f,\tau \}\circ g+\{g,\tau \}$ with $g=f^{-1}$, we arrive at the Schwarzian action
\begin{align}\label{eq:schwarzian}
    \Gamma ^E\big \vert _{\mathrm{pEOM}}=S_{\mathrm{Sch}}[f^{-1}]=-\frac{k\hat{y}}{2\pi }\int\limits_0^\beta  \,\dd \tau \Big( \frac{2\pi ^2}{\beta ^2}(f^{-1})'{}^2+\{f^{-1},\tau \} \Big) ~.
\end{align}
One can check that its equation of motion for $f^{-1}$ reproduces the last unimposed boundary equation of motion \eqref{eq:boundaryeom3}.

\section{Target space diffeomorphisms and Wick rotations}\label{sec:Wick}

Target space diffeomorphisms may map the PSMs describing Euclidean dilaton gravities to those describing Lorentzian dilaton gravities and back. As an example, let us start with Euclidean JT with target space coordinates $(X,X^{\underline{1}},X^{\underline{2}})$ and consider the map
\begin{equation}\label{WickX}
    \bar X=-X^{\underline{1}} \qquad \qquad \bar X^+=\frac 1{\sqrt{2}} (X^{\underline{2}}+X) \qquad \qquad \bar X^-=\frac 1{\sqrt{2}} (X^{\underline{2}}-X) ~.
\end{equation}
The Poisson tensor expressed in new coordinates according to \eqref{PSM05} reads
\begin{equation}
    \bar{P}^{\bar{X}+}=-\bar{X}^+ \qquad \qquad \bar{P}^{\bar{X}-}=\bar{X}^- \qquad \qquad \bar{P}^{+-}=-\bar X
\end{equation}
where we immediately recognize the Poisson tensor of the JT model in Lorentzian signature. The transformation $A_I\to \bar{A}_{\bar{I}}$
\begin{equation}
    \bar \omega =-e^{\underline{1}} \qquad \qquad \bar{e}^+=\frac 1{\sqrt{2}} (e^{\underline{2}}+\omega ) \qquad \qquad \bar{e}^-=\frac 1{\sqrt{2}} (e^{\underline{2}}-\omega ) \label{Wicke}
\end{equation}
mixes up the zweibein with the spin connection. This transformation is globally defined, i.e., it maps the whole target space of Euclidean JT to the whole target space of Lorentzian JT. The sign of the Casimir function is inverted,
\begin{equation}
    \bar{\mathcal{C}}=-\mathcal{C}
\end{equation}
and the Euclidean horizons are \emph{not} mapped to Lorentzian horizons. 

To avoid confusion, we stress that the Euclidean metric is defined as $g_{\mu\nu}=\delta_{ab}e_\mu^a e_\nu^b$ while for the Lorentzian model $\bar{g}_{\mu\nu}=\bar{e}_\mu^+ \bar{e}_\nu^- + \bar{e}_\mu^- \bar{e}_\nu^+$.

The transformation \eqref{WickX}, \eqref{Wicke} is, of course, not the standard Wick rotation, which is customarily understood as a continuation of one of the coordinates to the imaginary axis. Since we work with non-static and even non-stationary metrics, such a transformation would inevitably lead to complex metric components, which is inconvenient and excludes any interpretation in terms of real Poisson geometry. However, the transformation  \eqref{WickX}, \eqref{Wicke} does exactly what we need to globally relate the target spaces. It also maps all fields in a Euclidean/Lorentzian theory to the fields in its Lorentzian/Euclidean counterpart. Thus, if we know the boundary action, the asymptotic conditions, the asymptotic symmetries, etc., we also know all these objects in the other theory. As a sanity check, let us see what happens with the leading term of the dilaton $x^+e^\rho$ in the asymptotic conditions \eqref{eq:asymp_cond_Eucl_JT}. After \eqref{WickX} the leading term in $\bar X$ remains $x^+e^\rho$, so that an asymptotic region is mapped to an asymptotic region.

The transformation \eqref{WickX}, \eqref{Wicke} does not map the asymptotic conditions for Euclidean JT obtained in section \ref{sec:EJT} to the asymptotic conditions of Lorentzian JT described in section \ref{sec:JT} since these sets of conditions were written in different gauges. This is, however, a feature rather than a bug. With our transformations, one can formulate Euclidean asymptotics in Bondi gauge and Lorentzian asymptotics in the Fefferman--Graham one.

The simplicity of this non-standard Wick rotation in the JT model is explained by the linearity of the Poisson tensor in all target space coordinates. The general case is much more complicated. If one wants to map a Euclidean dilaton gravity to a Lorentzian dilaton gravity it may be easier to first map both models to Euclidean/Lorentzian JT and then use the transformation \eqref{WickX}, \eqref{Wicke}.

\section{Target space diffeomorphisms in the Euclidean case}\label{sec:Eucl_td}

By arguing similarly to the Lorentzian case we conclude that the target space diffeomorphism that preserves the Euclidean dilaton gravity interpretation can be written as
\begin{eqnarray}
\bar{X}	&=&	\bar{X}\left(X,Y\right)\label{T01a_Eucl}\\
\bar{X}_{\mathrm{E}}^{+} &=&	X_{\mathrm{E}}^{+}f^{\left(+\right)}\left(X,Y\right)\label{T01b_Eucl}\\
\bar{X}_{\mathrm{E}}^{-} &=& X_{\mathrm{E}}^{-}f^{\left(-\right)}\left(X,Y\right)  \label{T01c_Eucl}  
\end{eqnarray}
where $Y\equiv X_{\mathrm{E}}^{+}X_{\mathrm{E}}^{-}$ and $f^{\left(\pm\right)}$ are undetermined functions such that
$\bar{X}\left(X,Y\right)$ is a real function, while 
\begin{equation}
    f^{\left(-\right)}=f^{\left(+\right)*} ~. \label{realf}
\end{equation}
The transformed Poisson tensor has the form \eqref{complexP} for some dilaton potential $\bar{\mathcal{V}}$ iff the following differential equations are satisfied
\begin{align}
\left(\partial_{X}-\mathcal{V}\partial_{Y}\right)\bar{X}	&=	1 \label{T03aE}\\
\left(\partial_{X}-\mathcal{V}\partial_{Y}\right)\tilde{f}	&=	-\bar{\mathcal{V}}\left(\bar{X},\bar{Y}\right)    \label{T03bE}
\end{align}
where $\tilde{f}\equiv Yf^{(+)}f^{(-)}$. 
A general solution for \eqref{T03aE} depends on an arbitrary function $g_1$ and reads
\begin{equation}\label{T04E}
\bar{X}=X+g_{1}\left(\mathcal{C}\right) ~.    
\end{equation}
To solve \eqref{T03bE} we substitute
\begin{equation}
\tilde{f}\left(X,Y\right)=e^{-\bar{Q}\left(\bar{X}\right)}\left(\bar{w}\left(\bar{X}\right)+\tilde{G}\left(X,Y\right)\right)    
\end{equation}
which reduces \eqref{T03bE} to the equation
\begin{equation}
\left(\partial_{X}-\mathcal{V}\partial_{Y}\right)\tilde{G}\left(X,Y\right)	= 0    
\end{equation}
solved by
\begin{equation}
    \tilde{G}\left(X,Y\right)=-g_{2}\left(\mathcal{C}\right) ~.
\end{equation}
Returning to our original notations obtains
\begin{equation}
   \frac 12 \left( (\bar X^{\underline{1}})^2 + (\bar X^{\underline{2}})^2 \right)= \bar{X}^{+}_{\mathrm{E}}\bar{X}^{-}_{\mathrm{E}}=e^{-\bar{Q}\left(\bar{X}\right)}\left(\bar{w}\left(\bar{X}\right)-g_{2}\left(\mathcal{C}\right)\right).  \label{T07b}
\end{equation}
This equation is equivalent to the condition
\begin{equation}
\bar{\mathcal{C}}=g_{2}\left(\mathcal{C}\right).    
\end{equation}

The formula \eqref{T07b} does not restrict the angle between $\bar X^{\underline{1}}$ and $\bar X^{\underline{2}}$, which may be gauge-fixed to any convenient value.  The absolute value $|\bar X^a|$ is well defined as long as $\bar w(\bar X)\geq \bar{\mathcal{C}}$. (This region also corresponds to the allowed values of $\bar X$ for classical solutions with a given value of Casimir function $\bar{\mathcal{C}}$.) However, to avoid a discontinuity at  
$\bar w(\bar X) = \bar{\mathcal{C}}$ one has to impose the condition
\begin{equation}\label{g2E}
g_{2}\left(w\left(X_{h}\right)\right)	=	\bar{w}\left(X_{h}+g_{1}\left(w\left(X_{h}\right)\right)\right)    
\end{equation}
where $X_{h}$ is a solution of the equation $w\left(X_{h}\right)=\mathcal{C}$. This relation can be interpreted as a shift of the horizon point,
\begin{equation}
    \bar{X}_h=X_h +g_1(w(X_h)) ~.\label{shiftXh}
\end{equation}
To write explicit formulas for transformations of all target space coordinates, we need to fix the freedom of rotations in the $(\bar{X}^{\underline{1}},\bar{X}^{\underline{2}})$ plane. A convenient choice is to take $f^{(+)}=f^{(-)}$. Then
\begin{equation}
\bar{X}=X+g_{1}\left(\mathcal{C}\right)\qquad \qquad\bar{X}^{a}=X^{a}\sqrt{\frac{1}{Y}e^{-\bar{Q}\left(\bar{X}\right)}\left(\bar{w}\left(\bar{X}\right)-g_{2}\left(\mathcal{C}\right)\right)} ~.    
\end{equation}
The function $g_2$ has to be invertible, so that $\mathcal{C}=g_{2}^{-1}\left(\bar{\mathcal{C}}\right)$. An inverse of the target space diffeomorphism described above in this section can be written as
\begin{equation}
   X=\bar{X}-g_{1}\left(g_{2}^{-1}\left(\bar{\mathcal{C}}\right)\right)\qquad \qquad X^{a}=\bar{X}^{a}\sqrt{\frac{1}{\bar{Y}}e^{-Q\left(X\right)}\left(w\left(X\right)-g_{2}^{-1}\left(\bar{\mathcal{C}}\right)\right)} ~.
\end{equation}
As in the Lorentzian case, when mapping the symplectic leaves containing horizon points, one has to map a horizon to a horizon. 

\subsection{Maps between Euclidean JT and other models}\label{sec:maps}
\begin{figure}
 \begin{center}
\begin{tikzpicture}
    \node[anchor=south west,inner sep=0] (image) at (0,0) {\includegraphics[width=11cm]{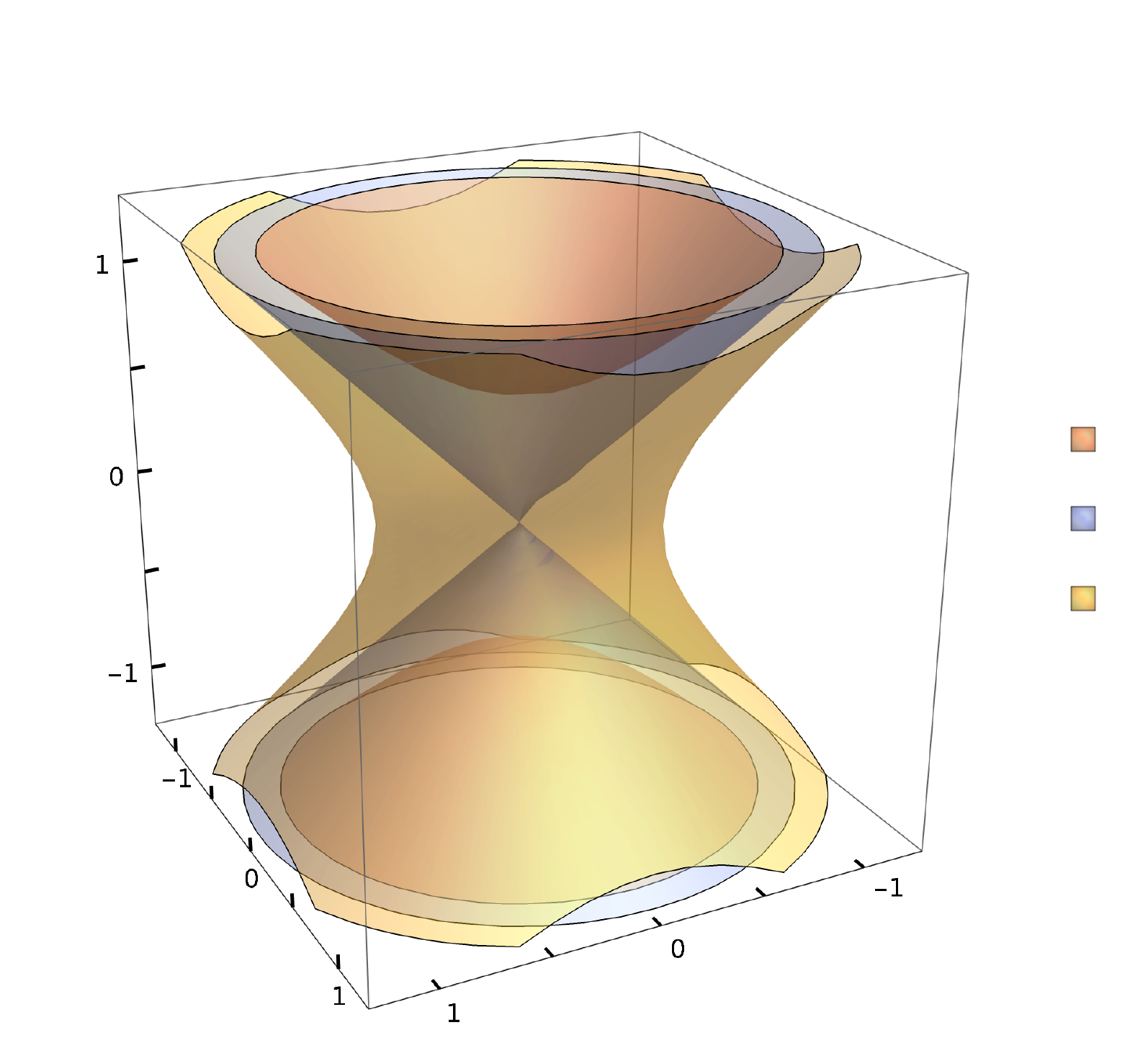}};
    \begin{scope}[x={(image.south east)},y={(image.north west)}]
        \node[label=above:$X^1$] (A) at (0.16,0.12) {};
         \node[label=above:$X^2$] (A) at (0.62,0.01) {};
          \node[label=above:$X$] (A) at (0.04,0.5) {};
          \node[label=above:$\mathcal{C}>0$] (A) at (1.01,0.538) {};
           \node[label=above:${\mathcal{C}=0}$] (A) at (1.01,0.46) {};
          \node[label=above:$\mathcal{C}<0$] (A) at (1.01,0.384) {};
    \end{scope}
\end{tikzpicture}
\caption{Symplectic leaves of the Euclidean JT model. The structure is very similar to the Lorentzian case (see figure \ref{fig:JT_targetspace}), but the leaves describing black holes are now two-sheet hyperboloids. The family we are interested in are the upper halves of these, i.e. the sector with $X>0$ and $\mathcal{C}>0$.}
\label{fig:EJT_targetspace}
\end{center}
 \end{figure}
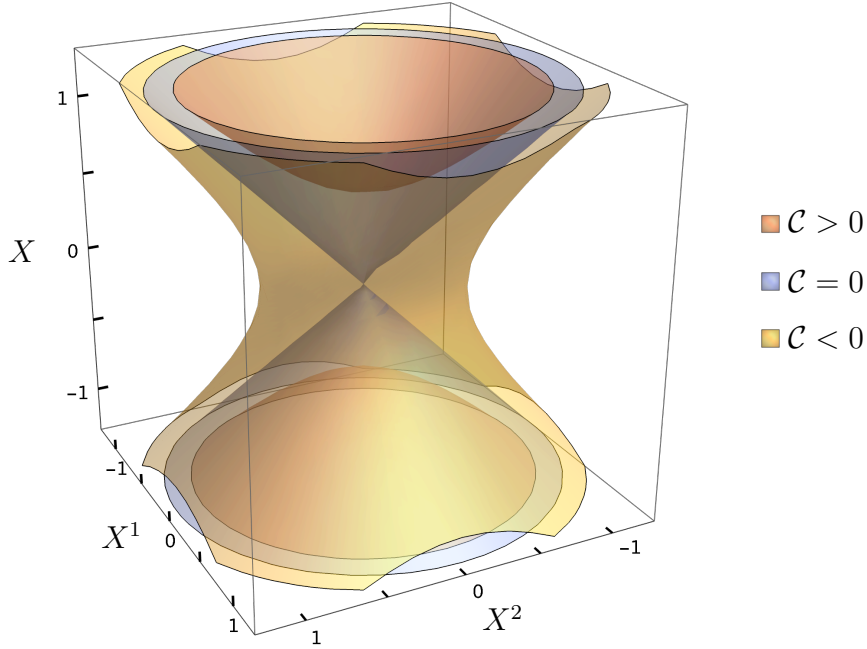

The structure of the symplectic foliation of the Euclidean JT target space coincides with that of its Lorentzian signature counterpart, see section \ref{sec:td}. However, there is an important difference in its physical interpretation. The symplectic leaves corresponding to black hole configurations, $\mathcal{C}>0$, are now two-sheet hyperboloids as can be seen in figure \ref{fig:EJT_targetspace}. One-sheet hyperboloids correspond to no-black hole configurations. Therefore, most of the statements made above regarding automorphisms of the JT model and global equivalences to the JT model can be translated to the Euclidean case with very few changes. However, in Euclidean signature, a symplectic leaf contains at most one horizon point which makes constructing the target space diffeomorphisms somewhat easier.

Let us consider a family of symplectic leaves in the target space of the Euclidean JT model with $\mathcal{C}>0$ and (to be specific) $X_h(\mathcal{C})=+\sqrt{2\mathcal{C}}$. (The family with $X_h(\mathcal{C})=-\sqrt{2\mathcal{C}}$ can be considered along the same lines.)  Let us take another Euclidean dilaton gravity with the potentials $\bar{U}$ and $\bar{V}$ which has a similar family of symplectic leaves. Namely, this has to be a continuous family of symplectic leaves with $\bar{\mathcal{C}}$ in a semi-infinite interval $(\bar{\mathcal{C}}_0,+\infty)$ or $(-\infty,\bar{\mathcal{C}}_0)$ and with all leaves being diffeomorphic to 2-planes. The function $\bar{X}_h(\bar{\mathcal{C}})=\bar{w}^{-1}(\bar{\mathcal{C}})$ has to be a smooth monotonous function from this semi-infinite interval to $(\bar{X}_{h,0},\infty)$. Also, $\bar{w}$ has to be a smooth monotonous function from $(\bar{X}_{h,0},\infty)$ to $(\bar{\mathcal{C}}_0,+\infty)$ or $(-\infty,\bar{\mathcal{C}}_0)$. This is almost all we need. Take \emph{any} function $g_1$ such that \eqref{shiftXh} defines a smooth invertible map between $(0,\infty)$ and $(\bar{X}_{h,0},\infty)$. Such a function exists iff every symplectic leaf extends over the whole interval $\bar X\in [\bar{X}_h,\infty)$, and this is the last condition which we have to impose. Then, the function $g_2$ is uniquely determined by \eqref{g2E}. The pair of functions $g_1$, $g_2$ defines a Poisson diffeomorphism between two families of symplectic leaves which have been described above.

Let us illustrate this procedure with  an example of a family of dilaton potentials 
\begin{equation}
    \bar{U}(\bar{X})=-\frac{\mathrm{a}}{\bar{X}}\qquad \qquad \bar{V}(\bar{X})=-\frac{B}{2}\, \bar{X}^{\mathrm{a}+\mathrm{b}}\label{abfam}
\end{equation}
where $\mathrm{a}$ and $\mathrm{b}$ are real numbers and $B$ is a scale parameter. For these models,
\begin{equation}
    \bar{w}(\bar{X}) =\frac{B}{2(\mathrm{b}+1)}\, \bar{X}^{\mathrm{b}+1}\,\qquad \qquad
    \bar{\mathcal{C}}=\frac{B}{2(\mathrm{b}+1)}\, \bar{X}^{\mathrm{b}+1} -\bar{X}^{-\mathrm{a}}\bar{X}^+_{\mathrm{E}}\bar{X}^-_{\mathrm{E}} ~.
\end{equation}
It is easy to check that the conditions formulated in the previous paragraph are satisfied in two regions in the parameter space,
\begin{equation}
    \mathrm{b}+1>0 \qquad \qquad B>0\qquad \mbox{with}\quad\bar{\mathcal{C}}>0, \label{1stfam1}\end{equation}
and
\begin{equation}
\mathrm{b}+1<0 \qquad\qquad B>0\qquad \mbox{with}\quad  \bar{\mathcal{C}}<0 ~. \label{1stfam2}
\end{equation}
The region \eqref{1stfam1} includes many interesting dilaton gravity models including, for example, spherically reduced Einstein gravities from $d\geq 4$ dimensions with
\begin{equation}
    \mathrm{a}=\frac{d-3}{d-2} \qquad \qquad \mathrm{b}=-\frac{1}{d-2} ~, \label{SRG}
\end{equation}
JT gravity ($\mathrm{a=0}$, $\mathrm{b=1}$) and the matterless CGHS model, also known as Witten black hole ($\mathrm{a=1}$, $\mathrm{b=0}$) \cite{Mandal:1991tz,Elitzur:1991cb,Witten:1991yr,Callan:1992rs} which is a formal limit $d\to\infty$ of the model \eqref{SRG}.

\begin{figure}
 \begin{center}
\begin{tikzpicture}
    \node[anchor=south west,inner sep=0] (image) at (0,0) {\includegraphics[width=11cm]{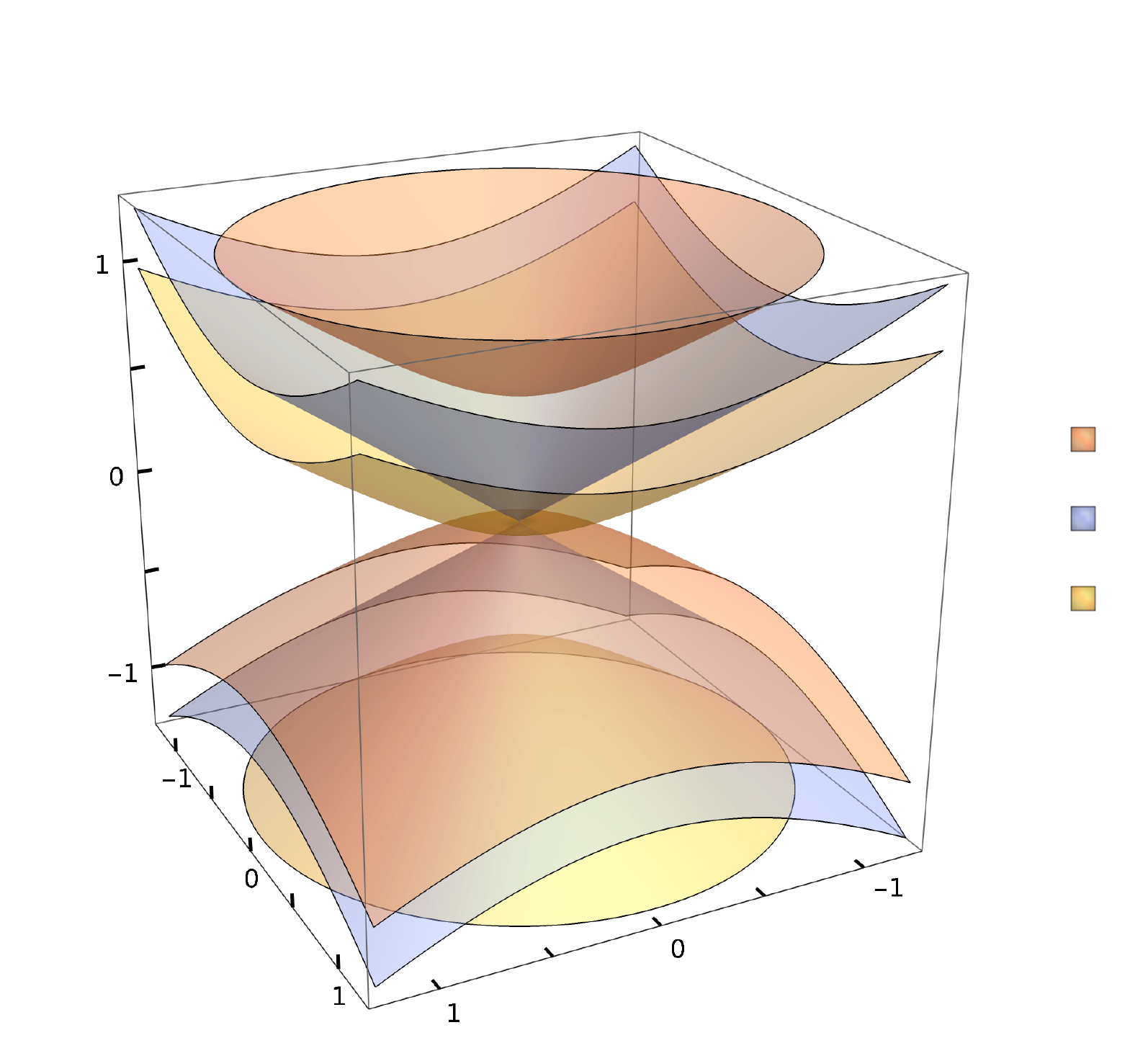}};
    \begin{scope}[x={(image.south east)},y={(image.north west)}]
        \node[label=above:$\bar X^1$] (A) at (0.16,0.12) {};
         \node[label=above:$\bar X^2$] (A) at (0.62,0.01) {};
          \node[label=above:$\bar X$] (A) at (0.04,0.5) {};
          \node[label=above:$\Bar{\mathcal{C}}>0$] (A) at (1.01,0.538) {};
           \node[label=above:${\Bar{\mathcal{C}}=0}$] (A) at (1.01,0.46) {};
          \node[label=above:$\Bar{\mathcal{C}}<0$] (A) at (1.01,0.384) {};
    \end{scope}
\end{tikzpicture}
\caption{Symplectic leaves of the Euclidean CGHS model. For every value of $\Bar{\mathcal{C}}$ there are again two branches. For $\Bar{\mathcal{C}}>0$ one of them ($\Bar{X}<0$) always contains a curvature singularity while the other one $(\Bar{X}>X_h)$ describes a Euclidean black hole.}
\label{fig:ECGHS_targetspace}
\end{center}
 \end{figure}
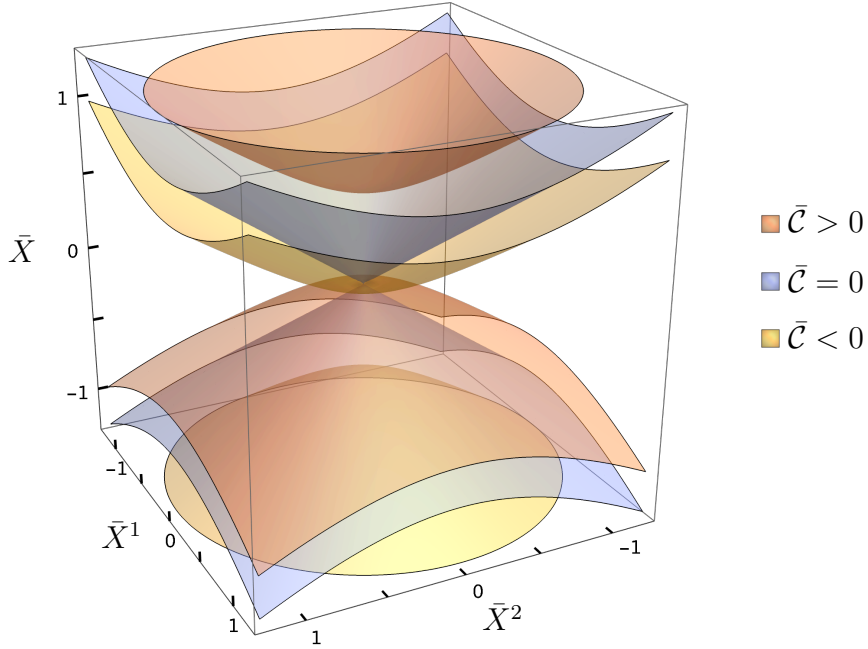

Let us write more explicit  formulas for the case of CGHS. A convenient choice for the scale parameter is $B=2$. With this choice, $\bar{w}(\bar{X})=\bar{X}$ and $\bar{X}_h(\bar{\mathcal{C}})=\bar{\mathcal{C}}$. Thus, all conditions on monotonicity, smoothness, domain, and range of $\bar{w}$ are trivially satisfied. The equation for constant $\bar{\mathcal{C}}$ surfaces reads
\begin{equation}
    \bar{X}_{\mathrm{E}}^+\bar{X}_{\mathrm{E}}^-= \bar{X} (\bar{X}-\bar{\mathcal{C}} )
\end{equation}
and has solutions for any $\bar{X}\geq \bar{X}_h$ with the corresponding symplectic leaves depicted in figure \ref{fig:ECGHS_targetspace}.

Therefore, the conditions for the existence of Poisson diffeomorphisms between the selected family of symplectic leaves with $\mathcal{C}>0$ and $X>0$ in JT and the symplectic leaves in CGHS which contain horizon points at $\Bar{X}_h>0$ are satisfied as well. To construct such maps, we need two functions, $g_1$ and $g_2$, satisfying the equations \eqref{g2E} and \eqref{shiftXh},
\begin{equation}
    \bar{X_h}=X_h+g_1\left( \tfrac 12 X_h^2\right)=g_2\left( \tfrac 12 X_h^2\right)
\end{equation}
with a smooth monotonously increasing $g_2$. Clearly, there are infinitely many choices for such functions. One of them,
\begin{equation}
    g_1=0 \qquad \qquad g_2(\mathcal{C})=\sqrt{2\mathcal{C}} \label{g10}
\end{equation}
gives particularly simple transformation rules.

Let us find the asymptotic conditions for CGHS that are obtained from asymptotic conditions in JT through the action of a target space diffeomorphism with the functions $g_1$ and $g_2$ as in \eqref{g10}. For our purposes, it is enough to restrict ourselves to a subset of the asymptotic conditions \eqref{eq:EJTsol1}, \eqref{eq:EJTsol2} with the redefinitions \eqref{eq:COS_fields}. For this we fix
\begin{equation}
     \mathcal{P}=0 \qquad \qquad \Theta =0 \label{L0Lp}
\end{equation}
which reduces the asymptotic symmetries to a subalgebra depending on a single free function $\varepsilon $ with
\begin{equation}
    \eta =\varepsilon'' \qquad \qquad \gamma =-\varepsilon ' ~.
\end{equation}
The field $\mathcal{T}$ transforms as a chiral half of a CFT$_2$ stress tensor
\begin{equation}
\delta_{\varepsilon}\mathcal{T}	=	\varepsilon \mathcal{T}'+2\varepsilon'\mathcal{T}   +\varepsilon'''
\end{equation}
thus revealing a Virasoro asymptotic symmetry algebra which is just the Euclidean counterpart of the third restriction mentioned in the item list in section \ref{sec:res_gauge_lorntz}. 

By transforming these asymptotic conditions of JT gravity with a target space diffeomorphism defined with the choice \eqref{g10} one obtains the following asymptotic conditions for CGHS
\begingroup
\allowdisplaybreaks
\begin{subequations}\label{eq:CGHS_bcs}
\begin{align}
\bar{\omega}_{\rho}	&=	-\frac{y^{0}}{y^{+}}\left[e^{-\rho}-\frac{2\sqrt{2\mathcal{C}}}{y^+}\,e^{-2\rho }+\mathcal{O}\left(e^{-3\rho}\right)\right] \label{CGHSa1}\\
\bar{\omega}_{\tau}	&=\frac{1}{y^{+}}\left[2\sqrt{2\mathcal{C}}+\frac{\mathcal{T}(y^{+}){}^{2}-6y^{+}y^{-}+2(y^{0}){}^{2}}{y^{+}}e^{-\rho}+\mathcal{O}\left(e^{-2\rho}\right)\right]	\\
\bar{e}_{1\rho}	&=	-\frac{1}{\sqrt{2}}\frac{y^{0}}{y^{+}}\left[e^{-\rho}-\frac{3\sqrt{2\mathcal{C}}}{2y^+}\,e^{-2\rho }+\mathcal{O}\left(e^{-3\rho}\right)\right] \\
\bar{e}_{1\tau}	&=\frac{1}{\sqrt{2}}\frac{1}{y^{+}}\left[2\sqrt{2\mathcal{C}}-\frac{2\left(2y^{+}y^{-}-(y^{0}){}^{2}\right)}{y^{+}}e^{-\rho}+\mathcal{O}\left(e^{-2\rho}\right)\right]\\
\bar{e}_{2\rho} &= \frac{1}{\sqrt{2}}\left[1+\frac{\sqrt{2\mathcal{C}}}{2y^{+}}e^{-\rho}+\mathcal{O}\left(e^{-2\rho}\right)\right] \\
\bar{e}_{2\tau}&=-\frac{1}{\sqrt{2}}\frac{y^{0}}{y^{+}}\left[1-\frac{3\sqrt{2\mathcal{C}}}{2y^{+}}e^{-\rho}+\mathcal{O}\left(e^{-2\rho}\right)\right]\\
 \bar{X}	&=y^{+}e^\rho +y^{-}e^{-\rho }+\mathcal{O}(e^{-2\rho })\\	
   \bar{X}^{\underline{1}}	&=	\sqrt{2}y^{+}\left[-e^{\rho}+\frac{\sqrt{2\mathcal{C}}}{2y^{+}}+\mathcal{O}\left(e^{-2\rho}\right)\right] \\
\bar{X}^{\underline{2}}	&=	-\sqrt{2}y^{0}\left[1-\frac{\sqrt{2\mathcal{C}}}{2y^{+}}e^{-\rho}+\mathcal{O}\left(e^{-2\rho}\right)\right]
\label{CGHSa9}
\end{align}
\end{subequations}
\endgroup
where $\sqrt{2\mathcal{C}}=\sqrt{4y^{+}y^{-}-(y^{0}){}^{2}}$. This square root is real as long as the fluctuations of the $X^I$ belong asymptotically to the selected class of symplectic leaves of JT with $\mathcal{C}>0$. Apart from that, all the functions are allowed to fluctuate freely except for the zero mode of $1/y^+$, as a consequence of \eqref{eq:zeromode:bc}. For comparison, we write down the corresponding metric boundary conditions,
\begin{equation}
    \overline{\dd s}^2 =\Bar{g}_{\rho \rho }\,\dd \rho ^2+2\Bar{g}_{\tau \rho }\,\dd \tau \dd \rho +\Bar{g}_{\tau \tau }\,\dd \tau ^2 
\end{equation}
which by construction allow for the Witten black hole as a solution.
The different components read
\begin{align}
    \Bar{g}_{\rho \rho }&= \frac{1}{2}+\frac{\sqrt{2\mathcal{C}}}{2y^+}e^{-\rho }-\frac{(y^0)^2}{2(y^+)^2}e^{-2\rho }+\mathcal{O}(e^{-3\rho })\\
    \Bar{g}_{\tau \rho }&=-\frac{y^0}{2y^+}-\frac{\sqrt{2\mathcal{C}}\, y^0}{2(y^+){}^2}e^{-\rho }+\frac{y^0\big(30\mathcal{C}+3(y^0){}^2+2\mathcal{T}(y^+){}^2\big)}{8(y^+){}^3}e^{-2\rho }+\mathcal{O}(e^{-3\rho })\\   
   \Bar{g}_{\tau \tau }&=\frac{8\mathcal{C}+(y^0){}^2}{2(y^+){}^2}-\frac{\sqrt{2\mathcal{C}}\,\big(8\mathcal{C}-(y^0){}^2\big)}{2(y^+){}^3}e^{-\rho }\\[.5em]
   &\qquad \qquad +\frac{16\mathcal{C}^2-22\mathcal{C}(y^0)^2-(y^0)^4-2\mathcal{T}(y^+)^2(8\mathcal{C}+(y^0)^2)}{4(y^+)^4}e^{-2\rho }+\mathcal{O}(e^{-2\rho }) \nonumber ~.
\end{align}
Evaluating this for a static configuration in JT, where the on-shell conditions \eqref{eq:boundaryeom1}-\eqref{eq:boundaryeom3} imply $y^+=\hat{y}$, $y^0=0$, $y^-=\frac{\mathcal{C}}{2\hat{y}}$ and $\mathcal{T}=\frac{\mathcal{C}}{\hat{y}^2}$ yields the metric
\begin{equation}\label{eq:stat_CGHS}
     \overline{\dd s}^2 \big \vert _{\mathrm{stat.}}=\Big(\frac{1}{2}+\frac{\sqrt{2\mathcal{C}}\hat{y}e^\rho }{2\hat{y}^2e^{2\rho }+\mathcal{C}}\Big)\dd \rho ^2+\Big(1-\frac{2\sqrt{2\mathcal{C}}\hat{y}e^\rho }{2\hat{y}^2e^{2\rho }+\mathcal{C}}\Big)\dd \tau ^2 ~.
\end{equation}
The geometry has a tip at $e^{\rho _h}=\frac{\sqrt{2\mathcal{C}}}{2\hat{y}}$ and describes a smooth world-sheet with the topology of a disk for $\tau \sim \tau +\beta $ with periodicity $\beta =\frac{2\pi \hat{y}}{\sqrt{2\mathcal{C}}}$. This coincides with the periodicity fixed by \eqref{eq:const_rep} describing a smooth black hole on the JT side.

The gauge parameters for the CGHS model can be obtained directly from \eqref{barlambda} and the restricted form of \eqref{eq:Eucl_PSM_parameters} after the change of slicing \eqref{eq:COS_params}-\eqref{eq:COS_fields}. We obtain
\begin{align}\label{eq:cov_params}
\bar{\lambda}_{X} &= -\frac{2\sqrt{2\mathcal{C}}}{y^{+}}\,\varepsilon  +\left(\frac{\varepsilon\left(6\mathcal{C}-(y^{0}){}^{2}\right)-2y^+y^0\varepsilon '}{2(y^{+}){}^{2}}\,-\mathcal{T}\varepsilon-\varepsilon '' \right)e^{-\rho}+\mathcal{O}\left(e^{-2\rho}\right)  \\
\bar{\lambda}_{1} &= -\frac{2\sqrt{\mathcal{C}}}{y^+}\varepsilon +\frac{\varepsilon \left(2\mathcal{C}-(y^0)^2\right)-y^+y^0\varepsilon '}{\sqrt{2}(y^+)^2}e^{-\rho }+\mathcal{O}\left(e^{-2\rho}\right)\\
\bar{\lambda}_{2} &= \frac{y^+\varepsilon '+\varepsilon y^0}{\sqrt{2}y^+}+\frac{\sqrt{\mathcal{C}}\big(\varepsilon 'y^+-3\varepsilon y^0\big)}{2(y^+)^2}e^{-\rho }+\mathcal{O}\left(e^{-2\rho}\right) ~.
\end{align}
It can be checked that they preserve the asymptotic conditions \eqref{eq:CGHS_bcs} under PSM gauge transformations given by the new generating set \eqref{eq:new_gaugetrafo3}-\eqref{eq:new_gaugetrafo4} and lead to the exact same transformation behavior of the field variables $(\mathcal{T},y^+,y^-,y^0)$. 

It is interesting to compare this to the behavior of the first-order dilaton gravity gauge transformations given by Euclidean local Lorentz transformations parametrized by $\sigma $ and diffeomorphisms $\xi ^\mu $. On-shell they are related to PSM gauge transformations by \eqref{eq:FO_gaugetrafos}. One can, however, also directly determine them off-shell by just demanding the boundary conditions to be preserved. For the restricted JT boundary conditions from above this leads to 
\begin{align}\label{JTdifL}
    \xi ^\tau &=\varepsilon -\frac{\varepsilon ''}{2}e^{-2\rho }+\mathcal{O}(e^{-3\rho }) & \xi ^\rho &=-\varepsilon ' & \sigma &=-\varepsilon '' e^{-\rho }+\mathcal{O}(e^{-3\rho }) ~.
\end{align}
Doing the same for the CGHS boundary conditions \eqref{eq:CGHS_bcs} on the other hand leads to
\begin{align}
    \Bar{\xi }^\tau &=\Bar{\varepsilon }-\frac{\Bar{\varepsilon }''}{2}e^{-2\rho }+\mathcal{O}(e^{-3\rho }) \label{CGHSxi1}\\
    \Bar{\xi }^\rho &=-\Bar{\varepsilon }'-\frac{\Bar{\varepsilon }''}{2}\big (y^+{}'+y^0\big )e^{-2\rho }+\mathcal{O}(e^{-3\rho }) \label{CGHSxi2}\\
    \Bar{\sigma }&=-\Bar{\varepsilon }'' e^{-\rho }+\mathcal{O}(e^{-3\rho })
\end{align}
with another free function $\Bar{\varepsilon }(\tau)$. One can see that as opposed to JT there are subleading orders proportional to the boundary EOM appearing in \eqref{CGHSxi2} for CGHS. The asymptotic Killing vectors thus only match on-shell upon imposing \eqref{eq:boundaryeom1}-\eqref{eq:boundaryeom3} (assuming $\Bar{\varepsilon }=\varepsilon $). We conclude from this that off-shell there is no relation between first-order dilaton gravity gauge parameters under target space diffeomorphisms. The off-shell description only works when sticking to the PSM formulation.

\subsection{Schwarzian holographic action for other models than JT}
Like in the Lorentzian case the target space diffeomorphism maps the variational principle of JT to the new model. We do not repeat these expressions here as everything works very similarly to section \ref{sec:var_prin}. However, we want to point out one neat feature appearing in the Euclidean case. As we know how target space diffeomorphisms act on off-shell configurations and off-shell asymptotic symmetries, we are not constrained to just mapping the full on-shell action to another model but can perform the map partially off-shell as well. We, therefore, find that the Schwarzian action \eqref{eq:schwarzian} does not only serve as a holographic action for Euclidean JT but also for any other model related to it by a target space diffeomorphism, i.e.
\begin{equation}
     \Gamma ^E\big \vert _{\mathrm{pEOM}}=-\frac{k\hat{y}}{2\pi }\int\limits_0^\beta  \,\dd \tau \Big( \frac{2\pi ^2}{\beta ^2}(f^{-1})'{}^2+\{f^{-1},\tau \} \Big) =\Bar{\Gamma }^E\big \vert _{\mathrm{pEOM}} ~.
\end{equation}
Here, pEOM denotes a partial on-shell evaluation, i.e. using all but one boundary EOM. From the perspective of the new model, it would be highly non-trivial to choose the right boundary conditions like \eqref{eq:CGHS_bcs} such that these symmetries are realized. Moreover, the extraction of the holographic action would not be as straightforward as in the JT case. This is because the split between boundary and bulk parts of the new PSM action works differently such that the Schwarzian in general would not appear as just a boundary action ``ready to be read off''. Although the asymptotic symmetry algebra contains an $\mathfrak{sl}(2,\mathbb{R})$ subalgebra this subalgebra coincides with the local isometries only in the case of JT gravity. This does not affect the interpretation of asymptotic symmetries as dynamical symmetries of the boundary action for general dilaton gravities. The $SL(2,\mathbb{R})$ stabilizer has to be seen as an abstract way of selecting a certain coadjoint orbit that describes
the phase space.

We want to emphasize that the periodicity of all the fields featuring in the Schwarzian is still the one of the JT model, which in turn is associated with the JT Hawking temperature. From the perspective of the new model, however, it will in general not be true that these two notions of temperature coincide as the definition of a Hawking temperature $\Bar{T}$ is always tied to a certain choice of asymptotic frame. But such a choice is not target space covariant in the same way as asymptotic Killing vectors are not target space covariant, so there is no reason to expect that the Hawking temperature as it is defined in JT gravity will transform in any definite way. 

The condition \eqref{g2E}, however, still assures that the geometry in the new model is smooth which we have seen explicitly in the example of the CGHS model with the solution \eqref{eq:stat_CGHS}. This smoothness condition can also be understood from a thermodynamical perspective: For general dilaton gravity models $X_h$ defines the entropy of the black hole while the conserved quantity $\mathcal{C}$ is the black hole mass \cite{Grumiller:2007ju}. There is a thermodynamic relation $w(X_h)=\mathcal{C}$ between these two quantities that has to be respected by every model. As $g_1$ effectively changes the entropy and $g_2$ the mass we cannot choose these two functions freely, we need to satisfy $\Bar{w}(\Bar{X}_h)=\Bar{\mathcal{C}}$.

\section{Discussion}\label{sec:final}

Let us summarize the main points of this work. Our setting is the reformulation of 2D dilaton gravity as a classically equivalent PSM with the main information about the model encoded in the Poisson structure on a three-dimensional target space manifold. Local Lorentz transformations and diffeomorphisms of the world-sheet spacetime are then given as a subset of PSM gauge transformations. Using diffeomorphisms on target space it is possible to map between different dilaton gravity models. The idea is then to take a comparably well-understood model like the JT model with one's favorite boundary conditions and map the whole phase space together with its symmetries to a different, possibly less explored model. This provides a powerful tool to obtain consistent sets of boundary conditions and holographic descriptions for a rather large class of 2D dilaton gravity theories.  

As the global structure of target space differs considerably between the Lorentzian and Euclidean theories we split this work into three parts, devoting sections \ref{sec:dil}-\ref{sec:U0} to maps between Lorentzian theories and sections \ref{sec:Edil}-\ref{sec:Eucl_td} to maps between Euclidean theories. Section \ref{sec:Wick} provides a map between both.

We provide now a more detailed summary of the key steps. We started by summarizing the main points about PSM gauge transformations in section \ref{sec:PSM_gauge_trafos}, emphasizing that in their usual form, they constitute a specific generating set \cite{Henneaux:1992} of all PSM gauge transformations. In section \ref{sec:fate}, the non-uniqueness of generating sets helped us to understand how asymptotic symmetries behave under target space diffeomorphisms as the latter do not leave generating sets invariant. The remainder of section \ref{sec:dil} described the JT model with general boundary conditions in the PSM formulation to make the analysis self-contained. Imposing various restrictions on these boundary conditions led to well-known asymptotic symmetry algebras like Virasoro \cite{Brown:1986nw}, warped conformal \cite{Hofman:2011zj,Detournay:2012pc}, twisted warped \cite{Afshar:2015wjm,Afshar:2019tvp,Afshar:2019axx,Godet:2020xpk} or BMS$_2$ \cite{Afshar:2019axx,Afshar:2021qvi}.   

In section \ref{sec:td}, we explicitly constructed target space diffeomorphisms between two dilaton gravity models of power-counting renormalizable models. An important restriction comes from requiring that after the target space diffeomorphism, the Poisson tensor is still of a form that allows an interpretation as a 2D dilaton gravity theory. Additionally demanding regularity of the diffeomorphism at horizons fixes the possible maps between two given models to a subset of all target space diffeomorphisms parametrized by one arbitrary function of the Casimir. This is a new requirement as compared to the previous work \cite{Valcarcel:2022zqm} by two of the authors which makes it possible to extend the target space diffeomorphism away from the asymptotic region. 
In section \ref{sec:aut}, we analyzed target space diffeomorphisms leaving the Poisson tensor invariant and, therefore, mapping back to the same model. We showed that these can be roughly classified by the first Poisson cohomology. Once given the form of the target space diffeomorphism the well-defined variational principle of JT is mapped to a well-defined variational principle in the new model as shown in section \ref{sec:var_prin}. Moreover, the asymptotic symmetries of JT can be translated directly into asymptotic symmetries of the new model given by the same group. However, as explained in section \ref{sec:fate}, they are given by a different generating set of PSM gauge transformations. A powerful result here is that this new generating set \eqref{eq:new_gaugetrafo3}-\eqref{eq:new_gaugetrafo4} closes off-shell for any Poisson tensor. For the standard generating set \eqref{PSM04a} this is only true if the Poisson tensor is linear in the target space coordinates which coincides with the possibility to write the model as a BF theory. 

Section \ref{sec:U0} is concerned with the global aspects of maps between two models. We found in particular that depending on the topology of the symplectic leaves one can in general only map diffeomorphically certain regions such as the $X>0$ sectors of each leaf. However, for a special class of $U=0$ models given by an admissible potential function $w(X)$, one can construct a global target space diffeomorphism as shown in section \ref{sec:glob_JT}. As an example, we describe the classical equivalence between the model studied in \cite{Witten:2020wvy} and the JT model. Compared to that work, this result is complementary in the sense that the model is solved quantum-mechanically, though perturbatively there while we provide an exact, though classical description.

In Euclidean signature we did a similar analysis; the JT model can be solved for general boundary conditions which lead to the same asymptotic symmetry algebra as in the Lorentzian case for a certain slicing of phase space. The gravitational dynamics can be described by an effective Schwarzian boundary action which we re-derived in section \ref{sec:schwarzian}.\footnote{In the language of \cite{Joung:2023doq}, we used the approach where metric and dilaton vary at the boundary \cite{Grumiller:2015vaa, Grumiller:2017qao} rather than the ``wiggly boundary''-approach \cite{Maldacena:2016upp,Cvetic:2016eiv}.} 

In section \ref{sec:Eucl_td}, we constructed target space diffeomorphisms between Euclidean dilaton gravity models. In this case, there were fewer restrictions on the global properties coming from the topology of the symplectic leaves as we only restrict to the black hole sector $\mathcal{C}>0$ in which case one always finds a disk. The results of this section, therefore, hold for a broad class of models, including $U\neq 0$. By explicitly looking at a map between JT and the CGHS model we showed how the boundary conditions and asymptotic symmetries were related. The main result was that the effective Schwarzian action even applies to other models than JT, provided one maps the boundary conditions accordingly. 

We conclude with avenues for future work. Using the connection between JT and CGHS we provide an effective description of 2D asymptotically flat spacetimes, given by the CGHS solutions. It could be rewarding to further study this from the view of the holographic principle. In particular, one could ask how many of the existing entries in the well-known ``JT/SYK'' dictionary (see, e.g.~\cite{Sarosi:2017ykf} and references therein) can be directly recycled into entries of a ``CGHS/SYK'' dictionary. This question naturally generalizes to an infinitely larger class of all 2D dilaton gravity models related to JT by a target space diffeomorphism. 

As a generalization, one could investigate Poisson diffeomorphisms for larger target space manifolds such as for the Cangemi--Jackiw or $\widehat{\textrm{CGHS}}$ model \cite{Afshar:2019axx,Kar:2022sdc}. In this case, the target space is four-dimensional with the fourth dimension corresponding to an additional (topological) Maxwell field. A concrete question would be if there are additional restrictions on the form of the target space diffeomorphism coming from regularity conditions or if there are just more free functions available. 
 
It could also be interesting to study the quantum group symmetries recently discovered in Liouville gravity \cite{Fan:2021bwt} under Poisson diffeomorphisms. 

An important problem that remains to be addressed is the transformation properties of the quantum gravity partition function under target space diffeomorphisms. On the one hand, we know how off-shell configurations and the variational principle transform so the transformation property of the integrand of the path integral is determined. On the other hand, the transformation of the path integral measure is a more subtle issue and it might not be possible to simply take the exact partition function of JT \cite{Stanford:2017thb} and use it for other models related by a target space diffeomorphism.

\acknowledgments
We are grateful to Jakob Salzer, Adrien Fiorucci, and Romain Ruzziconi for discussions.
FE and DG were supported by the Austrian Science Fund (FWF), projects P~32581, P~33789, P~36619, and W~1252. 
The work of D.V. was supported in parts by the S\~ao Paulo Research Foundation (FAPESP), grant 2021/10128-0, and by the National Council for Scientific and Technological Development (CNPq), grant 304758/2022-1.

\appendix

\section{PSM Hamiltonian analysis}\label{app:gauge_structure}

In this appendix, we summarize an analysis of the gauge structure of PSMs using the Hamiltonian framework. For a start we are just interested in the bulk theory, i.e., we assume that the boundary conditions of the fields are chosen such that no boundary contributions have to be taken into account. Starting from the PSM action
\begin{equation}
I\mathrm{_{PSM}}=\frac{k}{2\pi}\int_{\mathcal{M}}\left(A_{J}\wedge\mathrm{d}X^{J}+\frac{1}{2}P^{JK}A_{J}\wedge A_{K}\right)  
\end{equation}
we restrict to a topology $\mathcal{M}=I\times S^1$ with coordinates chosen as $(\rho ,\tau )$. The action can then be rewritten in Hamiltonian form
\begin{align}
    I\mathrm{_{PSM}}=\frac{k}{2\pi}\int_{I\times S^1}\dd \rho \,\dd \tau \, \Big(A_{J\rho }\partial _\tau X^J-A_{J\tau }\big(\partial _\rho X^J+P^{JK}A_{K\rho }\big)  \Big) 
\end{align}
which allows reading off the symplectic form on field space
\begin{align}
    \Omega =\frac{k}{2\pi }\int _I\dd \rho \, \delta A_{J\rho }\wedge \delta X^J ~.
\end{align}
This form induces the standard Poisson bracket of a symplectic pair on an equal time slice
\begin{align}
    \{X^J(x),A_{K\rho '}(x')\}=\delta ^J_K \delta (\rho -\rho ')
\end{align}
where $x=(\rho ,\tau )$ and $x'=(\rho ',\tau )$. Using this bracket, one can identify a system of first-class constraints 
\begin{align}\label{eq:constraints}
    G_\rho ^J=-\partial _\rho X^J-P^{JK}A_{K\rho } 
\end{align}
satisfying
\begin{align}
    \{G^J_\rho ,G^K_{\rho '}\}=C^{JK}{}_L\,G^L_\rho \delta (\rho -\rho ') 
\end{align}
where the structure functions $C^{JK}{}_L$ are given by
\begin{align}
    C^{JK}{}_L=-\partial _LP^{JK} ~.
\end{align}
One can see that the components $A_{J\tau }$ are Lagrange multipliers and carry no dynamical information in this setup. Associating canonical gauge generators to the constraints \eqref{eq:constraints},
\begin{align}
    G[\lambda _J]=\frac{k}{2\pi }\int _I\dd \rho \,\lambda _J\, G^J_\rho 
\end{align}
one can determine the transformation behavior of $A_{J\tau }$ by demanding gauge invariance of the action. This leads to
\begin{align}
    \delta _\lambda A_{J\tau }=\{G[\lambda ],A_{J\tau }\}=-\partial _\tau \lambda _J-\partial _JP^{KL}A_{K\tau }\lambda _L ~.
\end{align}
Gauge transformations generate Hamiltonian vector fields $V_\lambda $ on field space
\begin{align}
    i_{V_\lambda }\Omega =-\delta G[\lambda _J] ~.
\end{align}
On fields $\phi =(X^J,A_{J })$ these vector fields act like $i_{V_G}\delta \phi =\delta _\lambda \phi $. 
For non-constant $C^{JK}{}_L$ these vector fields are only in involution when restricted to the constraint surface which can be seen in the strong equalities \eqref{eq:gaugecomm1}-\eqref{eq:gaugecomm2}.

If we introduce a boundary and pick boundary conditions like in the main text, some of the gauge transformations become physical. They are generated by charges which do not vanish on the constraint surface due to boundary terms. A simple way to arrive at the relevant expressions is by using the covariant phase space formalism, see e.g.~\cite{Fiorucci:2021pha} for a pedagogical introduction. It tells us that on-shell there is a relation between the
symplectic current $\omega (\delta \phi ,\delta \phi )$ and the variation of codimension-2 charges,
\begin{align}
    \omega (\delta _\lambda \phi ,\delta \phi )=\frac{k}{2\pi }\Big(\delta _\lambda A_J\delta X^J-\delta A_J\delta _\lambda X^J\Big)=-\dd \slashed \delta Q _\lambda [\phi ,\delta \phi ] 
\end{align}
with a boundary condition preserving gauge transformation $\lambda _J$. Direct computation leads to
\begin{align}
    \slashed \delta Q _\lambda [\phi ,\delta \phi ]=\frac{k}{2\pi }\lambda _J\,\delta X^J 
\end{align}
which only holds up to the addition of codimension-$3$ terms. However, as we are in two spacetime dimensions this ambiguity is not present.

\section{Coadjoint representation of \texorpdfstring{$\hat{\mathcal{G}}$}{Ghat}}\label{app:coadjoint}

This appendix discusses the asymptotic symmetry group for the general JT boundary conditions presented in section \ref{sec:EJT}. 
Analyzing the bracket relations between the modes \eqref{eq:fourier_modes} yields at maximum six non-trivial cocycles,
\begin{subequations}\label{eq:algebra}
\begin{align}
    [T_n,T_m]^\ast&=(n-m)T_{n+m}+\frac{c}{12}(n^3-n)\delta _{n+m,0} & [P_n,P_m]^\ast&=\nu \, n\delta _{n+m,0} \\
    [T_n,P_m]^\ast&=-mP_{m+n} +\kappa (n^2-n)\delta _{n+m,0}& [Q_n,Q_m]^\ast&=0\\
    [T_n,Q_m]^\ast&=-(m+n)Q_{m+n}-(\lambda n+a)\delta _{n+m,0} & [Q_n,P_m]^\ast&=\mu \,\delta _{n+m,0} ~.
\end{align}
\end{subequations}
Comparing this with the transformation of the gravitational variables 
\begin{subequations}\label{eq:grav_trafo}
\begin{align}
    \delta _\lambda \mathcal{T}&=\varepsilon \mathcal{T}'+2\varepsilon '\mathcal{T}+\mathcal{P}\gamma '-\Theta '\eta +\eta '\\
    \delta _\lambda \mathcal{P}&=\varepsilon \mathcal{P}'+\varepsilon '\mathcal{P}-\eta -\gamma '\\
    \delta _\lambda \Theta &=\varepsilon \Theta '+\varepsilon '+\gamma 
\end{align}
\end{subequations}
we anticipate that $c=\kappa =a=0$. This follows from the relations $\mathcal{T}\leftrightarrow T_n$, $\mathcal{P}\leftrightarrow P_n$ and $\Theta \leftrightarrow Q_n$ between the gravitational fields and the symmetry generators based on a comparison of conformal weights\footnote{The modes of a chiral primary $\phi (z)$ of weight $h$ fulfill $[T_n,\phi _m]=((h-1)n-m)\phi _{n+m}$.}. One can compare this to Ref.~\cite{Afshar:2021qvi} where the equivalent to the cocycle $a$ crucially was non-zero. The only difference to those boundary conditions is that we allow $\Theta \neq 0$ which makes a new tower of generators $Q_n$ appear and sets $a=0$.

Let us focus on the algebra cocycles which are non-zero, i.e., $\lambda $, $\mu $, and $\nu $. Denoting the algebra of the group $\mathcal{G}$ by $\mathfrak{g}$, the centrally extended algebra is given by $\hat{\mathfrak{g}}=\mathfrak{g}\oplus \mathbb{R}^3$. Writing a generic element as $(\epsilon ,\rho ,\zeta ;\mathbf{z})$ with $\mathbf{z}=z_1\,\mathbf{a}_1+z_2\,\mathbf{a}_2+z_3\,\mathbf{a}_3 \in \mathbb{R}^3$ for properly normalized basis vectors $\mathbf{a}_i$ the commutation relations are of the form 
\begin{align}
    &[(\epsilon _1,\rho _1,\zeta _1;\mathbf{z}_1),(\epsilon _2,\rho _2,\zeta _2;\mathbf{z}_2)]=\big([\epsilon _1,\epsilon _2],\epsilon _1\rho _2'-\epsilon _2\rho _1',(\epsilon _1\zeta _2)'-(\epsilon _2\zeta _1)'; \mathbf{z}_{12}\big)
\end{align}
where the three algebra cocycles are given by
\begin{align}
\mathbf{z}_{12}=
    \mathbf{a}_1\frac{1 }{\beta }\int\limits_0^\beta \dd \tau \; (\epsilon _1\zeta _2'-\epsilon _2\zeta _1')
    &+\mathbf{a}_2\frac{1 }{\beta }\int\limits_0^\beta \dd \tau \left(\zeta _1\rho _2-\zeta _2\rho _1\right) \\
    & +\mathbf{a}_3  \frac{1 }{\beta }\int\limits_0^\beta \dd \tau \left(\rho _1\rho _2'-\rho _2\rho _1'\right)~.
\end{align}
It is possible to integrate these to the corresponding group cocycles (assuming a sufficiently connected group, see \cite{Oblak:2016eij}). For two group elements $(x,y)=(e^{tX},e^{sY})\big \vert _{s,t=1}$ with corresponding algebra elements $(X,Y)$ one can relate a Lie algebra cocycle $c(X,Y)$ to the Lie group cocycle $C(x,y)$ by
\begin{align}
    c(X,Y)=-\frac{\dd ^2}{\dd t \; \dd s}\left(C(e^{tX},e^{sY})-C(e^{sY},e^{tX})\right)\Big \vert _{t,s=0} ~.
\end{align}
Applying this to the three cocycles from before yields 
\begin{align}
    C_\lambda (f_1,h_2)&=\frac{1 }{\beta }\int\limits_0^\beta \, \dd \tau \, \log (f_1'(\tau ))h_2(\tau )\\
    C_\mu (f_1,g_1,h_2)&=\frac{1 }{\beta }\int\limits_0^\beta \, \dd \tau \, g_1(\tau )\big (h_2\circ f_1^{-1}\big ) (f_1^{-1})'\\
    C_\nu (f_1,g_1,g_2)&=\frac{1 }{\beta }\int\limits_0^\beta \dd \tau \, g_1'(\tau )(g_2\circ f_1^{-1}) ~.
\end{align}
The first and third are known from the literature \cite{Afshar:2015wjm, Afshar:2021qvi} and the second one is new. From the non-triviality of the algebra cocycles, it then follows that these group cocycles are non-trivial as well. 
The centrally extended group $\hat{\mathcal{G}}=\mathcal{G}\times \mathbb{R}^3$ has the modified group product
\begin{align}
    (f_1,g_1,h_1;\nu _1,\nu _2,\nu _3)&\cdot (f_2,g_2,h_2;\bar{\nu }_1,\bar{\nu }_2,\bar{\nu }_3)=\\[.5em]
    &=\big (f_1\circ f_2,g_1+g_2\circ f_1^{-1},h_1+(f_1^{-1})^\ast h_2\\
    &\hspace{2cm};\nu _1+\bar{\nu }_1+C_\lambda ,\nu _2+\bar{\nu }_2+C_\mu ,\nu _3+\bar{\nu }_3+C_\nu  \big ) \nonumber
\end{align}
that is used to arrive at the adjoint representation 
\begin{align}
    \text{Ad}_{(f,g,h)}(\epsilon ,\rho ,\zeta ;\mathbf{z})&=\frac{\dd }{\dd t}\Big [(f,g,h)\cdot (e^{t\epsilon },t\rho ,t\zeta ;t\mathbf{z})\cdot(f,g,h)^{-1}\Big ]\Big \vert _{t=0} ~.
\end{align}
We use the symbols $\epsilon$ and $\zeta$ for both the components and the tensors themselves from now on as the distinction is clear in each case. 
Also, the central terms in $(f,g,h;\nu_1,\nu_2,\nu_3)$ have been left out here because they only play a passive role. Explicitly we get
\begin{align}
     \text{Ad}_{(f,g,h)}(\epsilon ,\rho ,\zeta ;\mathbf{z})=\Big (f_\ast \epsilon , \rho \circ f^{-1}+\mathcal{L}_{f_\ast \epsilon }g,&(f^{-1})^\ast \zeta +\mathcal{L}_{f_\ast \epsilon }h\\
     &;\mathbf{z}+\tilde{\mathbf{C}}(f,g,h,\epsilon ,\rho ,\zeta )\Big ) \nonumber
\end{align}
where
\begin{align}
    \tilde{C}_1&=\frac{1}{\beta }\int\limits_0^\beta \dd \tau \Big (\zeta \log (f')+\epsilon f'(h\circ f)'\Big)\\
    \tilde{C}_2&=\frac{1 }{\beta }\int\limits_0^\beta \dd \tau \Big (\zeta (g\circ f)-\rho (h\circ f)f'-\epsilon (g\circ f)'(h\circ f)f'\Big ) \\
    \Tilde{C}_3&=\frac{1 }{\beta }\int\limits_0^\beta \dd \tau \,2\rho \big (g\circ f\big)'+\epsilon \Big [(g\circ f)'\Big ]^2~.
\end{align}
As a crosscheck, we compute the differential of the adjoint action at the identity which reproduces the bracket relations of the centrally extended algebra,
\begin{align}
[(\epsilon _1,\rho _1,\zeta _1;\mathbf{z}_1),(\epsilon _2,\rho _2,\zeta _2;\mathbf{z}_2)]&=-\frac{\dd }{\dd s}\text{Ad}_{(e^{s\epsilon _1},s\rho _1,s\zeta _1;s\mathbf{z}_1)}\big (\epsilon _2,\rho _2,\zeta _2;\mathbf{z}_2\big )\Big \vert _{s=0}\\
&=:\text{ad}_{(\epsilon _1,\rho _1,\zeta _1;\mathbf{z}_1)}(\epsilon _2,\rho _2,\zeta _2;\mathbf{z}_2) 
\end{align}
indeed matching with \eqref{eq:algebra} for $c=\kappa =a=0$ upon expanding in Fourier modes.
Let us now define a pairing between this algebra and its dual by
\begin{align}
    \big \langle (\mathcal{T},\mathcal{P},\Theta ;\mathbf{c}),(\epsilon ,\rho ,\zeta ;\mathbf{z})\big \rangle =\frac{1}{\beta }\int\limits_0^\beta \dd \tau \big (\mathcal{T}\epsilon+\mathcal{P} \rho +\Theta \zeta \big )+c_iz_i ~.
\end{align}
The quantities $(\mathcal{T},\mathcal{P},\Theta )$ are the components of appropriate densities with weights $(2,1,0)$, respectively. The dual vector $\mathbf{c}$ contains the central charges. The coadjoint action of the group is defined by 
\begin{align}
    \big \langle \text{Ad}^\ast _{(f,g,h)}(\mathcal{T},\mathcal{P},\Theta ;\mathbf{c}),(\epsilon ,\rho ,\zeta ;\mathbf{z})\big \rangle =\big \langle (\mathcal{T},\mathcal{P},\Theta ;\mathbf{c}),\text{Ad}_{(f,g,h)^{-1}}(\epsilon ,\rho ,\zeta ;\mathbf{z})\big \rangle 
\end{align}
such that it leaves the central terms invariant. The coadjoint action is then determined as
\begin{align}
    \Tilde{\mathcal{T}}(\tau )&=(f^{-1})'^2\big (\mathcal{T}\circ f^{-1}-(\mathcal{P}\circ f^{-1})(g\circ f)'\circ f^{-1}+(\Theta 'f')\circ f^{-1}h\big )\\[.5em]
    &\quad -c_1 \Big (h'-h\frac{(f^{-1})''}{(f^{-1})'}\Big )-c_2 g'h+c_3 (g')^2 \nonumber \\[.5em]
    \Tilde{\mathcal{P}}(\tau )&=(f^{-1})'\,\mathcal{P}\circ f^{-1}+c_2 h-2c_3 g'\\[.5em]
    \Tilde{\Theta }(\tau )&=\Theta \circ f^{-1} +c_1 \log (f^{-1})'-c_2 g~.
\end{align}
It is sometimes convenient to rewrite this evaluated on $f(\tau)$,
\begin{align}
    \Tilde{\mathcal{T}}(f(\tau ))&=\frac{1}{(f')^2}\Big (\mathcal{T}-\mathcal{P}(g\circ f)'+\Theta 'f'(h\circ f)\label{eq:coadj1}\\
    &\qquad \qquad-c_1 \Big ((h\circ f)f'\Big )'-c_2 f'(g\circ f)'(h\circ f)+c_3 (g\circ f)'^2  \Big )\nonumber \\
    \Tilde{\mathcal{P}}(f(\tau ))&=\frac{1}{f'}\Big (\mathcal{P}+ c_2 (h\circ f)f'-2c_3 (g\circ f)'\Big )\label{eq:coadj2}\\
    \Tilde{\Theta }(f(\tau ))&=\Theta -c_1 \ln f'-c_2 g\circ f\label{eq:coadj3} ~.
\end{align}
Infinitesimally this is 
\begin{align}
   \text{ad}^\ast _{(\epsilon ,\rho ,\zeta ;\mathbf{z})}\mathcal{T}&=\epsilon \mathcal{T}'+2\epsilon '\mathcal{T}+\mathcal{P}\rho '-\Theta '\zeta  +c_1 \zeta '\\
  \text{ad}^\ast _{(\epsilon ,\rho ,\zeta ;\mathbf{z})}\mathcal{P}&=\epsilon \mathcal{P}'+\epsilon '\mathcal{P} -c_2 \zeta +2c_3 \rho '\\
  \text{ad}^\ast _{(\epsilon ,\rho ,\zeta ;\mathbf{z})}\Theta &=\epsilon \Theta '+c_1 \epsilon '+c_2 \rho 
\end{align}
which matches with the gravitational transformations \eqref{eq:trafo1}-\eqref{eq:trafo3} if
\begin{align}\label{eq:central_ch}
    c_1 =1 && c_2 =1 && c_3 =-\frac{1}{2} ~.
\end{align}

\providecommand{\href}[2]{#2}\begingroup\raggedright\endgroup


\begin{thebibliography}{10}

\bibitem{Maldacena:2016hyu}
J.~Maldacena and D.~Stanford, \emph{{Remarks on the Sachdev-Ye-Kitaev model}},
  \href{https://doi.org/10.1103/PhysRevD.94.106002}{\emph{Phys. Rev. D}
  {\bfseries 94} (2016) 106002}
  [\href{https://arxiv.org/abs/1604.07818}{{\ttfamily 1604.07818}}].

\bibitem{Maldacena:2016upp}
J.~Maldacena, D.~Stanford and Z.~Yang, \emph{{Conformal symmetry and its
  breaking in two dimensional Nearly Anti-de-Sitter space}},
  \href{https://doi.org/10.1093/ptep/ptw124}{\emph{PTEP} {\bfseries 2016}
  (2016) 12C104} [\href{https://arxiv.org/abs/1606.01857}{{\ttfamily
  1606.01857}}].

\bibitem{Jensen:2016pah}
K.~Jensen, \emph{{Chaos in AdS$_2$ Holography}},
  \href{https://doi.org/10.1103/PhysRevLett.117.111601}{\emph{Phys. Rev. Lett.}
  {\bfseries 117} (2016) 111601}
  [\href{https://arxiv.org/abs/1605.06098}{{\ttfamily 1605.06098}}].

\bibitem{Sachdev:1992fk}
S.~Sachdev and J.~Ye, \emph{{Gapless spin fluid ground state in a random,
  quantum Heisenberg magnet}},
  \href{https://doi.org/10.1103/PhysRevLett.70.3339}{\emph{Phys. Rev. Lett.}
  {\bfseries 70} (1993) 3339}
  [\href{https://arxiv.org/abs/cond-mat/9212030}{{\ttfamily
  cond-mat/9212030}}].

\bibitem{Kitaev:15ur}
A.~Kitaev, ``{A simple model of quantum holography}.'' \href{http://online.kitp.ucsb.edu/online/entangled15/}{http://online.kitp.ucsb.edu/online/entangled15/} and \href{http://online.kitp.ucsb.edu/online/entangled15/kitaev2/}{http://online.kitp.ucsb.edu/online/entangled15/kitaev2/}
.

\bibitem{Saad:2019lba}
P.~Saad, S.H.~Shenker and D.~Stanford, \emph{{JT gravity as a matrix
  integral}},  \href{https://arxiv.org/abs/1903.11115}{{\ttfamily 1903.11115}}.

\bibitem{Jackiw:1984}
R.~Jackiw, \emph{{Liouville field theory: A two-dimensional model for
  gravity?}},  in \emph{Quantum Theory Of Gravity}, S.~Christensen, ed.,
  (Bristol), pp.~403--420, Adam Hilger (1984).

\bibitem{Teitelboim:1983ux}
C.~Teitelboim, \emph{Gravitation and {H}amiltonian structure in two space-time
  dimensions}, {\emph{Phys. Lett.} {\bfseries B126} (1983) 41}.

\bibitem{Grumiller:2002nm}
D.~Grumiller, W.~Kummer and D.V.~Vassilevich, \emph{Dilaton gravity in two
  dimensions}, {\emph{Phys. Rept.} {\bfseries 369} (2002) 327}
  [\href{https://arxiv.org/abs/http://arXiv.org/abs/hep-th/0204253}{{\ttfamily
  http://arXiv.org/abs/hep-th/0204253}}].

\bibitem{Ikeda:1993fh}
N.~Ikeda, \emph{{Two-dimensional gravity and nonlinear gauge theory}},
  \href{https://doi.org/10.1006/aphy.1994.1104}{\emph{Annals Phys.} {\bfseries
  235} (1994) 435} [\href{https://arxiv.org/abs/hep-th/9312059}{{\ttfamily
  hep-th/9312059}}].

\bibitem{Schaller:1994es}
P.~Schaller and T.~Strobl, \emph{Poisson structure induced (topological) field
  theories}, {\emph{Mod. Phys. Lett.} {\bfseries A9} (1994) 3129}
  [\href{https://arxiv.org/abs/http://arXiv.org/abs/hep-th/9405110}{{\ttfamily
  http://arXiv.org/abs/hep-th/9405110}}].

\bibitem{Izawa:1999ib}
K.I.~Izawa, \emph{On nonlinear gauge theory from a deformation theory
  perspective}, {\emph{Prog. Theor. Phys.} {\bfseries 103} (2000) 225}
  [\href{https://arxiv.org/abs/http://arXiv.org/abs/hep-th/9910133}{{\ttfamily
  http://arXiv.org/abs/hep-th/9910133}}].

\bibitem{Grumiller:2021cwg}
D.~Grumiller, R.~Ruzziconi and C.~Zwikel, \emph{{Generalized dilaton gravity in
  2d}}, \href{https://doi.org/10.21468/SciPostPhys.12.1.032}{\emph{SciPost
  Phys.} {\bfseries 12} (2022) 032}
  [\href{https://arxiv.org/abs/2109.03266}{{\ttfamily 2109.03266}}].

\bibitem{Valcarcel:2022zqm}
C.~Valc\'arcel and D.~Vassilevich, \emph{{Target space diffeomorphisms in
  Poisson sigma models and asymptotic symmetries in 2D dilaton gravities}},
  \href{https://doi.org/10.1103/PhysRevD.105.106016}{\emph{Phys. Rev. D}
  {\bfseries 105} (2022) 106016}
  [\href{https://arxiv.org/abs/2202.02603}{{\ttfamily 2202.02603}}].


\bibitem{Callan:1992rs}
C.G.~Callan, Jr., S.B.~Giddings, J.A.~Harvey and A.~Strominger,
  \emph{Evanescent black holes}, {\emph{Phys. Rev.} {\bfseries D45} (1992)
  1005} [\href{https://arxiv.org/abs/hep-th/9111056}{{\ttfamily
  hep-th/9111056}}].



\bibitem{Henneaux:1992}
M.~Henneaux and C.~Teitelboim, \emph{{Quantization of Gauge Systems}},
  Princeton University Press, Princeton, New Jersey (1992).

\bibitem{Barnich:2010xq}
G.~Barnich, \emph{{A Note on gauge systems from the point of view of Lie
  algebroids}}, \href{https://doi.org/10.1063/1.3527427}{\emph{AIP Conf. Proc.}
  {\bfseries 1307} (2010) 7} [\href{https://arxiv.org/abs/1010.0899}{{\ttfamily
  1010.0899}}].

\bibitem{Bojowald:2003pz}
M.~Bojowald and T.~Strobl, \emph{{Classical solutions for Poisson sigma models
  on a Riemann surface}}, {\emph{JHEP} {\bfseries 07} (2003) 002}
  [\href{https://arxiv.org/abs/hep-th/0304252}{{\ttfamily hep-th/0304252}}].

\bibitem{Fukuyama:1985gg}
T.~Fukuyama and K.~Kamimura, \emph{{Gauge Theory of Two-dimensional Gravity}},
  \href{https://doi.org/10.1016/0370-2693(85)91322-X}{\emph{Phys. Lett. B}
  {\bfseries 160} (1985) 259}.

\bibitem{Isler:1989hq}
K.~Isler and C.A.~Trugenberger, \emph{A gauge theory of two-dimensional quantum
  gravity}, {\emph{Phys. Rev. Lett.} {\bfseries 63} (1989) 834}.

\bibitem{Chamseddine:1989yz}
A.H.~Chamseddine and D.~Wyler, \emph{Gauge theory of topological gravity in
  (1+1)-dimensions}, {\emph{Phys. Lett.} {\bfseries B228} (1989) 75}.

\bibitem{Ruzziconi:2020wrb}
R.~Ruzziconi and C.~Zwikel, \emph{{Conservation and Integrability in
  Lower-Dimensional Gravity}},
  \href{https://doi.org/10.1007/JHEP04(2021)034}{\emph{JHEP} {\bfseries 04}
  (2021) 034} [\href{https://arxiv.org/abs/2012.03961}{{\ttfamily
  2012.03961}}].

\bibitem{Bergamin:2007sm}
L.~Bergamin, D.~Grumiller, R.~McNees and R.~Meyer, \emph{{Black Hole
  Thermodynamics and Hamilton-Jacobi Counterterm}},
  \href{https://doi.org/10.1088/1751-8113/41/16/164068}{\emph{J. Phys. A}
  {\bfseries 41} (2008) 164068}
  [\href{https://arxiv.org/abs/0710.4140}{{\ttfamily 0710.4140}}].

\bibitem{Bergamin:2005pg}
L.~Bergamin, D.~Grumiller, W.~Kummer and D.V.~Vassilevich,
  \emph{{Physics-to-gauge conversion at black hole horizons}},
  \href{https://doi.org/10.1088/0264-9381/23/9/019}{\emph{Class. Quant. Grav.}
  {\bfseries 23} (2006) 3075}
  [\href{https://arxiv.org/abs/hep-th/0512230}{{\ttfamily hep-th/0512230}}].

\bibitem{Harlow:2019yfa}
D.~Harlow and J.-Q.~Wu, \emph{{Covariant phase space with boundaries}},
  \href{https://doi.org/10.1007/JHEP10(2020)146}{\emph{JHEP} {\bfseries 10}
  (2020) 146} [\href{https://arxiv.org/abs/1906.08616}{{\ttfamily
  1906.08616}}].

\bibitem{Afshar:2021qvi}
H.~Afshar and B.~Oblak, \emph{{Flat JT gravity and the BMS-Schwarzian}},
  \href{https://doi.org/10.1007/JHEP11(2022)172}{\emph{JHEP} {\bfseries 11}
  (2022) 172} [\href{https://arxiv.org/abs/2112.14609}{{\ttfamily
  2112.14609}}].

\bibitem{Afshar:2015wjm}
H.~Afshar, S.~Detournay, D.~Grumiller and B.~Oblak, \emph{{Near-Horizon
  Geometry and Warped Conformal Symmetry}},
  \href{https://doi.org/10.1007/JHEP03(2016)187}{\emph{JHEP} {\bfseries 03}
  (2016) 187} [\href{https://arxiv.org/abs/1512.08233}{{\ttfamily
  1512.08233}}].

\bibitem{Godet:2020xpk}
V.~Godet and C.~Marteau, \emph{{New boundary conditions for AdS$_{2}$}},
  \href{https://doi.org/10.1007/JHEP12(2020)020}{\emph{JHEP} {\bfseries 12}
  (2020) 020} [\href{https://arxiv.org/abs/2005.08999}{{\ttfamily
  2005.08999}}].

\bibitem{Hofman:2011zj}
D.M.~Hofman and A.~Strominger, \emph{{Chiral Scale and Conformal Invariance in
  2D Quantum Field Theory}},
  \href{https://doi.org/10.1103/PhysRevLett.107.161601}{\emph{Phys. Rev. Lett.}
  {\bfseries 107} (2011) 161601}
  [\href{https://arxiv.org/abs/1107.2917}{{\ttfamily 1107.2917}}].

\bibitem{Detournay:2012pc}
S.~Detournay, T.~Hartman and D.M.~Hofman, \emph{{Warped Conformal Field
  Theory}}, \href{https://doi.org/10.1103/PhysRevD.86.124018}{\emph{Phys.Rev.}
  {\bfseries D86} (2012) 124018}
  [\href{https://arxiv.org/abs/1210.0539}{{\ttfamily 1210.0539}}].

\bibitem{Afshar:2019tvp}
H.R.~Afshar, \emph{{Warped Schwarzian theory}},
  \href{https://doi.org/10.1007/JHEP02(2020)126}{\emph{JHEP} {\bfseries 02}
  (2020) 126} [\href{https://arxiv.org/abs/1908.08089}{{\ttfamily
  1908.08089}}].

\bibitem{Grumiller:2017qao}
D.~Grumiller, R.~McNees, J.~Salzer, C.~Valc\'arcel and D.~Vassilevich,
  \emph{{Menagerie of AdS$_{2}$ boundary conditions}},
  \href{https://doi.org/10.1007/JHEP10(2017)203}{\emph{JHEP} {\bfseries 10}
  (2017) 203} [\href{https://arxiv.org/abs/1708.08471}{{\ttfamily
  1708.08471}}].


\bibitem{Grumiller:2019fmp}
D.~Grumiller, A.~P\'erez, M.M.~Sheikh-Jabbari, R.~Troncoso and C.~Zwikel,
  \emph{{Spacetime structure near generic horizons and soft hair}},
  \href{https://doi.org/10.1103/PhysRevLett.124.041601}{\emph{Phys. Rev. Lett.}
  {\bfseries 124} (2020) 041601}
  [\href{https://arxiv.org/abs/1908.09833}{{\ttfamily 1908.09833}}].

\bibitem{Adami:2020ugu}
H.~Adami, M.M.~Sheikh-Jabbari, V.~Taghiloo, H.~Yavartanoo and C.~Zwikel,
  \emph{{Symmetries at null boundaries: two and three dimensional gravity
  cases}}, \href{https://doi.org/10.1007/JHEP10(2020)107}{\emph{JHEP}
  {\bfseries 10} (2020) 107}
  [\href{https://arxiv.org/abs/2007.12759}{{\ttfamily 2007.12759}}].

\bibitem{Adami:2021nnf}
H.~Adami, D.~Grumiller, M.M.~Sheikh-Jabbari, V.~Taghiloo, H.~Yavartanoo and
  C.~Zwikel, \emph{{Null boundary phase space: slicings, news \& memory}},
  \href{https://doi.org/10.1007/JHEP11(2021)155}{\emph{JHEP} {\bfseries 11}
  (2021) 155} [\href{https://arxiv.org/abs/2110.04218}{{\ttfamily
  2110.04218}}].

\bibitem{Adami:2022ktn}
H.~Adami, P.~Mao, M.M.~Sheikh-Jabbari, V.~Taghiloo and H.~Yavartanoo,
  \emph{{Symmetries at causal boundaries in 2D and 3D gravity}},
  \href{https://doi.org/10.1007/JHEP05(2022)189}{\emph{JHEP} {\bfseries 05}
  (2022) 189} [\href{https://arxiv.org/abs/2202.12129}{{\ttfamily
  2202.12129}}].

\bibitem{Geiller:2021vpg}
M.~Geiller, C.~Goeller and C.~Zwikel, \emph{{3d gravity in Bondi-Weyl gauge:
  charges, corners, and integrability}},
  \href{https://doi.org/10.1007/JHEP09(2021)029}{\emph{JHEP} {\bfseries 09}
  (2021) 029} [\href{https://arxiv.org/abs/2107.01073}{{\ttfamily
  2107.01073}}].

\bibitem{Crainic:2021}
M.~Crainic, R.L.~Fernandes and I.~M\u{a}rcu\c{t}, \emph{{Lectures on Poisson
  Geometry}}, American Mathematical Society, Providence (2021).

\bibitem{Cattaneo:2000iw}
A.S.~Cattaneo and G.~Felder, \emph{{Poisson sigma models and symplectic
  groupoids}}, \href{https://doi.org/10.1007/978-3-0348-8364-1_4}{\emph{Prog.
  Math.} {\bfseries 198} (2000) 61}
  [\href{https://arxiv.org/abs/math/0003023}{{\ttfamily math/0003023}}].

\bibitem{Baulieu:2001fi}
L.~Baulieu, A.S.~Losev and N.A.~Nekrasov, \emph{{Target space symmetries in
  topological theories. 1.}},
  \href{https://doi.org/10.1088/1126-6708/2002/02/021}{\emph{JHEP} {\bfseries
  02} (2002) 021} [\href{https://arxiv.org/abs/hep-th/0106042}{{\ttfamily
  hep-th/0106042}}].

\bibitem{Salzer:2018zlv}
J.~Salzer, \emph{{Asymptotic dynamics of two-dimensional dilaton gravity}},
  Ph.D. thesis, Vienna, Tech. U., 2018.

\bibitem{Liebl:1996ti}
H.~Liebl, D.V.~Vassilevich and S.~Alexandrov, \emph{{Hawking radiation and
  masses in generalized dilaton theories}},
  \href{https://doi.org/10.1088/0264-9381/14/4/007}{\emph{Class. Quant. Grav.}
  {\bfseries 14} (1997) 889}
  [\href{https://arxiv.org/abs/gr-qc/9605044}{{\ttfamily gr-qc/9605044}}].

\bibitem{Witten:2020wvy}
E.~Witten, \emph{{Matrix Models and Deformations of JT Gravity}},
  \href{https://doi.org/10.1098/rspa.2020.0582}{\emph{Proc. Roy. Soc. Lond. A}
  {\bfseries 476} (2020) 20200582}
  [\href{https://arxiv.org/abs/2006.13414}{{\ttfamily 2006.13414}}].

\bibitem{Momeni:2020tyt}
D.~Momeni, \emph{{Real classical geometry with arbitrary deficit parameter(s)
  $\alpha (_{I})$ in deformed Jackiw\textendash{}Teitelboim gravity}},
  \href{https://doi.org/10.1140/epjc/s10052-021-08985-1}{\emph{Eur. Phys. J. C}
  {\bfseries 81} (2021) 202}
  [\href{https://arxiv.org/abs/2010.00377}{{\ttfamily 2010.00377}}].

\bibitem{Johnson:2020lns}
C.V.~Johnson and F.~Rosso, \emph{{Solving Puzzles in Deformed JT Gravity: Phase
  Transitions and Non-Perturbative Effects}},
  \href{https://doi.org/10.1007/JHEP04(2021)030}{\emph{JHEP} {\bfseries 04}
  (2021) 030} [\href{https://arxiv.org/abs/2011.06026}{{\ttfamily
  2011.06026}}].

\bibitem{Turiaci:2020fjj}
G.J.~Turiaci, M.~Usatyuk and W.W.~Weng, \emph{{2D dilaton-gravity, deformations
  of the minimal string, and matrix models}},
  \href{https://doi.org/10.1088/1361-6382/ac25df}{\emph{Class. Quant. Grav.}
  {\bfseries 38} (2021) 204001}
  [\href{https://arxiv.org/abs/2011.06038}{{\ttfamily 2011.06038}}].

\bibitem{Alishahiha:2020jko}
M.~Alishahiha, A.~Faraji~Astaneh, G.~Jafari, A.~Naseh and B.~Taghavi,
  \emph{{Free energy for deformed Jackiw-Teitelboim gravity}},
  \href{https://doi.org/10.1103/PhysRevD.103.046005}{\emph{Phys. Rev. D}
  {\bfseries 103} (2021) 046005}
  [\href{https://arxiv.org/abs/2010.02016}{{\ttfamily 2010.02016}}].

\bibitem{Bergamin:2004pn}
L.~Bergamin, D.~Grumiller, W.~Kummer and D.V.~Vassilevich, \emph{{Classical and
  quantum integrability of 2-D dilaton gravities in Euclidean space}},
  \href{https://doi.org/10.1088/0264-9381/22/7/010}{\emph{Class. Quant. Grav.}
  {\bfseries 22} (2005) 1361}
  [\href{https://arxiv.org/abs/hep-th/0412007}{{\ttfamily hep-th/0412007}}].

\bibitem{Oblak:2016eij}
B.~Oblak, \emph{{BMS Particles in Three Dimensions}},
  \href{https://arxiv.org/abs/1610.08526}{{\ttfamily 1610.08526}}.

\bibitem{Mandal:1991tz}
G.~Mandal, A.M.~Sengupta and S.R.~Wadia, 
  \emph{Classical solutions of two-dimensional string theory}, \href{https://doi.org/10.1142/S0217732391001822}{{\emph{Mod. Phys. Lett. A} {\bfseries 6} (1991)
  1685--1692}}.

\bibitem{Elitzur:1991cb}
S.~Elitzur, A.~Forge and E.~Rabinovici,
  \emph{Some global aspects of string compactifications}, \href{https://doi.org/10.1016/0550-3213(91)90073-7}{{\emph{Nucl. Phys. B} {\bfseries 359} (1991)
  581--610}}.

 \bibitem{Witten:1991yr}
E.~Witten,
  \emph{On string theory and black holes}, \href{https://doi.org/10.1103/PhysRevD.44.314}{{\emph{Phys. Rev. D} {\bfseries 44} (1991)
  1005}}. 


\bibitem{Grumiller:2007ju}
D.~Grumiller and R.~McNees, \emph{Thermodynamics of black holes in two (and
  higher) dimensions}, {\emph{JHEP} {\bfseries 04} (2007) 074}
  [\href{https://arxiv.org/abs/hep-th/0703230}{{\ttfamily hep-th/0703230}}].

\bibitem{Brown:1986nw}
J.D.~Brown and M.~Henneaux, \emph{{Central Charges in the Canonical Realization
  of Asymptotic Symmetries: An Example from Three-Dimensional Gravity}},
  {\emph{Commun. Math. Phys.} {\bfseries 104} (1986) 207}.

\bibitem{Afshar:2019axx}
H.~Afshar, H.A.~Gonz\'alez, D.~Grumiller and D.~Vassilevich, \emph{{Flat space
  holography and the complex Sachdev-Ye-Kitaev model}},
  \href{https://doi.org/10.1103/PhysRevD.101.086024}{\emph{Phys. Rev. D}
  {\bfseries 101} (2020) 086024}
  [\href{https://arxiv.org/abs/1911.05739}{{\ttfamily 1911.05739}}].

\bibitem{Joung:2023doq}
E.~Joung, P.~Narayan and J.~Yoon, \emph{{Gravitational Edge Mode in
  Asymptotically AdS$_2$: JT Gravity Revisited}},
  \href{https://arxiv.org/abs/2304.06088}{{\ttfamily 2304.06088}}.

\bibitem{Grumiller:2015vaa}
D.~Grumiller, J.~Salzer and D.~Vassilevich, \emph{{AdS$_{2}$ holography is
  (non-)trivial for (non-)constant dilaton}},
  \href{https://doi.org/10.1007/JHEP12(2015)015}{\emph{JHEP} {\bfseries 12}
  (2015) 015} [\href{https://arxiv.org/abs/1509.08486}{{\ttfamily
  1509.08486}}].

\bibitem{Cvetic:2016eiv}
M.~Cveti\v{c} and I.~Papadimitriou, \emph{{AdS$_{2}$ holographic dictionary}},
  \href{https://doi.org/10.1007/JHEP12(2016)008}{\emph{JHEP} {\bfseries 12}
  (2016) 008} [\href{https://arxiv.org/abs/1608.07018}{{\ttfamily
  1608.07018}}].

\bibitem{Sarosi:2017ykf}
G.~S\'arosi, \emph{{AdS$_{2}$ holography and the SYK model}},
  \href{https://doi.org/10.22323/1.323.0001}{\emph{PoS} {\bfseries Modave2017}
  (2018) 001} [\href{https://arxiv.org/abs/1711.08482}{{\ttfamily
  1711.08482}}].

\bibitem{Kar:2022sdc}
A.~Kar, L.~Lamprou, C.~Marteau and F.~Rosso, \emph{{A Matrix Model for Flat
  Space Quantum Gravity}},  \href{https://arxiv.org/abs/2208.05974}{{\ttfamily
  2208.05974}}.

\bibitem{Fan:2021bwt}
Y.~Fan and T.G.~Mertens, \emph{{From quantum groups to Liouville and dilaton
  quantum gravity}}, \href{https://doi.org/10.1007/JHEP05(2022)092}{\emph{JHEP}
  {\bfseries 05} (2022) 092}
  [\href{https://arxiv.org/abs/2109.07770}{{\ttfamily 2109.07770}}].

\bibitem{Stanford:2017thb}
D.~Stanford and E.~Witten, \emph{{Fermionic Localization of the Schwarzian
  Theory}}, \href{https://doi.org/10.1007/JHEP10(2017)008}{\emph{JHEP}
  {\bfseries 10} (2017) 008}
  [\href{https://arxiv.org/abs/1703.04612}{{\ttfamily 1703.04612}}].

\bibitem{Fiorucci:2021pha}
A.~Fiorucci, \emph{{Leaky covariant phase spaces: Theory and application to
  $\Lambda$-BMS symmetry}}, Ph.D. thesis, Brussels U., Intl. Solvay Inst.,
  Brussels, 2021.
\newblock \href{https://arxiv.org/abs/2112.07666}{{\ttfamily 2112.07666}}.

\end{thebibliography}
\end{document}